\documentclass[12pt]{article}
\usepackage{jheppub}
\usepackage{physics}
\usepackage{comment}
\usepackage{slashed}

\newcommand{\pic}[2]{\vcenter{\hbox{\includegraphics[height=#1]{#2}}}}

\graphicspath{ {./images/} }
\newcommand{\pick}[2]{\vcenter{\hbox{\includegraphics[width=#1]{#2}}}}

\usepackage{soul}

\newcommand{\p}{\partial}
\newcommand{\calI}{\mathcal{I}}

\usepackage{tikz,tikz-3dplot}
\usepackage{pgfplots}
\pgfplotsset{compat=1.17}
\usetikzlibrary{shapes,arrows,cd,chains,decorations.markings,decorations.pathmorphing,calc}
\tikzset{
	->-/.style args={#1rotate#2}{decoration={markings, mark=at position #1 with {\arrow[scale=1.5,rotate = #2 ]{stealth}}}, postaction={decorate}}
}
\tikzstyle{GraphNode}=[circle, draw=black, fill=black, inner sep=2pt, minimum size=5pt]
\tikzstyle{GraphEdge}=[black]
\pgfmathsetmacro{\gS}{1}

\title{Feynman Diagrams in Four-Dimensional Holomorphic Theories and the Operatope}
\abstract{We study a class of universal Feynman integrals which appear in four-dimensional holomorphic theories. We recast the integrals as the Fourier transform of a certain polytope in the space of loop momenta (aka the ``Operatope''). We derive a set of quadratic recursion relations which appear to fully determine the final answer. Our strategy can be applied to a very general class of twisted supersymmetric quantum field theories.}

\author[\pick{1.5ex}{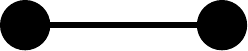},\pic{1.2ex}{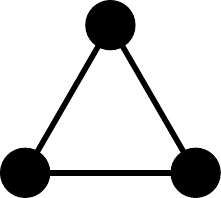}]{Kasia Budzik,}
\author[\pick{1.5ex}{segment}]{Davide Gaiotto,}
\author[\pick{1.5ex}{segment},\pic{1.2ex}{triangle}]{Justin Kulp,}
\author[\pic{1ex}{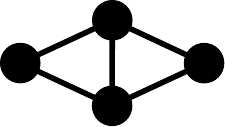}]{Jingxiang Wu,}
\author[\pick{1.5ex}{segment},\pic{1.2ex}{triangle}]{and Matthew Yu}

\affiliation[\pick{1.5ex}{segment}]{Perimeter Institute for Theoretical Physics, Waterloo, ON N2L 2Y5, Canada}
\affiliation[\pic{1.2ex}{triangle}]{Department of Physics \& Astronomy, University of Waterloo, Waterloo, ON N2L 3G1, Canada}
\affiliation[\pic{1ex}{bitriangle}]{Mathematical Institute, University of Oxford,
Andrew-Wiles Building, Woodstock Road, Oxford, OX2 6GG, UK}

\emailAdd{kbudzik@perimeterinstitute.ca}
\emailAdd{dgaiotto@perimeterinstitute.ca}
\emailAdd{jkulp@perimeterinstitute.ca}
\emailAdd{Jingxiang.Wu@maths.ox.ac.uk}
\emailAdd{myu@perimeterinstitute.ca}

\begin{document}
\maketitle

\section{Introduction}

The main subject of this note is a class of four-dimensional holomorphic quantum field theories which emerge from the holomorphic twist of ${\cal N}=1$ supersymmetric gauge theories \cite{Johansen:1994aw, Costello:2011np} (see also \cite{nik,Losev:1995cr,Losev:1996up}). Local operators in these holomorphic theories are endowed with the structure of a holomorphic factorization algebra: a higher-dimensional analogue of the vertex algebras which occur in two-dimensional holomorphic theories \cite{Costello:2016vjw,Williams:2018ows,Gwilliam:2018lpo,Saberi:2019ghy}. 

The holomorphic factorization algebra includes a BRST differential as well as a collection of operations defined with the help of a descent procedure \cite{Beem:2018fng,Oh:2019mcg}. A companion paper \cite{factor} proposes a perturbative Feynman diagram expansion for these structures.\footnote{An alternative perspective is that the Feynman integrals simply compute the holomorphic factorization algebra structure for the free theory. Indeed, the factorization algebra includes infinite towers of operations which control the perturbative deformation of the algebra itself.} As usual, the contribution of a Feynman diagram combines theory-specific combinatorial data, such as field content and a collection of interaction vertices, with some universal Feynman integrals. In this paper we study these integrals in detail.

The operations which arise from the factorization algebra satisfy certain associativity axioms. For example, the BRST differential should be nilpotent. One of our main results is to derive an infinite collection of quadratic relations which are satisfied by the Feynman integrals under consideration. These relations imply that the integrals do indeed provide the coefficients for a well-defined factorization algebra structure, as long as the interactions satisfy a collection of anomaly-cancellation conditions. 

The separation of the factorization algebra data into theory-specific combinatorial structures and universal coefficients which satisfy appropriate quadratic relations is strongly analogous to what happens in two-dimensional TQFT examples \cite{Kontsevich:1997vb,Gaiotto:2015zna, Gaiotto:2015aoa} which inspired much of our analysis. Although the details are different, the quadratic relations in this paper and in these works follow from the same underlying geometric structure, which we dub the ``Operatope.'' 

As a bonus, we will find evidence that the quadratic relations can be employed to evaluate the Feynman integrals recursively, providing an alternative to direct integration. Although the details of our analysis are specific to four-dimensional holomorphic systems, our computational strategy applies to a wider class of examples. 

The Feynman diagrams which we study in this note have the following noteworthy features:
\begin{itemize}
	\item The external momenta associated to a vertex $v$, which we will often denote as $\lambda_v$, are purely holomorphic. 
	\item The arguments of the position space propagators, labelled by edges $e$ of a graph, include certain extra holomorphic shifts $z_e$. In applications, these allow us to study local operators which are formally bi-local or multi-local as functions of the holomorphic coordinates and play the role of generating functions for infinite towers of local operators built out of an arbitrary number of holomorphic derivatives. 
	\item The superfields we employ can be canonically identified with differential forms, as the superspace coordinates have the same quantum numbers as the anti-holomorphic coordinates on spacetime. As a consequence, manipulations involving supergraphs map to standard manipulations of differential forms. In particular, the equations of motion involve the $\bar \partial$ differential operator acting on forms, and the (super-)propagators invert $\bar \partial$.
	\item The Feynman diagrams we are interested in involve a sum over all possible ways to ``cut'' a propagator in a more conventional Feynman diagram, by acting with $\bar \partial$ on it. We implement and simplify this sum using a trick involving differential forms on the space of Schwinger parameters.
	\item Although we need IR and UV regulators at intermediate steps of the calculation, the final answer is finite and does not require counterterms. 
\end{itemize}

We denote the final answer associated to a given Feynman graph $\Gamma$ as ${\cal I}_\Gamma(\lambda_v;z_e)$. This should be interpreted as a generating function, i.e. we are interested in the coefficients of the expansion of ${\cal I}_\Gamma(\lambda_v;z_e)$ as a power series in $z$ and $\lambda$. 

The quadratic relations we derive are also labelled by a graph $\Gamma$. They involve a sum over all possible ways to obtain $\Gamma$ by ``nesting'' a Feynman diagram $\Gamma[S]$ inside a second Feynman diagram $\Gamma(S)$. Here $S$ denotes a collection of vertices of $\Gamma$ which induce the subgraph $\Gamma[S]$, and can be collapsed to obtain $\Gamma(S)$. The relations take the schematic form
\begin{equation}
    \boxed{\hphantom{.}\vphantom{\sum^.} \sum_{S} \sigma(\Gamma,S) \, {\cal I}_{\Gamma[S]}[\lambda + \partial_{z'};z] \, {\cal I}_{\Gamma(S)}[\lambda';z'] = 0 \hphantom{.}} \,\,.
\end{equation}
The precise meaning of these terms will be explained in Section \ref{sec:graphsInsideGraphs}.

\subsection{Structure of the Paper}
Section \ref{sec:defs} contains the definition of the integrand and integration contour for the Feynman diagrams, as well as a coordinate change which trivializes the integrand and makes the integral manifestly finite at the price of making the integration region complicated. Section \ref{sec:graphsInsideGraphs} derives a geometric identity satisfied by the integration regions which implies an infinite collection of quadratic relations for the integrals. Section \ref{sec:Examples} contains examples of calculations up to three loops. We demonstrate the use of the quadratic relations to bootstrap higher loop answers from the explicit 1-loop integral. Section \ref{sec:op} briefly discusses how the Feynman integrals appear in factorization algebra operations.  Section \ref{sec:twod} discusses analogous calculations for other dimensions and twists. Appendix \ref{append:1loopintegrand} collects some extra details on the calculation of the 1-loop triangle Feynman diagram. Appendix \ref{app:gauge} discusses the gauge dependence of our Feynman integrals. 

\section{Definitions and Properties of Feynman Diagrams} \label{sec:defs}
In this section, we introduce the basic ingredients which are required to build the relevant Feynman diagrams for our analysis. We also detail their basic properties. We start by studying the propagator and its ``cut'' form, and combine them to make the integrand manifestly UV-finite. A judicious change of variables then maps the integral to the Fourier transform of a complicated curvilinear polytope in the space of holomorphic loop momenta, aka the Operatope. This sets the stage for a recursive ``bootstrap'' approach for their calculation in Section \ref{sec:graphsInsideGraphs}. We refer the reader to our companion paper \cite{factor} for a full derivation of the Feynman diagrams we introduce in this section. 

\subsection{Basic Definitions}
Very schematically, the superfields $\phi^a$ of the holomorphic theory are functions of a superspace with coordinates $(x^\alpha, \bar x^\alpha, d\bar x^\alpha)$: these are holomorphic and anti-holomorphic coordinates on $\mathbb{R}^4$ as well as odd coordinates $d\bar x^\alpha$. They are thus identified with differential forms with anti-holomorphic indices only. The free equations of motion take the form $\bar \partial \phi^a=0$, and the superspace propagator can be taken to be proportional to the Bochner-Martinelli kernel
\begin{equation}
	P(x, \bar x, d \bar x) \equiv \frac{\bar x^2 d \bar x^1 - \bar x^1 d \bar x^2}{|x|^4} \,,
\end{equation}
where $|x|^2 = x^1 \,\overline{x}^1 + x^2 \,\overline{x}^2 $. The propagator is the Green's function for $\overline{\p} = d\overline{x}^1 \frac{\p}{\p \overline{x}^1}+d\overline{x}^2 \frac{\p}{\p \overline{x}^2}$, i.e. we have
\begin{equation}
   \overline{ \p}	P(x, \bar x, d \bar x) = \pi^2 d\overline{x}^1 d\overline{x}^2  \delta^4(x)\,.
\end{equation}
Momentarily, we will introduce a UV-regulated version $P_\epsilon(x, \bar x, d \bar x)$ of the propagator with a smeared source.

The Feynman graph\footnote{Which we assume is connected, has at least two vertices, and has no edges joining a vertex with itself.} $\Gamma$, represents the pattern of Wick contractions in a given calculation: each vertex represents an operator of the theory (which may be the interaction Lagrangian), and each edge represents a contraction leading to a propagator factor. As the propagator is a 1-form, the overall sign of a product of propagators will depend on the order in which we multiply them. Henceforth, all occurrences of a graph $\Gamma$ will come equipped with some choice of relative order among the edges, so that a permutation of the order of the edges of a graph will change the sign of 
${\cal I}_\Gamma(\lambda_v;z_e)$ by the sign of the permutation.

In addition, we will denote the set of edges and vertices of the graph by $\Gamma_1$ and $\Gamma_0$ respectively. We will denote a generic edge in $\Gamma$ by $e$ and a generic vertex by $v$. We will assign an orientation to edges so that $e(0)$ is the first vertex of the edge $e$, and $e(1)$ is the second. The (wedge) product of two propagators between the same pair of vertices vanishes, so there will be at most a single edge between any two vertices. As a result, we will often refer to an edge by its endpoint vertices $e(0)$ and $e(1)$ since there can be no ambiguity.\footnote{If we pick some ordering of the vertices and orient the edges so that $e(0)<e(1)$, we can assign a lexicographic order to the edges. This is the ordering of edges we typically employ in examples.}

The positions of vertices in spacetime are given by $(x_v, \bar x_v)$ and associated holomorphic momenta are given by $\lambda_v$. The auxiliary holomorphic shifts in the propagator arguments will be denoted by $z_e$, with the caveat that changing the orientation convention for an edge should change the sign of the corresponding shifts, i.e. $z_{v v'} = - z_{v' v}$.

The integrand of our Feynman diagrams contains a total derivative of the form 
\begin{equation}
    \bar \partial \left[ \prod_{e \in \Gamma_1} P_\epsilon[e] \right]\,,
\end{equation}
where $P_\epsilon[e]$ is the regulated propagator associated to the edge $e$ of the graph $\Gamma$. Using the Leibniz rule, this becomes a sum over ``cut'' edges, where the differential acts on the edge which is cut:
\begin{equation} \label{eq:sumcut}
    \sum_{e\in \Gamma_1} (-1)^{e} \bar \partial P_\epsilon[e] \left[ \prod_{e' \in \Gamma_1}^{e' \neq e} P_\epsilon[e'] \right]\,.
\end{equation}
Our first task will be to rearrange this expression into a more convenient form. We will then combine it with a holomorphic measure containing the external momenta $\lambda_v$ and define the full integrand.

\subsection{Propagator Manipulations}
For our propagators we employ a Schwinger parameterization
\begin{equation}
	P_\epsilon(x, \bar x) \equiv \int_{\epsilon}^\infty \frac{dt}{t} \, (\bar x^2 d \bar x^1 - \bar x^1 d \bar x^2) K_t(x)\,,
\end{equation}
using the heat kernel
\begin{equation}
	K_t(x) \equiv  \frac{1}{t^2}  e^{-\frac{|x|^2}{t}} \,.
\end{equation}
The heat kernel satisfies
\begin{equation}
    \Delta K_t(x) = \p_t K_t(x)\,, \quad \Delta = \p_{x^1}\p_{\overline{x}^1}+\p_{x^2}\p_{\overline{x}^2}\,,
\end{equation}
so that if we define the operator 
\begin{equation}
    \overline{\p}^* = d\overline{x}^1 \p_{x^2}-d\overline{x}^2 \p_{x^1}
\end{equation}
such that $\p \overline{\p}^* =  d \overline{x}^1 \wedge  d \overline{x}^2 \,\Delta$\,, then 
\begin{equation}
   P_\epsilon(x,\overline{x}) = \int_\epsilon^\infty dt\, \overline{\p}^* K_t(x)  \,.
\end{equation}
As a result, the ``cut'' propagator is then the heat kernel itself
\begin{equation}
	\bar \partial P_\epsilon(x, \bar x)= K_\epsilon(x)\, d^2 \overline{x} \,,
\end{equation}
where $d^2 \overline{x}= d \bar x^1 \wedge  d \overline{x}^2$. Notice the normalization:
\begin{equation}
    \int_{\mathbb{C}^2} K_t(x) d^2 \overline{x} \frac{d^2 x}{(2 \pi i)^2} = 1\,.
\end{equation}

The propagator and cut propagator can be envisioned respectively as the integral of a 1-form integrated on the $[\epsilon,\infty)$ half-line in Schwinger time, or as the evaluation of a $0$-form at $t=\epsilon$. We can thus combine them into a single form\footnote{This crucial step was suggested to us by K. Costello.}
\begin{equation}\label{eq:propagatorcombined}
	{\cal P}(x, \bar x, t) \equiv  \frac{dt}{t} \, (\bar x^2 d \bar x^1 - \bar x^1 d \bar x^2) K_t(x)  + d^2 \bar x K_t(x) \,,
\end{equation}
and encode the choice between the two into a choice of integration contour in the $t$ space. 

The sum over cut propagators in equation (\ref{eq:sumcut})
can thus be written as an integral of $\prod_{e \in \Gamma_1} {\cal P}[e]$ over a sum of co-dimension 1 contours in the space of Schwinger times:
\begin{equation}
\sum_{e\in \Gamma_1} (-1)^{e} \left[t_e=\epsilon\right] \times \prod_{e' \in \Gamma_1}^{e' \neq e} \left[\epsilon \leq t_e'<\infty \right]\,.
\end{equation}
By also introducing an IR cutoff $L$, we denote the combined integration cycle as the ``UV'' component $\partial_\epsilon [\epsilon,L]^{|\Gamma_1|}$ of the boundary of the hypercube 
$[\epsilon,L]^{|\Gamma_1|}$ in the space of Schwinger parameters, i.e. the union of the facets sitting at $t_e=\epsilon$ for some $e$. 

We also record two useful facts: 
\begin{equation}
	(d_t + \bar \partial) {\cal P}(x, \bar x, t) =0 \,,
\end{equation}
and
\begin{equation}
i_{t \partial_t + \bar x \cdot \bar \partial_x} {\cal P}(x, \bar x, t) =0 \,.
\end{equation}
\subsection{Auxiliary Coordinates}
The combined propagator \eqref{eq:propagatorcombined} can be written concisely as
\begin{equation}\label{eq:simprop}
	{\cal P}(x, \bar x, t) = e^{- y \cdot x} d^2 y \,,
\end{equation}
in terms of auxiliary variables 
\begin{equation}
	y = t^{-1}  \bar x \,.
\end{equation}

This expression will greatly simplify many of our calculations. Indeed, as all the position space integrals are Gaussian, we can easily trade the $\bar x = x^*$ physical integration cycle for an integration cycle with real $\bar x$ and imaginary $x$'s. Then we can trade the intricate functional dependence and form manipulations in the $\bar x$ and $t$ coordinates for much simpler expressions in the $y$ coordinates, at the price of a more intricate integration region.
\subsection{Translation Invariance}
When one considers scattering amplitudes, all vertices of a Feynman diagram are integrated over spacetime, leading to a momentum-conservation delta function. In contrast, our calculations are closer to that of a form factor: one of the vertices is kept at a fixed position, say $x_{v_0}=0$, and all other vertices are integrated over, with an extra measure factor of $e^{\lambda_v \cdot x_v} d^2 x_v$. We thus have a holomorphic momentum $\lambda_v$ for all $v \neq v_0$, and we find it convenient to {\it define} $\lambda_{v_0} \equiv - \sum_{v \neq v_0} \lambda_v$
so that we have a momentum-conservation identity
\begin{equation}\label{eq:balance}
	\sum_v \lambda_v=0\,.
\end{equation}

With this definition, translation symmetry makes the choice of which vertex that is not integrated over immaterial: the Feynman integral for a different choice of fixed vertex, $v_1$ say, will be obtained by replacing 
\begin{equation}
    \lambda_{v_1} = - \sum_{v\neq v_1} \lambda_v
\end{equation}
in the expression for the Feynman integral with fixed $v_0$.\footnote{Indeed, set $x'_v = x_v - x_{v_1}$ so that $x'_{v_1}=0$ and $x'_{v_0}= - x_{v_1}$. Then \eqref{eq:balance} implies that 
\begin{equation}
	\sum_v x_v \lambda_v = \sum_v x'_v \lambda_v\,.
\end{equation}}

\subsection{Balancing Conditions}
Each propagator contributes an anti-holomorphic 1-form to the position integral, while the cut propagator contributes an anti-holomorphic 2-form. The product of all propagators is thus a form of degree $(0,|\Gamma_1|+1)$ in position space, which has to be combined with the holomorphic integration measure 
\begin{equation}
	\prod_{v \neq v_0} e^{\lambda_v \cdot x_v} \frac{d^2 x_v}{(2\pi i)^2}\,,
\end{equation}
and integrated over $\mathbb{R}^{4 |\Gamma_0|-4}$; an $\mathbb{R}^4$ for each vertex position degree of freedom, minus an overall translational symmetry. A diagram will thus be non-vanishing only if the holomorphic form degree and antiholomorphic form degree match, giving us a balancing condition:
\begin{equation}
    2 |\Gamma_0| = |\Gamma_1|+3\,.
\end{equation}

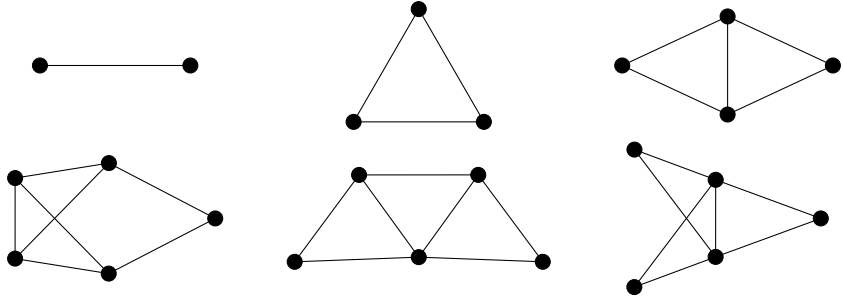
\begin{figure}
\centering
\setlength\tabcolsep{12pt} 
\renewcommand{\arraystretch}{3} 
\begin{tabular}{ccc}
   \begin{tikzpicture}
        [
    	baseline={(current bounding box.center)},
    	line join=round
    	]
    	\coordinate (pd1) at (1.*\gS,0.*\gS);
    	\coordinate (pd2) at (-1.*\gS,0.*\gS);
    	\draw (pd1) node[GraphNode] {} ;
    	\draw (pd2) node[GraphNode] {};
        \draw[GraphEdge] (pd1) -- (pd2);
    \end{tikzpicture}&
    \begin{tikzpicture}
		[
		baseline={(current bounding box.center)},
		line join=round
		]
		\coordinate (pd1) at (-0.866*\gS,-0.5*\gS);
		\coordinate (pd2) at (0.866*\gS,-0.5*\gS);
		\coordinate (pd3) at (0.*\gS,1.*\gS);
		\draw (pd1) node[GraphNode] {};
		\draw (pd2) node[GraphNode] {};
		\draw (pd3) node[GraphNode] {};
		\draw[GraphEdge] (pd1) -- (pd2);
		\draw[GraphEdge] (pd1) -- (pd3);
		\draw[GraphEdge] (pd2) -- (pd3);
	\end{tikzpicture}&
	\begin{tikzpicture}
    	[
    	baseline={(current bounding box.center)},
    	line join=round
    	]
        \def\gS{1.5}
    	\coordinate (pd1) at (1.8671*\gS,0.435*\gS);
    	\coordinate (pd2) at (0.9338*\gS,0.87*\gS);
    	\coordinate (pd3) at (0.9339*\gS,0.*\gS);
    	\coordinate (pd4) at (0.*\gS,0.435*\gS);
    	\draw (pd1) node[GraphNode] {} ;
    	\draw (pd2) node[GraphNode] {}  ;
    	\draw (pd3) node[GraphNode] {}  ;
    	\draw (pd4) node[GraphNode] {}  ;
    	\draw[GraphEdge] (pd1) -- (pd2)  ;
    	\draw[GraphEdge] (pd1) -- (pd3)  ;
    	\draw[GraphEdge] (pd2) -- (pd3)  ;
    	\draw[GraphEdge] (pd2) -- (pd4)  ;
    	\draw[GraphEdge] (pd3) -- (pd4)  ;
	\end{tikzpicture}\\
	\begin{tikzpicture}
		[
    	baseline={(current bounding box.center)},
	    line join=round
	    ]
        \def\gS{1.5};
    	\coordinate (pd1) at (1.7738*\gS,0.4892*\gS);
    	\coordinate (pd2) at (0.8297*\gS,0.*\gS);
    	\coordinate (pd3) at (0.0002*\gS,0.8472*\gS);
    	\coordinate (pd4) at (0.*\gS,0.1329*\gS);
    	\coordinate (pd5) at (0.8314*\gS,0.9802*\gS);
    	\draw (pd1) node[GraphNode] {} ;
    	\draw (pd2) node[GraphNode] {} ;
    	\draw (pd3) node[GraphNode] {} ;
    	\draw (pd4) node[GraphNode] {} ;
    	\draw (pd5) node[GraphNode] {} ;
    	\draw[GraphEdge] (pd1) -- (pd2) ;
    	\draw[GraphEdge] (pd1) -- (pd5) ;
    	\draw[GraphEdge] (pd2) -- (pd3);
    	\draw[GraphEdge] (pd2) -- (pd4);
    	\draw[GraphEdge] (pd3) -- (pd4) ;
    	\draw[GraphEdge] (pd3) -- (pd5) ;
    	\draw[GraphEdge] (pd4) -- (pd5) ;
	\end{tikzpicture}&
    \begin{tikzpicture}
        [
    	baseline={(current bounding box.center)},
    	line join=round
    	]
        \def\gS{1.5};
    	\coordinate (pd1) at (2.1987*\gS,0.0003*\gS);
    	\coordinate (pd2) at (1.627*\gS,0.7711*\gS);
    	\coordinate (pd3) at (1.0995*\gS,0.0413*\gS);
    	\coordinate (pd4) at (0.5706*\gS,0.7707*\gS);
    	\coordinate (pd5) at (0.*\gS,0.*\gS);
    
    	\draw (pd1) node[GraphNode] {};
    	\draw (pd2) node[GraphNode] {} ;
    	\draw (pd3) node[GraphNode] {} ;
    	\draw (pd4) node[GraphNode] {} ;
    	\draw (pd5) node[GraphNode] {} ;
    
    	\draw[GraphEdge] (pd1) -- (pd2) ;
    	\draw[GraphEdge] (pd1) -- (pd3);
    	\draw[GraphEdge] (pd2) -- (pd3) ;
    	\draw[GraphEdge] (pd2) -- (pd4) ;
    	\draw[GraphEdge] (pd3) -- (pd4) ;
    	\draw[GraphEdge] (pd3) -- (pd5) ;
    	\draw[GraphEdge] (pd4) -- (pd5) ;
    \end{tikzpicture}&
    \begin{tikzpicture}
        [
    	baseline={(current bounding box.center)},
    	line join=round
    	]
        \def\gS{1.5};
    	\coordinate (pd1) at (0.0009*\gS,1.2196*\gS);
    	\coordinate (pd2) at (0.7208*\gS,0.9517*\gS);
    	\coordinate (pd3) at (0.*\gS,0.*\gS);
    	\coordinate (pd4) at (0.7203*\gS,0.2677*\gS);
    	\coordinate (pd5) at (1.6524*\gS,0.6091*\gS);
    	\draw (pd1) node[GraphNode] {};
    	\draw (pd2) node[GraphNode] {} ;
    	\draw (pd3) node[GraphNode] {} ;
    	\draw (pd4) node[GraphNode] {} ;
    	\draw (pd5) node[GraphNode] {} ;
    	\draw[GraphEdge] (pd1) -- (pd2) ;
    	\draw[GraphEdge] (pd1) -- (pd4) ;
    	\draw[GraphEdge] (pd2) -- (pd3) ;
    	\draw[GraphEdge] (pd2) -- (pd5) ;
    	\draw[GraphEdge] (pd3) -- (pd4) ;
    	\draw[GraphEdge] (pd4) -- (pd5) ;
    	\draw[GraphEdge] (pd2) -- (pd4) ;
    \end{tikzpicture}
\end{tabular}
    \caption{The six simplest Laman graphs. These are the graphs which will appear at tree level (the segment), 1-loop (the triangle), two loops (the bitriangle) and three loops in the four-dimensional holomorphic theories we consider. Note that Laman graphs do not have to be planar.}
    \label{fig:sixLamanGraphs}
\end{figure}

We can actually impose a stronger constraint. Consider an arbitrary subset $S$ of vertices of $\Gamma$, and denote the corresponding induced subgraph\footnote{A subgraph induced by a subset $S$ of the vertices of a graph is the graph whose vertex set is $S$ and whose edge set consists of all the edges that have both endpoints in $S$. } by $\Gamma[S]$. Since $\Gamma$ is connected, the total form degree of the propagators in $\Gamma[S]$ must be smaller than $2 |\Gamma[S]_0|-2$, i.e. 
\begin{equation}
	2 |\Gamma_0[S]| \geq |\Gamma_1[S]|+3\,.
\end{equation}
Graphs with this property are called {\it Laman graphs} \cite{laman1970graphs} (which also appeared in the earlier references \cite{henneberg1911graphische, pollaczek1927gliederung}). We draw the first few in Figure \ref{fig:sixLamanGraphs}.

Laman graphs can be built recursively from a sequence of two types of (Henneberg) moves:
\begin{enumerate}
	\item Add a new vertex to the graph, together with edges connecting it to two distinct previously existing vertices.
	\item Subdivide an edge of the graph, and add an edge connecting the newly formed vertex to a third previously existing vertex (distinct from the two endpoints of the original (now subdivided) edge).
\end{enumerate}

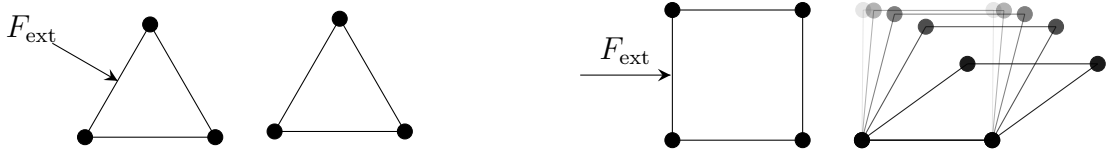
\begin{figure}
	\begin{minipage}{.47\textwidth}
		\centering
		\begin{tikzpicture}
			[
			baseline={(current bounding box.center)},
			line join=round
			]
			
			\coordinate (pd1) at (-0.866*\gS,-0.5*\gS);
			\coordinate (pd2) at (0.866*\gS,-0.5*\gS);
			\coordinate (pd3) at (0.*\gS,1.*\gS);
			
			\draw (pd1) node[GraphNode] {};
			\draw (pd2) node[GraphNode] {};
			\draw (pd3) node[GraphNode] {};
			
			\draw[GraphEdge] (pd1) -- (pd2);
			\draw[GraphEdge] (pd1) -- (pd3);
			\draw[GraphEdge] (pd2) -- (pd3);
			
			\draw[->- = 1 rotate 0]  (-1.3*\gS,0.75*\gS) -- (-0.433*\gS,0.25*\gS) node[above left, near start] {$F_{\mathrm{ext}}$};
		\end{tikzpicture}
		\hspace{3mm}
		\begin{tikzpicture}
			[
			baseline={(current bounding box.center)},
			line join=round
			]
			
			\coordinate (pd1) at (-0.866*\gS,-0.5*\gS);
			\coordinate (pd2) at (0.866*\gS,-0.5*\gS);
			\coordinate (pd3) at (0.*\gS,1.*\gS);
			
			\draw (pd1) node[GraphNode] {};
			\draw (pd2) node[GraphNode] {};
			\draw (pd3) node[GraphNode] {};
			
			\draw[GraphEdge] (pd1) -- (pd2);
			\draw[GraphEdge] (pd1) -- (pd3);
			\draw[GraphEdge] (pd2) -- (pd3);
		\end{tikzpicture}
	\end{minipage}
	\hspace{1cm}
	\begin{minipage}{.47\textwidth}
		\centering
		\begin{tikzpicture}
			[
			line join=round
			]
			
			\coordinate (pd1) at (0.*\gS,0.*\gS);
			\coordinate (pd2) at (1.732*\gS,0.*\gS);
			\coordinate (pd3) at (1.732*\gS,1.732*\gS);
			\coordinate (pd4) at (0.*\gS,1.732*\gS);
			
			\draw[GraphEdge] (pd1) -- (pd2);
			\draw[GraphEdge] (pd1) -- (pd4);
			\draw[GraphEdge] (pd2) -- (pd3);
			\draw[GraphEdge] (pd3) -- (pd4);
			
			\draw (pd1) node[GraphNode] {};
			\draw (pd2) node[GraphNode] {};
			\draw (pd3) node[GraphNode] {};
			\draw (pd4) node[GraphNode] {};
			
			\draw[->- = 1 rotate 0]  (-1.\gS-\gS/8,0.866*\gS) -- (0.\gS-\gS/8,0.866*\gS) node[midway, above] {$F_{\mathrm{ext}}$};
			
		\end{tikzpicture}
		\hspace{3mm}
		\begin{tikzpicture}
		[
		line join=round
		]
		\def\maxN{10};
		
		\foreach \i in {1,3,5,7,9}
		{
			\draw[opacity = \i/\maxN] (0,0) 
			-- ({sqrt(3)},0);
			\draw[opacity = \i/\maxN] ({sqrt(3)},0) 
			-- ({sqrt(3)*(1+\i*\i/(\maxN*\maxN))},{1*sqrt(3)*sqrt(1-\i*\i*\i*\i/(\maxN*\maxN*\maxN*\maxN))});
			\draw[opacity = \i/\maxN] ({sqrt(3)*(1+\i*\i/(\maxN*\maxN))},{1*sqrt(3)*sqrt(1-\i*\i*\i*\i/(\maxN*\maxN*\maxN*\maxN))})
			-- ({1*sqrt(3)*\i*\i/(\maxN*\maxN)},{sqrt(3)*sqrt(1-\i*\i*\i*\i/(\maxN*\maxN*\maxN*\maxN))});
			\draw[opacity = \i/\maxN] ({1*sqrt(3)*\i*\i/(\maxN*\maxN)},{sqrt(3)*sqrt(1-\i*\i*\i*\i/(\maxN*\maxN*\maxN*\maxN))})
			-- (0,0);
		}			
		\foreach \i in {1,3,5,7,9}
		{
			\draw (0,0) node[opacity = \i/\maxN, GraphNode] {};
			\draw ({sqrt(3)},0) node[opacity = \i/\maxN, GraphNode] {};
			\draw ({sqrt(3)*(1+\i*\i/(\maxN*\maxN))},{1*sqrt(3)*sqrt(1-\i*\i*\i*\i/(\maxN*\maxN*\maxN*\maxN))}) node[opacity = \i/\maxN, GraphNode] {};
			\draw ({1*sqrt(3)*\i*\i/(\maxN*\maxN)},{sqrt(3)*sqrt(1-\i*\i*\i*\i/(\maxN*\maxN*\maxN*\maxN))}) node[opacity = \i/\maxN, GraphNode] {};
		}			
	\end{tikzpicture}
	\end{minipage}
	\caption{In the classical mechanical interpretation of Laman graphs as minimally rigid graphs, each edge is imagined as a rigid rod and each vertex as a hinge. The triangle (left) is a Laman graph: it is not deformable by an external force $F_{\mathrm{ext}}$, but would be if any edge were removed. The square (right) is deformable.}\label{fig:fallingBox}
\end{figure}

Laman graphs were originally introduced as {\it minimally rigid} graphs drawn on the plane. Non-technically, this means if we imagine replacing each edge with a rigid rod of a fixed length, and each vertex by a hinge, then it will retain its shape when pressed on. That is, it has no degrees of freedom besides rigid translations and rotations in the plane. This property, though, would fail if we removed any edge. For example, if ones pushes on a triangle made from rods and hinges, it would still retain it's shape (and may only slide or rotate). Meanwhile, a horizontal force on a square, which is not a Laman graph, would cause it to collapse.\footnote{One may also notice that the square is only one edge away from being a Laman graph in two different senses. On one hand, we could reinforce a diagonal of the square to form a ``bitriangle'' graph, which would satisfy the Laman criteria. On the other hand, we could take any edge in the square and contract it down to a point to produce the triangle graph, which is Laman. We will return to these points in Section \ref{sec:graphsInsideGraphs}.}  This is drawn in Figure \ref{fig:fallingBox}.

The rigidity property of Laman graphs also holds if we fix the slopes of the edges in the plane to generic values, instead of the lengths. The three surviving degrees of freedom are then translations and scale transformations. An important feature of the edge slope constraints is that they are linear in the positions of the vertices and thus have a unique solution (if any). In particular, we can parameterize the shape of a Laman graph in the plane by the slopes of its edges, the position of a vertex, and the length of an edge. This is a very convenient parameterization \cite{Kontsevich:1997vb} which we will employ extensively to simplify our calculations.  

\subsection{Putting Everything Together}
At this point we have assembled all the ingredients for a generic integrand:
\begin{equation}
	\Omega_\Gamma[\lambda;z] \equiv \left[\prod_{e \in \Gamma_1} {\cal P}(x_{e(0)} - x_{e(1)}+z_e, \bar x_{e(0)} - \bar x_{e(1)},t_e)\right] \left[ \prod_{v \in \Gamma_0|v \neq v_0} e^{\lambda_v \cdot x_v} \frac{d^2 x_v}{(2\pi i)^2} \right]\,.
\end{equation}
Each edge $e$ is associated to a propagator 1-form, with an extra holomorphic shift $z_e \in \mathbb{C}^2$ and Schwinger time $t_e$. Each vertex $v \neq v_0$ is associated to a position $(x_v, \bar x_v) \in \mathbb{R}^4$ and to a holomorphic momentum $\lambda_v \in \mathbb{C}^2$. We define $\lambda_{v_0}$ so that the total momentum vanishes.

The Feynman integrals are:
\begin{equation}
{\cal I}_\Gamma[\lambda;z] \equiv \int_{\mathbb{R}^{4 |\Gamma_0|-4} \times \partial_\epsilon \left[[\epsilon,L]^{|\Gamma_1|}\right]} \Omega_\Gamma[\lambda;z] \,.
\end{equation}
As discussed before, the overall sign of ${\cal I}_\Gamma[\lambda;z]$ depends on a choice of ordering of the edges of $\Gamma$, now encoded in the definition of the 
hypercube $[\epsilon,L]^{|\Gamma_1|}$.\footnote{As the integrand is a closed form, we could alternatively integrate along the ``IR'' components at $t_e=L$ of the boundary of the hypercube to get the same answer (up to a sign). This is more cumbersome in practice, as the IR cutoff $L$ should be sent to infinity at the end of the calculation. }

A standard strategy to evaluate such an integral would be to first perform the Gaussian integral over the $(x_v, \bar x_v)$'s to obtain a form $\omega_\Gamma[\lambda;z]$ in the space of Schwinger times $t_e$'s. The form $\omega_\Gamma$ is closed and has some interesting properties. In particular, it is scale-invariant in the space of Schwinger times and can be written as the pull-back of a top form on the corresponding real projective space $\mathbb{RP}^{|\Gamma_1|-1}$. 

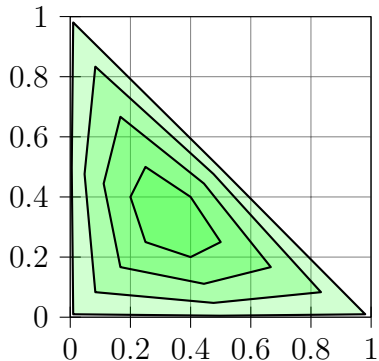
\begin{figure}
    \centering
    \begin{tikzpicture}
    [
    line join=round,
    scale=4
    ]
    
    \draw[gray,very thin] (0,0) grid (1,1)
    [step=0.20cm] (0,0) grid (1,1);
    \draw[black,thin] (0,0) grid (1,1);
    
    \foreach \pos in {0,0.2,0.4,0.6,0.8,1}
    {
        \draw[shift={(\pos,0)}] (0pt,1pt) -- (0pt,-1pt) node[below] {$\pos$};
        \draw[shift={(0,\pos)}] (1pt,0pt) -- (-1pt,0pt) node[left] {$\pos$};
    }
    
    \foreach \eL in {100,10,4,2}
    {
        \draw[fill=green!90, opacity=0.2]   
            ({\eL/(1+2*\eL)},{\eL/(1+2*\eL)}) -- 
            ({\eL/(2+\eL)},{1/(2+\eL)}) -- 
            ({\eL/(1+2*\eL)},{1/(1+2*\eL)}) --
            ({1/(2+\eL)},{1/(2+\eL)}) --
            ({1/(1+2*\eL)},{\eL/(1+2*\eL)}) --
            ({1/(2+\eL)},{\eL/(2+\eL)}) -- cycle;
    }
    \foreach \eL in {2,4,10,100}
    {
        \draw[thick, black]   
            ({\eL/(1+2*\eL)},{\eL/(1+2*\eL)}) -- 
            ({\eL/(2+\eL)},{1/(2+\eL)}) -- 
            ({\eL/(1+2*\eL)},{1/(1+2*\eL)}) --
            ({1/(2+\eL)},{1/(2+\eL)}) --
            ({1/(1+2*\eL)},{\eL/(1+2*\eL)}) --
            ({1/(2+\eL)},{\eL/(2+\eL)}) -- cycle;
    }
   				
    \end{tikzpicture}
    \caption{For the one-loop calculation, the integration region can be identified with the positive part of $\mathbb{RP}^{2}$ as $L/\epsilon \to \infty$. We plot this region for $L/\epsilon=2,4,10,100$ in the $t_1,t_2$-plane, in the Feynman gauge $t_1+t_2+t_3=1$.}
    \label{fig:fill}
\end{figure}

Remarkably, as we send $L\to \infty$, the integration region $\partial_\epsilon \left[[\epsilon,L]^{|\Gamma_1|}\right]$ can be identified with the positive part $\mathbb{RP}_{>}^{|\Gamma_1|-1}$ of $\mathbb{RP}^{|\Gamma_1|-1}$.\footnote{This statement is reminiscent of the Cheng-Wu theorem \cite{chengWu}.}\footnote{Essentially, 
the region of $\mathbb{RP}^{|\Gamma_1|-1}$ where the $e$'th $t_e$ is smaller than all other $t$'s can be identified with the component $t_e=\epsilon$, $t_{e'} \geq \epsilon$ in the original integral. Incidentally, we could have picked independent UV cutoffs for each propagator and integrated over $\partial\left[\prod_{e\in \Gamma_1} [\epsilon_e,\infty)\right] $. Then each facet of the UV boundary could be identified with the region of $\mathbb{RP}^{|\Gamma_1|-1}$ where the $e$'th $t_e/\epsilon_e$ is smaller than all other $t/\epsilon$'s, giving the same $\mathbb{RP}_{>}^{|\Gamma_1|-1}$ answer.} See Figure \ref{fig:fill} for an example of the integration region of the one loop graph in the region of $t$'s. Furthermore, $\omega_\Gamma[\lambda;z]$ appear to be  non-singular in $\mathbb{RP}_{>}^{|\Gamma_1|-1}$ and the integral is finite and well-defined.\footnote{Convergence at the corners of $\mathbb{RP}_{>}^{|\Gamma_1|-1}$ is not immediately obvious, but the change of variables we define momentarily makes finiteness manifest.} 

Despite these simplifications, $\omega_\Gamma[\lambda;z]$ is an intricate function of the $t_e$ variables and direct integration is challenging beyond one loop. The curious reader may consult Appendix \ref{append:1loopintegrand} for the explicit derivation of $\omega_\Gamma$ at one loop. 

It turns out that the functional form of $\omega_\Gamma$ can be drastically simplified by a judicious change of variables, at the price of making the integration region more intricate. 
The change of variables can be described directly in terms of the $t_e$'s, but it's more elegant to step back to the original integral and do the coordinate change there.

\subsection{A Convenient Coordinate Change}
We are interested in ${\cal I}_\Gamma[\lambda;z]$ as a power series in $z_e$ and $\lambda_v$ around the origin. Without loss of generality, we can take the $\lambda_v$ momenta to be real and the $z_e$ shifts to be purely imaginary. The Gaussian integral over complex conjugate $\bar x = x^*$ is then equivalent to an integral done over real $\bar x_v$ and imaginary $x_v$'s.  Once the integration contours for $\bar x_v$ and $x_v$'s are independent, we can trade the $\bar x_v$ and $t_e$ coordinates for the combinations
\begin{equation}\label{eq:ydef}
	y_e = t_e^{-1}  (\bar x_{e(0)}-\bar x_{e(1)})\,,
\end{equation}
which simplify the combined propagators as in equation (\ref{eq:simprop}). Notice that the $y_e$'s change signs in the same manner as the $z_e$'s if we change our choice of orientation for an edge. The balancing condition guarantees that the number of $y_e$ coordinates is:
\begin{equation}
    2|\Gamma_1| = (|\Gamma_1|-1) + 2 (|\Gamma_0| -1)\,,
\end{equation}
and thus is the same as the number of independent $\bar x_v$ and $t_e$ coordinates. 

The $y_e$ defined by (\ref{eq:ydef}) are global coordinates on the integration cycle $\mathbb{R}^{2 |\Gamma_0|-2} \times \partial_\epsilon \left[[\epsilon,L]^{|\Gamma_1|}\right]$ and realize it as a polytope in $\mathbb{R}^{2|\Gamma_1|}$. This is easy to show: if we are given the $y_e$ we can determine the $\bar x_e$ up to translations and scale transformations by the collection of linear equations $y_e \wedge (\bar x_{e(0)}-\bar x_{e(1)})=0$. The $t_e$ are then immediately determined modulo a scale transformation and gauge-fixed uniquely to $\partial_\epsilon \left[[\epsilon,L]^{|\Gamma_1|}\right]$ as discussed before. Once we set one of the $t_e$ to $\epsilon$, the scale of the $\bar x_e$ is also fixed and we can translate them to $\bar x_{v_0}=0$. 

The immediate consequence is that we can perform the integral in terms of the $(x_v,y_e)$ coordinates over $\mathbb{R}^{2 |\Gamma_0|-2}$ times the image of $\mathbb{R}^{2 |\Gamma_0|-2} \times \partial_\epsilon \left[[\epsilon,L]^{|\Gamma_1|}\right]$ under (\ref{eq:ydef}). In these new coordinates, the integrand simplifies to 
\begin{equation}\label{eq:integrandnew}
	\Omega_\Gamma[\lambda;z] \equiv 
	    \left[
	        \prod_{e \in \Gamma_1}  e^{- y_e \cdot (x_{e(0)} - x_{e(1)}+z_e)} d^2 y_e 
	        \vphantom{\prod_{v \in \Gamma_0|v \neq v_0} \frac{d^2 x_v}{(2\pi i)^2}} \right] 
	    \left[ 
	        \prod_{v \in \Gamma_0|v \neq v_0} e^{\lambda_v \cdot x_v} \frac{d^2 x_v}{(2\pi i)^2} 
	    \right]\,,
\end{equation}
or equivalently 
\begin{equation}
	\Omega_\Gamma[\lambda;z] \equiv 
	    \left[
	        \prod_{e \in \Gamma_1}  e^{- y_e \cdot z_e} d^2 y_e 
	        \vphantom{\prod_{v \in \Gamma_0|v \neq v_0} \frac{d^2 x_v}{(2 \pi i)^2} }
        \right] 
        \left[ 
            \prod_{v \in \Gamma_0|v \neq v_0} e^{\left(\lambda_v- \sum_{e|e(0)=v} y_e + \sum_{e|e(1)=v} y_e \right)\cdot x_v} \frac{d^2 x_v}{(2 \pi i)^2} 
        \right]\,.
\end{equation}

At this point, we can safely remove the IR regulator and do the integral directly on the image $\Delta_\Gamma$ of $\mathbb{R}^{2 |\Gamma_0|-2} \times \partial_\epsilon \left[[\epsilon,\infty]^{|\Gamma_1|}\right]$ under (\ref{eq:ydef}):
\begin{equation}
    {\cal I}_\Gamma[\lambda;z] \equiv \int_{(i\mathbb{R})^{2 |\Gamma_0|-2} \times \Delta_\Gamma} \Omega_\Gamma[\lambda;z] \,.
\end{equation}

The integration cycle $\Delta_\Gamma$ has a nice geometric interpretation. Given a $y_e \in \Delta_\Gamma$, the (\ref{eq:ydef}) relations fix the slope of $\bar x_{e(1)} - \bar x_{e(0)}$. The $\bar x_v$ thus present $\Gamma$ as a graph in the plane, with fixed slopes for the edges. The residual translation symmetry is fixed by $\bar x_{v_0}=0$ and the overall scale can be fixed by setting the overall scale of the $t_e$, say by $\sum_e t_e=1$. The magnitude of an individual $y_e$ is unconstrained by (\ref{eq:ydef}), so the shape of $\Delta_\Gamma$ is determined by the range of possible slopes which arise from embedding  $\Gamma$ in the plane. It is specifically this polytope, as well as its generalizations, that we assign the name ``Operatope.''

The integration cycle $\Delta_\Gamma$ comes with an orientation induced by the orientation of $\mathbb{R}^{2 |\Gamma_0|-2} \times \partial_\epsilon \left[[\epsilon,L]^{|\Gamma_1|}\right]$.
The first factor has a natural orientation, the second has an orientation determined by the ordering we chose for the edges of $\Gamma$. We can also think about it as an orientation of $\mathbb{RP}^{|\Gamma_1|-1}$. 

Note that the $x_v$ integral simply imposes a vertex constraint:
\begin{equation} \label{eq:vertex}
	\lambda_v= \left[\sum_{e|e(0)=v} - \sum_{e|e(1)=v}\right]y_e\,,
\end{equation}
which identifies the $y_e$ as the holomorphic momenta along the edges of $\Gamma$. Hence if we do the $x_v$ integral first, 
we are left with what is essentially the Fourier transform 
\begin{equation}
    {\cal I}_\Gamma[\lambda;z] \equiv \int_{\Delta_\Gamma}\left[\prod_{e \in \Gamma_1} e^{- y_e \cdot z_e} d^2 y_e \right] \left[ \prod_{v \in \Gamma_0|v \neq v_0} \delta\left(\lambda_v- \sum_{e|e(0)=v} y_e + \sum_{e|e(1)=v} y_e \right) \right] 
\end{equation}
of a $\lambda$-dependent region in $\mathbb{R}^{2|\Gamma_1|}$ cut out by (\ref{eq:vertex}) and (\ref{eq:ydef}). 

The vertex constraints (\ref{eq:vertex}) can be solved by writing $y_e$ as linear combination of $\lambda_v$'s and some loop variables $Y_\ell$. The definition (\ref{eq:ydef}) implies that a positive linear combination of $y_e$'s along any loop vanishes, i.e.
\begin{equation}
    \sum_{e \in \ell} t_e y_e=0\,.
\end{equation}
This constrains $Y_\ell$ to live in some bounded region parameterized by the $\lambda_v$. The Fourier transform of such a bounded region is a smooth function, analytic in $z_e$ around the origin. 

Incidentally, these linear constraints on $Y_\ell$ can be inverted to give functions $Y_\ell(\lambda,t)$. This is precisely the aforementioned coordinate change which brings $\omega_\Gamma[\lambda;z]$ to a simple form.

\subsection{Some Symmetries}\label{section:shiftsymmetry}
Basic scaling consideration 
show that ${\cal I}_\Gamma[\lambda;z]$ has weight $2 |\Gamma_1| - 2 |\Gamma_0|+2$ if $z_e$ is given weight $-1$ and $\lambda_v$ is given weight $1$. 

A shift $x_v \to x_v + \delta_v$  shows that 
\begin{equation} \label{eq:shift}
    {\cal I}_\Gamma[\lambda, z_e - \delta_{e(0)} + \delta_{e(1)}] = e^{\sum_{v} \lambda_v \cdot \delta_v} {\cal I}_\Gamma[\lambda, z_e]\,.
\end{equation}
We can use this freedom to express ${\cal I}_\Gamma[\lambda, z_e]$ as an exponential times a function of the linear combinations of the $z_e$ which are invariant under the shift  $z_e\to z_e - \delta_{e(0)} + \delta_{e(1)}$. There is such a linear combination $Z_\ell$ for each independent loop $\ell$ in $\Gamma$. 

A useful perspective on this parameterization is that if we expand out the $y_e$ as linear combinations of $\lambda_v$ and $Y_\ell$ and reorganize the coupling to $z_e$ accordingly
\begin{equation}
      \sum_e y_e \cdot z_e = \sum \lambda_v \cdot \tilde z_v + \sum_\ell Y_\ell \cdot Z_\ell \,.
\end{equation}
If we shift $z_e\rightarrow z_e -\delta_{e(0)} + \delta_{e{(1)}}$, we have
\begin{equation}
    \sum_e y_e \cdot z_e \rightarrow \sum_e y_e \cdot z_e -\sum_v \lambda_v \cdot \delta_v\,.
\end{equation}
If we set $\delta_v = \tilde z_v$ we arrive at the desired   
\begin{equation}
    \sum_e y_e \cdot z_e \rightarrow \sum_\ell Y_\ell \cdot Z_\ell\,.
\end{equation}

\section{Graphs Within Graphs}\label{sec:graphsInsideGraphs}
Next we review a construction, inspired by \cite{Kontsevich:1997vb, Gaiotto:2015zna, Gaiotto:2015aoa}, which gives a collection of quadratic relations satisfied by the $\Delta_\Gamma$ configuration spaces. In turn, these relations will imply quadratic relations for the ${\cal I}_\Gamma$ integrals and, in our companion paper \cite{4dcohomology}, associativity of
the holomorphic factorization algebras built from the ${\cal I}_\Gamma$.  

As anticipated in the introduction, the quadratic relations should be associated to the possible ways one can produce some overall graph $\tilde \Gamma$ by replacing a vertex of a Feynman diagram with a second Feynman diagram. We expect to find one relation for each such $\tilde \Gamma$. 

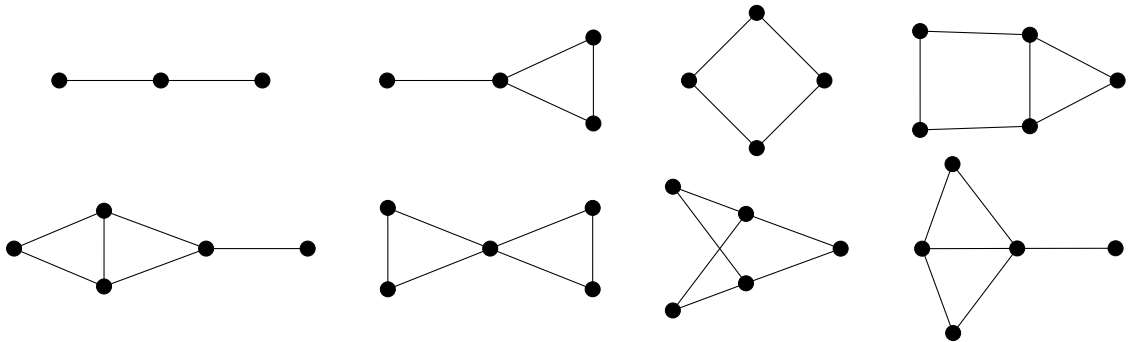
\begin{figure}
\centering
\setlength\tabcolsep{12pt} 
\renewcommand{\arraystretch}{3} 
\begin{tabular}{cccc}
   \begin{tikzpicture}
        [
    	baseline={(current bounding box.center)},
    	line join=round
    	]
        \def\gS{1.35};
    	\coordinate (pd1) at (0.*\gS,0.*\gS);
    	\coordinate (pd2) at (1.*\gS,0.*\gS);
    	\coordinate (pd3) at (2.*\gS,0.*\gS);
    
    	\draw (pd1) node[GraphNode] {} ;
    	\draw (pd2) node[GraphNode] {} ;
    	\draw (pd3) node[GraphNode] {} ;
    
    	\draw[GraphEdge] (pd1) -- (pd2) ;
    	\draw[GraphEdge] (pd2) -- (pd3) ;
    \end{tikzpicture}
&
    \begin{tikzpicture}
        [
    	baseline={(current bounding box.center)},
    	line join=round
    	]
        \def\gS{1.35};
    	\coordinate (pd1) at (1.1141*\gS,0.4234*\gS);
    	\coordinate (pd2) at (2.0313*\gS,0.*\gS);
    	\coordinate (pd3) at (2.0313*\gS,0.8471*\gS);
    	\coordinate (pd4) at (0.*\gS,0.4235*\gS);
    
    	\draw (pd1) node[GraphNode] {};
    	\draw (pd2) node[GraphNode] {} ;
    	\draw (pd3) node[GraphNode] {} ;
    	\draw (pd4) node[GraphNode] {} ;
    
    	\draw[GraphEdge] (pd1) -- (pd2) ;
    	\draw[GraphEdge] (pd1) -- (pd3);
    	\draw[GraphEdge] (pd1) -- (pd4) ;
    	\draw[GraphEdge] (pd2) -- (pd3) ;
    \end{tikzpicture}
&
    \begin{tikzpicture}
        [
    	baseline={(current bounding box.center)},
    	line join=round
    	]
        \def\gS{0.9};
    	\coordinate (pd1) at (-1.*\gS,0.*\gS);
    	\coordinate (pd2) at (0.*\gS,1.*\gS);
    	\coordinate (pd3) at (0.*\gS,-1.*\gS);
    	\coordinate (pd4) at (1.*\gS,0.*\gS);
    
    	\draw (pd1) node[GraphNode] {};
    	\draw (pd2) node[GraphNode] {} ;
    	\draw (pd3) node[GraphNode] {} ;
    	\draw (pd4) node[GraphNode] {};
    
    	\draw[GraphEdge] (pd1) -- (pd2) ;
    	\draw[GraphEdge] (pd1) -- (pd3) ;
    	\draw[GraphEdge] (pd2) -- (pd4);
    	\draw[GraphEdge] (pd3) -- (pd4) ;
    \end{tikzpicture}
&
    \begin{tikzpicture}
        [
    	baseline={(current bounding box.center)},
    	line join=round
    	]
        \def\gS{1.35};
    	\coordinate (pd1) at (1.9449*\gS,0.4869*\gS);
    	\coordinate (pd2) at (1.0809*\gS,0.0359*\gS);
    	\coordinate (pd3) at (1.081*\gS,0.9376*\gS);
    	\coordinate (pd4) at (0.0001*\gS,0.*\gS);
    	\coordinate (pd5) at (0.*\gS,0.9737*\gS);
    
    	\draw (pd1) node[GraphNode] {} ;
    	\draw (pd2) node[GraphNode] {} ;
    	\draw (pd3) node[GraphNode] {} ;
    	\draw (pd4) node[GraphNode] {} ;
    	\draw (pd5) node[GraphNode] {} ;
    
    	\draw[GraphEdge] (pd1) -- (pd2) ;
    	\draw[GraphEdge] (pd1) -- (pd3) ;
    	\draw[GraphEdge] (pd2) -- (pd3);
    	\draw[GraphEdge] (pd2) -- (pd4) ;
    	\draw[GraphEdge] (pd3) -- (pd5) ;
    	\draw[GraphEdge] (pd4) -- (pd5) ;
	\end{tikzpicture}\\
    \begin{tikzpicture}
        [
    	baseline={(current bounding box.center)},
    	line join=round
    	]
        \def\gS{1.35};
    	\coordinate (pd1) at (0.*\gS,0.3733*\gS);
    	\coordinate (pd2) at (0.8859*\gS,0.7463*\gS);
    	\coordinate (pd3) at (0.8857*\gS,0.*\gS);
    	\coordinate (pd4) at (1.891*\gS,0.3731*\gS);
    	\coordinate (pd5) at (2.891*\gS,0.3732*\gS);
    
    	\draw (pd1) node[GraphNode] {} ;
    	\draw (pd2) node[GraphNode] {} ;
    	\draw (pd3) node[GraphNode] {} ;
    	\draw (pd4) node[GraphNode] {} ;
    	\draw (pd5) node[GraphNode] {};
    
    	\draw[GraphEdge] (pd1) -- (pd2) ;
    	\draw[GraphEdge] (pd1) -- (pd3) ;
    	\draw[GraphEdge] (pd2) -- (pd3) ;
    	\draw[GraphEdge] (pd2) -- (pd4);
    	\draw[GraphEdge] (pd3) -- (pd4) ;
    	\draw[GraphEdge] (pd4) -- (pd5) ;
    \end{tikzpicture}
&
    \begin{tikzpicture}
        [
    	baseline={(current bounding box.center)},
    	line join=round
    	]
        \def\gS{1.35};
    	\coordinate (pd1) at (2.0175*\gS,0.8022*\gS);
    	\coordinate (pd2) at (2.0173*\gS,0.0003*\gS);
    	\coordinate (pd3) at (1.0083*\gS,0.4013*\gS);
    	\coordinate (pd4) at (0.*\gS,0.8025*\gS);
    	\coordinate (pd5) at (0.0002*\gS,0.*\gS);
    
    	\draw (pd1) node[GraphNode] {} ;
    	\draw (pd2) node[GraphNode] {} ;
    	\draw (pd3) node[GraphNode] {} ;
    	\draw (pd4) node[GraphNode] {} ;
    	\draw (pd5) node[GraphNode] {} ;
    
    	\draw[GraphEdge] (pd1) -- (pd2) ;
    	\draw[GraphEdge] (pd1) -- (pd3) ;
    	\draw[GraphEdge] (pd2) -- (pd3) ;
    	\draw[GraphEdge] (pd3) -- (pd4) ;
    	\draw[GraphEdge] (pd3) -- (pd5) ;
    	\draw[GraphEdge] (pd4) -- (pd5) ;
    \end{tikzpicture}
&
    \begin{tikzpicture}
        [
    	baseline={(current bounding box.center)},
    	line join=round
    	]
        \def\gS{1.35};
    	\coordinate (pd1) at (0.0009*\gS,1.2196*\gS);
    	\coordinate (pd2) at (0.7208*\gS,0.9517*\gS);
    	\coordinate (pd3) at (0.*\gS,0.*\gS);
    	\coordinate (pd4) at (0.7203*\gS,0.2677*\gS);
    	\coordinate (pd5) at (1.6524*\gS,0.6091*\gS);
    
    	\draw (pd1) node[GraphNode] {};
    	\draw (pd2) node[GraphNode] {} ;
    	\draw (pd3) node[GraphNode] {} ;
    	\draw (pd4) node[GraphNode] {} ;
    	\draw (pd5) node[GraphNode] {} ;
    
    	\draw[GraphEdge] (pd1) -- (pd2) ;
    	\draw[GraphEdge] (pd1) -- (pd4) ;
    	\draw[GraphEdge] (pd2) -- (pd3) ;
    	\draw[GraphEdge] (pd2) -- (pd5) ;
    	\draw[GraphEdge] (pd3) -- (pd4) ;
    	\draw[GraphEdge] (pd4) -- (pd5) ;
    \end{tikzpicture}
&
    \begin{tikzpicture}
        [
    	baseline={(current bounding box.center)},
    	line join=round
    	]
        \def\gS{1.35};
    	\coordinate (pd1) at (0.3054*\gS,0.*\gS);
    	\coordinate (pd2) at (0.936*\gS,0.834*\gS);
    	\coordinate (pd3) at (0.*\gS,0.8315*\gS);
    	\coordinate (pd4) at (0.299*\gS,1.6647*\gS);
    	\coordinate (pd5) at (1.9049*\gS,0.8355*\gS);
    
    	\draw (pd1) node[GraphNode] {} ;
    	\draw (pd2) node[GraphNode] {} ;
    	\draw (pd3) node[GraphNode] {} ;
    	\draw (pd4) node[GraphNode] {} ;
    	\draw (pd5) node[GraphNode] {} ;
    
    	\draw[GraphEdge] (pd1) -- (pd2) ;
    	\draw[GraphEdge] (pd1) -- (pd3) ;
    	\draw[GraphEdge] (pd2) -- (pd3) ;
    	\draw[GraphEdge] (pd2) -- (pd4) ;
    	\draw[GraphEdge] (pd2) -- (pd5) ;
    	\draw[GraphEdge] (pd3) -- (pd4) ;
    \end{tikzpicture}
\end{tabular}
    \caption{The eight simplest sliding graphs. Unlike the Laman graphs, these graphs are not rigid and have ``moving parts'' with one degree of freedom. These will give constraints on amplitudes up to two loops through the quadratic relation(s) in \eqref{eq:DD} and \eqref{eq:II}.}\label{fig:slidingGraphs}
\end{figure}

In order for both building blocks to be Laman, $\tilde \Gamma$ must be a 
{\it sliding graph}: a graph $\tilde{\Gamma}$ such that  
\begin{equation}
    2 |\tilde{\Gamma}_0| = |\tilde{\Gamma}_1|+4\,,
\end{equation}
and $2 |\tilde{\Gamma}_0[S]| \geq |\tilde{\Gamma}_1[S]|+3$ for all induced subgraphs, see Figure \ref{fig:slidingGraphs}.

Recall the mechanical analogy where a Laman graph gives a minimally rigid configuration of rigid rods on the plane. ``Minimally'' here refers to the fact that the configuration is generically not over-constrained, so that any small perturbation of the edge lengths can be accommodated by a deformation of the graph. 

In the same analogy, a sliding graph $\tilde \Gamma$ gives a configuration of rigid rods with an intrinsic degree of freedom. Suppose that we are given two such Laman graphs $\Gamma$ and $\Gamma'$ of rigid rods on the plane and that the image of $\Gamma$ is much smaller than the edge lengths of $\Gamma'$. In such a situation, we could replace a vertex $v'$ in $\Gamma'$ with a copy of $\Gamma$ rotated in any way we wish, adjusting slightly the shape of $\Gamma'$ to restore the original edge lengths. The result will be a sliding graph $\tilde \Gamma$ of rigid rods, with a moduli space consisting essentially of rotations of the small subgraph $\Gamma$ relative to the ambient $\Gamma'$.

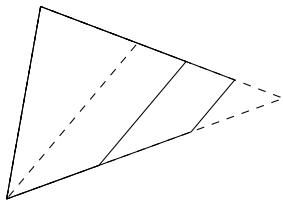
\begin{figure}
    \centering
        \begin{tikzpicture}
        [
    	baseline={(current bounding box.center)},
    	line join=round
    	]
        \def\gS{2};
        
        \def\rTL{1.3};
        \def\thTL{80};
        
        \def\rTR{0.8};
        \def\thTR{50};
        
        \def\thBR{20};
        
    	\coordinate (pBL) at (0*\gS,0*\gS);
    	\coordinate (pTL) at ({\rTL*\gS*cos(\thTL)},{\rTL*\gS*sin(\thTL)});
    	\coordinate (pBR) at (0*\gS,0*\gS);
    	\coordinate (pTR) at ({\gS*(0.869702)}, {\gS*(1.03647)});
        \draw (pBL) node {};
        \draw (pTL) node {};
        \draw (pBR) node {};
        \draw (pTR) node {};
        \draw[GraphEdge, dashed] (pBL) -- (pTL) -- (pTR) -- (pBR) -- cycle;
        
    	\coordinate (pBL) at (0*\gS,0*\gS);
    	\coordinate (pTL) at ({\rTL*\gS*cos(\thTL)},{\rTL*\gS*sin(\thTL)});
    	\coordinate (pBR) at (1.83925*\gS, 0.669433*\gS);
    	\coordinate (pTR) at (1.83925*\gS, 0.669433*\gS);
        \draw (pBL) node {};
        \draw (pTL) node {};
        \draw (pBR) node {};
        \draw (pTR) node {};
        \draw[GraphEdge, dashed] (pBL) -- (pTL) -- (pTR) -- (pBR) -- cycle;
        
        \foreach \rBR in {0.65243, 1.30486}
        {
        	\coordinate (pBL) at (0*\gS,0*\gS);
        	\coordinate (pTL) at ({\rTL*\gS*cos(\thTL)},{\rTL*\gS*sin(\thTL)});
        	\coordinate (pBR) at ({\rBR*\gS*cos(\thBR)},{\rBR*\gS*sin(\thBR)});
        	\coordinate (pTR) at ({\gS*(0.869702 + 0.495353*\rBR)}, {\gS*(1.03647 - 0.187523*\rBR)});
    	    \draw (pBL) node {};
    	    \draw (pTL) node {};
    	    \draw (pBR) node {};
    	    \draw (pTR) node {};
    	    \draw[GraphEdge] (pBL) -- (pTL) -- (pTR) -- (pBR) -- cycle;
        }
    \end{tikzpicture}
    \caption{We depict the deformation space modulo translations and scale transformations of a generic quadrilateral as the edge slopes are held fixed. We fix the leftmost edge location and length to fix the translation and scale symmetry. The deformation space has limiting ``ends'' where the quadrilateral degenerates to a triangle in two possible ways.}
    \label{fig:fixedSlopes}
\end{figure}

Something similar happens if we consider graphs in the plane with fixed edge slopes, rather than edge lengths. In this context, a Laman graph is rigid up to translations and scale transformations, while a sliding graph has an extra degree of freedom. The deformation theory in this setup is even simpler, as the slope constraints are the combination of a linear constraint and a linear inequality: the difference between the positions of two vertices connected by an edge must lie in a specific half-ray in $\mathbb{R}^2$. See Figure \ref{fig:fixedSlopes} for a simple example. 

Given two Laman graphs $\Gamma$ and $\Gamma'$ in the plane, with fixed edge slopes, we can now make $\Gamma$ arbitrarily small and replace a vertex $v'$ in $\Gamma'$ with a copy of $\Gamma$ of any sufficiently small size, adjusting slightly the shape of $\Gamma'$ to restore the original edge slopes. This produces a sliding graph $\tilde \Gamma$ with fixed edge slopes and a moduli space of deformations which locally consists of rescaling the small subgraph $\Gamma$ relative to the ambient $\Gamma'$. 

A crucial difference with the rigid rods case is that we cannot make $\Gamma$ arbitrarily large: as the relative size of $\Gamma$ grows, the procedure breaks down. We can still follow the moduli space of deformations of $\tilde \Gamma$, though. The crucial idea is that a 1-dimensional moduli space will generically have a second asymptotic end of the same type, where a different subgraph becomes arbitrarily small compared with the rest of $\tilde \Gamma$. 

Following \cite{Gaiotto:2015aoa}, we will argue that both endpoints of any such deformation space give rise to pairs $(\Gamma, \Gamma')$ of Laman graphs with fixed edge slopes. 
We will also show how to count with signs the pairs $(\Gamma, \Gamma')$ in such a way that 
the two pairs associated to a given deformation space cancel out in the sum. This will give us the desired quadratic identity.

\subsection{A Quadratic Identity}
Pick a sliding graph $\tilde \Gamma$ and use the map (\ref{eq:ydef}) to define a region $\Delta_{\tilde{\Gamma}}$ in the space of $y_e$, just as we did for Laman graphs. This map is now many-to-one: the pre-image $L[y]$ of a point $y$ in $\Delta_{\tilde{\Gamma}}$ is generically one-dimensional, as long as we pick some gauge-fixing condition for the $t_e$ scale, such as $\sum_e t_e=1$. Indeed, the $\bar x_v$ map $\tilde \Gamma$ to the plane, with edge slopes determined by the slope of the $y_e$, and overall size determined by the $t_e$  gauge-fixing condition. 

\begin{figure}
    \centering
    \begin{equation*}
    \begin{tikzpicture}
        [
    	baseline={(current bounding box.center)},
    	line join=round
    	]
        \def\gS{1.5};
    	\coordinate (pd1) at (1.9449*\gS,0.4869*\gS);
    	\coordinate (pd2) at (1.0809*\gS,0.0359*\gS);
    	\coordinate (pd3) at (1.081*\gS,0.9376*\gS);
    	\coordinate (pd4) at (0.0001*\gS,0.*\gS);
    	\coordinate (pd5) at (0.*\gS,0.9737*\gS);
    	\draw[GraphEdge,draw = red] (pd1) -- (pd2) ;
    	\draw[GraphEdge,draw = red] (pd1) -- (pd3) ;
    	\draw[GraphEdge,draw = red] (pd2) -- (pd3);
    	\draw[GraphEdge] (pd2) -- (pd4) ;
    	\draw[GraphEdge] (pd3) -- (pd5) ;
    	\draw[GraphEdge] (pd4) -- (pd5) ;
    	\draw (pd1) node[GraphNode] {} node[right] {$v_{1}$};
    	\draw (pd2) node[GraphNode] {} node[left,below] {$v_{4}$};
    	\draw (pd3) node[GraphNode] {} node[left, above] {$v_{5}$};
    	\draw (pd4) node[GraphNode] {} node[left] {$v_{2}$};
    	\draw (pd5) node[GraphNode] {} node[left] {$v_{3}$};
	\end{tikzpicture}
	\quad\longmapsto\quad
	\begin{tikzpicture}
		[
		baseline={(current bounding box.center)},
		line join=round
		]
		\coordinate (pd1) at (-0.5*\gS,-0.866*\gS);
		\coordinate (pd2) at (-0.5*\gS,0.866*\gS);
		\coordinate (pd3) at (1.*\gS,0.*\gS);
    	\draw (pd1) node[GraphNode] {} node[left] {$v_{2}$};
    	\draw (pd2) node[GraphNode] {} node[left] {$v_{3}$};
    	\draw (pd3) node[GraphNode] {} node[right] {$p_0$};
		\draw[GraphEdge] (pd1) -- (pd2);
		\draw[GraphEdge] (pd1) -- (pd3);
		\draw[GraphEdge] (pd2) -- (pd3);
	\end{tikzpicture}
	\quad
	\begin{tikzpicture}
		[
		baseline={(current bounding box.center)},
		line join=round
		]	
		\coordinate (pd1) at (-0.5*\gS,-0.866*\gS);
		\coordinate (pd2) at (-0.5*\gS,0.866*\gS);
		\coordinate (pd3) at (1.*\gS,0.*\gS);
		\draw[GraphEdge,draw = red] (pd1) -- (pd2);
		\draw[GraphEdge,draw = red] (pd1) -- (pd3);
		\draw[GraphEdge,draw = red] (pd2) -- (pd3);
    	\draw (pd1) node[GraphNode] {} node[left] {$v_{4}$};
    	\draw (pd2) node[GraphNode] {} node[left] {$v_{5}$};
    	\draw (pd3) node[GraphNode] {} node[right] {$v_{1}$};
	\end{tikzpicture}
	\end{equation*}
    \caption{We can take the left ``square-triangle'' sliding graph and shrink the subset given by the triangle (red). The result is another triangle graph $\tilde{\Gamma}(S)$, with a new vertex $p_0$ corresponding to the collapsed point of the triangle.}
    \label{fig:shrinkingGraph}
\end{figure}
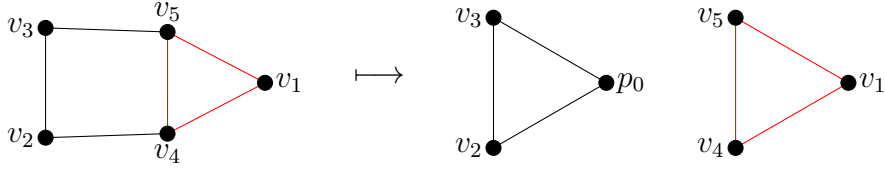

The pre-image $L[y]$ is thus precisely the moduli space of deformations we discussed above. We will now characterize the endpoints of this moduli space. First of all, we can write \eqref{eq:ydef} as 
\begin{equation}
    t_e y_e = \bar x_{e(0)}-\bar x_{e(1)}\,,
\end{equation}
and eliminate the $\bar x_v$ by summing over edges around loops, obtaining linear constraints on the $t_e$ alone. These give us a line\footnote{Recall that the number of loops is $|\tilde{\Gamma}_1|-|\tilde{\Gamma}_0|+1$.} in $\mathbb{RP}^{|\tilde{\Gamma}_1|-1}$ intersecting $\mathbb{RP}_{>}^{|\tilde{\Gamma}_1|-1}$ along a segment. The endpoints of the segment are configurations where a group of $t_e$ is much smaller than the rest. Correspondingly, a collection $S$ of $\bar x_v$ is much closer than the others in the plane. Given a subset $S$, we can define $\tilde{\Gamma}[S]$ as the induced subgraph of $\tilde{\Gamma}$, and $\tilde{\Gamma}(S)$ as the graph obtained from $\tilde{\Gamma}$ by collapsing $\tilde{\Gamma}[S]$ to a single new vertex $p_0$, see Figure \ref{fig:shrinkingGraph}. We will sometimes refer to the graph $\tilde{\Gamma}[S]$ as the ``cut'' diagram, in the sense that collapsing the diagram is like excising $S$ from the original $\tilde{\Gamma}$.
The linear nature of the problem makes it relatively easy to argue that a subset $S$ of the vertices of $\tilde \Gamma$ can shrink as we move along the pre-image of a generic $y_e$ only if both $\tilde{\Gamma}[S]$ and $\tilde{\Gamma}(S)$ are Laman \cite{Gaiotto:2015aoa}. We will denote such a subset $S$ as a ``Laman'' subset.

As we shrink $S$ towards zero size, the $t_e$ and $\bar x_v$ give us solutions of (\ref{eq:ydef}) for $\tilde{\Gamma}[S]$ and $\tilde{\Gamma}(S)$ with the same $y_e$ coordinates, where we identify edges of $\tilde{\Gamma}[S]$ and $\tilde{\Gamma}(S)$ with the corresponding edges of $\tilde \Gamma$. 
 
As a result, we have associated to each (generic) point $y_e \in \Delta_{\tilde{\Gamma}}$ a pair of collections $S_0[y]$ and $S_1[y]$ which can shrink as we move along $L[y]$, as well as a point $y_e \in \Delta_{\tilde{\Gamma}[S_0[y]]} \times \Delta_{\tilde{\Gamma}(S_1[y])}$ and a point $y_e \in \Delta_{\tilde{\Gamma}[S'_0[y]]} \times \Delta_{\tilde{\Gamma}(S'_1[y])}$. 
 
The construction can be inverted: for every Laman subset $S$ such that the graphs $\tilde{\Gamma}[S]$ and $\tilde{\Gamma}(S)$ admit edges with slopes inherited from some $y_e$, with vertices at positions $\bar x_{q}$ and $\bar x_{p}$ respectively, we can build a solution $\bar x_{v}$ by setting the positions of the vertices that are inherited from $\tilde{\Gamma}(S)$ to the corresponding $\bar x_{p}$ location, and the vertices inherited from $\tilde{\Gamma}[S]$ to the location $\bar x_{p_0} +\epsilon \bar x_{q}$, for small $\epsilon$. Candidate $t_e$ can be built analogously, rescaling the ones inherited from edges of $\tilde{\Gamma}[S]$ by $\epsilon$.

The slopes for the resulting $\tilde{\Gamma}$ embedding will be close to those of the original $y_e$ and for sufficiently small $\epsilon$ we can correct the $\bar x_v$ and $t_e$ to match the $y_e$. As a consequence, for each point $y_e$ in $\Delta_{\tilde{\Gamma}[S]} \times \Delta_{\tilde{\Gamma}(S)}$ we get a point with the same $y_e$ coordinates in $\Delta_{\tilde{\Gamma}}$, such that $S=S_0[y]$. We also get a second subset $S_1[y]$ from the other endpoint. Thus the collection of regions $\Delta_{\tilde{\Gamma}[S]} \times \Delta_{\tilde{\Gamma}(S)}$ in $\mathbb{R}^{2|\tilde \Gamma_1|}$ for all Laman subsets $S$ gives a double-cover of the region  $\Delta_{\tilde \Gamma}$. 

We can refine this statement by keeping track of the orientation of the neighbourhood of 
the point under consideration. For a Laman graph $\Gamma$, we equipped the $\Delta_\Gamma$ moduli spaces with an orientation $o_\Gamma$ which depends on the chosen ordering of the edges of $\Gamma$. Schematically, we can write 
\begin{equation}
    \prod_{e \in \Gamma_1} dt_e= d \rho_\Gamma \, o_\Gamma\,,
\end{equation}
where $o_\Gamma$ is the orientation on $\mathbb{RP}^{|\Gamma_1|-1}$, and thus $\rho_\Gamma$ is an overall scale.

Recall that, near an endpoint, a coordinate along $L[y]$ can be identified with $\rho_{\tilde \Gamma[S]}$. We can write 
\begin{equation}
    \prod_{e \in \tilde \Gamma_1} dt_e = \sigma(\tilde{\Gamma},S) \left[\prod_{e \in \tilde \Gamma[S]_1} dt_e\right]\left[\prod_{e \in \tilde \Gamma(S)_1} dt_e\right]\,,
\end{equation}
with $\sigma(\tilde{\Gamma},S)$ being the signature of the permutation required to match the ordering of edges of $\tilde{\Gamma}$ with the concatenation of the ordering of edges used in $\Delta_{\tilde{\Gamma}[S]}$ and in $\Delta_{\tilde{\Gamma}(S)}$. See Section \ref{sec:Examples} for examples of the signs.

Altogether, this implies\footnote{We are implicitly using the fact that Laman graphs have an odd number of edges.} the relation:
\begin{equation}
    \prod_{e \in \tilde \Gamma_1} dt_e = \sigma(\tilde{\Gamma},S) d \rho_{\tilde \Gamma[S]} d \rho_{\tilde \Gamma(S)} o_{\tilde \Gamma[S]} o_{\tilde \Gamma(S)}\,.
\end{equation}
We can roughly identify $\rho_{\tilde \Gamma(S)}$ with an overall scale and $\rho_{\tilde \Gamma[S]}$ as a coordinate along $L[y]$. The quantity
\begin{equation}
    \sigma(\tilde{\Gamma},S) o_{\tilde \Gamma[S]} o_{\tilde \Gamma(S)}
\end{equation}
gives us a local orientation on $\Delta_{\tilde{\Gamma}}$.

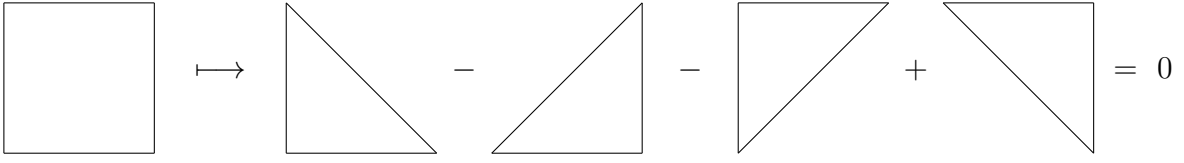
\begin{figure}
\centering
\begin{equation*}
   \begin{tikzpicture}
        [
    	baseline={(current bounding box.center)},
    	line join=round
    	]
        \def\gS{1};
    	\coordinate (pd1) at (-1.*\gS,1.*\gS);
    	\coordinate (pd2) at (1.*\gS,1.*\gS);
    	\coordinate (pd3) at (1.*\gS,-1.*\gS);
    	\coordinate (pd4) at (-1.*\gS,-1.*\gS);
    
    
    	\draw[GraphEdge] (pd1) -- (pd2) ;
    	\draw[GraphEdge] (pd2) -- (pd3);
    	\draw[GraphEdge] (pd3) -- (pd4) ;
    	\draw[GraphEdge] (pd1) -- (pd4) ;
    \end{tikzpicture}
\quad\longmapsto\quad
    \begin{tikzpicture}
		[
		baseline={(current bounding box.center)},
		line join=round
		]
		\def\gS{1};
    	\coordinate (pd1) at (-1.*\gS,1.*\gS);
    	\coordinate (pd2) at (1.*\gS,1.*\gS);
    	\coordinate (pd3) at (1.*\gS,-1.*\gS);
    	\coordinate (pd4) at (-1.*\gS,-1.*\gS);
        \draw[GraphEdge] (pd1) -- (pd3) ;
        \draw[GraphEdge] (pd3) -- (pd4);
        \draw[GraphEdge] (pd1) -- (pd4) ;
	\end{tikzpicture}
\,\,-\,\,
	\begin{tikzpicture}
		[
		baseline={(current bounding box.center)},
		line join=round
		]
		\def\gS{1};
    	\coordinate (pd1) at (-1.*\gS,1.*\gS);
    	\coordinate (pd2) at (1.*\gS,1.*\gS);
    	\coordinate (pd3) at (1.*\gS,-1.*\gS);
    	\coordinate (pd4) at (-1.*\gS,-1.*\gS);
    	\draw[GraphEdge] (pd2) -- (pd3) ;
    	\draw[GraphEdge] (pd3) -- (pd4);
    	\draw[GraphEdge] (pd2) -- (pd4) ;
	\end{tikzpicture}
\quad-\quad
	\begin{tikzpicture}
		[
		baseline={(current bounding box.center)},
		line join=round
		]
		\def\gS{1};
        \coordinate (pd1) at (-1.*\gS,1.*\gS);
    	\coordinate (pd2) at (1.*\gS,1.*\gS);
    	\coordinate (pd3) at (1.*\gS,-1.*\gS);
    	\coordinate (pd4) at (-1.*\gS,-1.*\gS);
    	\draw[GraphEdge] (pd1) -- (pd2);
    	\draw[GraphEdge] (pd2) -- (pd4);
    	\draw[GraphEdge] (pd4) -- (pd1);
	\end{tikzpicture}
\,\,+\,\,
	\begin{tikzpicture}
    	[
    	baseline={(current bounding box.center)},
    	line join=round
    	]
    	\def\gS{1};
        \coordinate (pd1) at (-1.*\gS,1.*\gS);
    	\coordinate (pd2) at (1.*\gS,1.*\gS);
    	\coordinate (pd3) at (1.*\gS,-1.*\gS);
    	\coordinate (pd4) at (-1.*\gS,-1.*\gS);
    	\draw[GraphEdge] (pd1) -- (pd2) ;
    	\draw[GraphEdge] (pd2) -- (pd3);
    	\draw[GraphEdge] (pd3) -- (pd1) ;
	\end{tikzpicture}
\,\,=\,\,
    0
\end{equation*}
\caption{A rough analogue of $\Delta_{\tilde \Gamma}$ consisting of a square region in the plane (left). The region can be decomposed into two triangular region in two ways by cutting along the diagonal. The difference between the two decompositions, which is a signed sum of triangular regions, obviously vanishes (right). }
\label{fig:quadexample}
\end{figure}

\begin{figure}
    \centering
    \begin{tikzpicture}
            \draw[] (1,1) node{};
            \draw[] (-1,1) node{};
            \draw[] (1,-1) node{};
            \draw[] (-1.3,-1.2) node{};
            \draw[] (1,1)--(-1,1);
             \draw[dotted,line width = .3mm] (-1,1)--(1,-1);
            \draw[] (1,-1)--(1,1);
            \draw[] (-1.3,-1.2)--(1,1);
            \draw[] (-1.3,-1.2)--(1,-1);
            \draw[] (-1.3,-1.2)--(-1,1);
            \draw[] (1,-2) node{};
            \draw[] (-.7,-2.1) node{};
            \draw[] (-1.3,-3) node{};
            \draw[] (.6,-3) node{};
            \draw[] (1,-2)--(-.7,-2.)--(-1.3,-3)--(.6,-3)--(1,-2);
            \draw[] (-.7,-2.)--(.6,-3);
            \draw[] (1,-2)--(-1.3,-3);
            \draw[] (-.3,-2.8)node{$\textcolor{red}{\bullet}$};
            \draw[] (-.3, 0) node{$\textcolor{red}{\bullet}$};
            \draw[] (-.3, -.8) node{$\textcolor{red}{\bullet}$};
            \draw[dotted, line width = .3mm,red] (-.3,-2.8)--(-.3, -.8);
            \draw[ line width = .3mm,red] (-.3, -.8)--(-.3, 0);
            \end{tikzpicture}
    \caption{We could also consider the four triangular regions as the projection on the plane of the facets of a tetrahedral region in space, with orientation induced by the projection. The preimage of a point in the square is a segment analogous to $L[y]$, with endpoints on two facets shown in red, which certifies the cancellation of a neighbourhood of the point in the signed sum.}
    \label{fig:tetrahedra}
\end{figure}
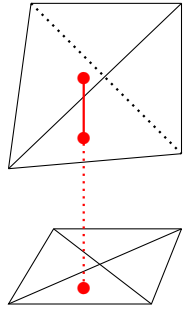

Crucially, the local orientations associated to two endpoints of the same $L[y]$ always disagree with each other, as $\rho_{\tilde \Gamma[S]}$ is a coordinate oriented away from the endpoint of the preimage. We thus derive the relation:
\begin{equation}\label{eq:DD}
	\boxed{\hphantom{.}\vphantom{\sum^.}\sum_{\mathrm{Laman}\, S} \sigma(\tilde{\Gamma},S) \Delta_{\tilde{\Gamma}[S]} \times \Delta_{\tilde{\Gamma}(S)} =0\hphantom{.}}\,,
\end{equation}
in the space of chains in $\mathbb{R}^{2|\tilde \Gamma_1|}$. Indeed, the terms in the sum cover the neighbourhood of any generic point of $\Delta_{\tilde \Gamma}$ twice with opposite local orientation and thus add up to zero. See Figure \ref{fig:quadexample} for a toy version of such a cover as well as Figure \ref{fig:tetrahedra} for another perspective of the double cover.

\subsection{A Quadratic Relation for the Feynman Integrals}

In order to promote the relation in \eqref{eq:DD} to a relation satisfied by the ${\cal I}$ integrals, we can consider the integral 
\begin{equation}
    \int_{\mathbb{R}^{2 |\Gamma_0|-2} \times \Delta_{\tilde{\Gamma}[S]} \times \Delta_{\tilde{\Gamma}(S)}} \Omega_{\tilde{\Gamma}}[\lambda;z] 
\end{equation}
for a sliding graph $\tilde{\Gamma}$ and Laman subset $S$. We will denote vertices associated to $\tilde{\Gamma}[S]$ by $v$, and vertices associated to $\tilde{\Gamma}(S)$ by $p$. Without loss of generality, we can pick the reference vertex $v_0$ to belong to $S$. We can take $v = v_0$ and $p_0$ to be the new vertex.  

We start with the second factor in $\Omega_{\tilde{\Gamma}}$, which factors nicely to:
\begin{equation}
    \prod_{v \in \tilde{\Gamma}_0[S]|v \neq v_0} e^{\lambda_{v} \cdot x_{v}} \frac{d^2 x_{v}}{(2\pi i)^2} \prod_{p \in \tilde{\Gamma}_0(S)|p \neq p_0} e^{\lambda_{p} \cdot x_{p}} \frac{d^2 x_{p}}{(2\pi i)^2}\,.
\end{equation}
However, the first factor is a bit more subtle: although we can factor it as
\begin{equation}
    \prod_{e \in \tilde{\Gamma}_1[S]}   e^{- y_{e} \cdot (x_{e(0)} - x_{e(1)}+z_{e})} d^2 y_{e} \prod_{\ell \in \tilde{\Gamma}_1(S)}  e^{- y_{\ell} \cdot (x_{\ell(0)} - x_{\ell(1)}+z_{\ell})} d^2 y_{\ell}\,,
\end{equation}
we need to be careful with the edges $\ell$ which end on vertices of $\tilde{\Gamma}[S]$, as the corresponding positions $x_{\ell(0)}$ or $x_{\ell(1)}$ differ from $x_{p_0}=0$. 

We can compensate for this difference by shifting $z_{\ell} \to z_{\ell}+ x_{\ell(0)}$ or $z_{\ell} \to z_{\ell}- x_{\ell(1)}$ if the edges end on $\tilde{\Gamma}[S]$. We can then convert these $x_{v}$ shifts to difference operators $e^{x_{\ell(0)} \cdot \partial_{z_{\ell}}}$ or $e^{-x_{\ell(1)} \cdot \partial_{z_{\ell}}}$ and combine them with the corresponding $e^{\lambda_{v} \cdot x_{v}}$ factors to $e^{(\lambda_{\ell(0)}+\partial_{z_{\ell}})\cdot x_{\ell(0)}}$ and $e^{(\lambda_{\ell(1)}-\partial_{z_{\ell}})\cdot x_{\ell(1)}}$. 

This allows us to factor $\Omega_{\tilde{\Gamma}}$ ``operatorially'' as 
\begin{equation}
    \Omega_{\tilde{\Gamma}}[\lambda;z] = \Omega_{\tilde{\Gamma}[S]}[\lambda_v + \partial_v;z_e] \Omega_{\tilde{\Gamma}(S)}[\lambda_p;z_\ell]\,,
\end{equation}
with 
\begin{equation}
    \partial_v \equiv \sum_{\ell| \ell(0) = v} \partial_{z_\ell}-\sum_{\ell| \ell(1) = v} \partial_{z_\ell}\,.
\end{equation}
Here $e$ and $v$ denote edges and vertices of $\tilde{\Gamma}$ inherited from $\tilde{\Gamma}[S]$ and $\ell$ and $p$ denote edges and vertices of $\tilde{\Gamma}$ inherited from $\tilde{\Gamma}(S)$.  

Integrating over $\mathbb{R}^{2 |\Gamma_0|-2} \times \Delta_{\tilde{\Gamma}[S]} \times \Delta_{\tilde{\Gamma}(S)}$ leads to the crucial identity:
\begin{equation}\label{eq:II}
\boxed{\hphantom{.}\vphantom{\sum^.}\sum_{\mathrm{Laman}\, S} \sigma(\tilde{\Gamma},S) \, {\cal I}_{\tilde{\Gamma}[S]}[\lambda_v + \partial_v;z_e] \, {\cal I}_{\tilde{\Gamma}(S)}[\lambda_p;z_\ell] =0\hphantom{.}}\,.
\end{equation}

We should observe that the shift symmetries of the ${\cal I}_\Gamma$ are necessary for this identity to be well-defined. Indeed, as the $\lambda_v$ for $v \in \tilde{\Gamma}[S]_0$ do not add up to $0$, the choice of reference vertex in ${\cal I}_{\tilde{\Gamma}[S]}[\lambda_v + \partial_v;z_e]$ would naively seem to matter. But 
\begin{equation}
    \sum_{v \in \tilde{\Gamma}[S]_0} (\lambda_v + \partial_v) =  \lambda_{p_0} + \sum_{\ell \in \tilde{\Gamma}(S)_1| \ell(0) = p_0} \partial_{z_\ell}-\sum_{\ell\in \tilde{\Gamma}(S)_1| \ell(1) = p_0} \partial_{z_\ell}
\end{equation}
annihilates ${\cal I}_{\tilde{\Gamma}(S)}[\lambda_p;z_\ell]$, by the infinitesimal version of the shift symmetry \eqref{eq:shift} at $p_0$. 

We should also observe that the simplest class of sliding graphs consists of two Laman graphs fused at one vertex. The quadratic identity for such ``butterfly'' sliding graphs has two terms, corresponding to the collapse of either of the two Laman graphs,  and is essentially trivial: if we pick the fused vertex as a reference for both graphs there are no shifts and we just get 
something like ${\cal I}_\Gamma {\cal I}_{\Gamma'} - {\cal I}_{\Gamma'} {\cal I}_\Gamma=0$. The relative sign follows from 
the fact that Laman graphs have an odd number of edges.

\section{Examples}\label{sec:Examples}
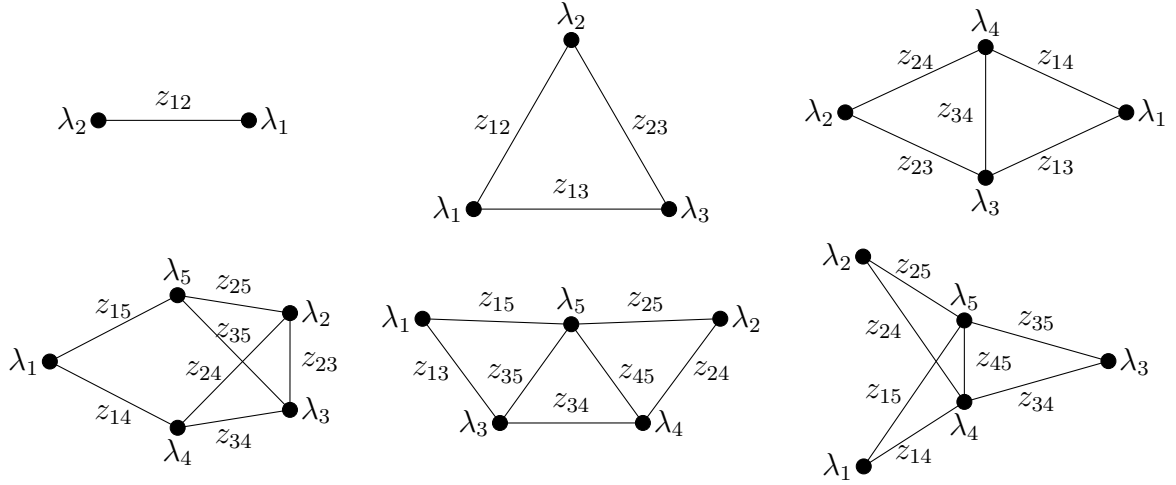
\begin{figure}
\centering
\setlength\tabcolsep{4pt} 
\renewcommand{\arraystretch}{3} 
\begin{tabular}{ccc}
   \begin{tikzpicture}
   [
	baseline={(current bounding box.center)},
	line join=round
	]
	\def\gS{1};
	\coordinate (pd1) at (1.*\gS,0.*\gS);
	\coordinate (pd2) at (-1.*\gS,0.*\gS);
	\draw (pd1) node[GraphNode] {} node[right] {$\lambda_{1}$};
	\draw (pd2) node[GraphNode] {} node[left] {$\lambda_{2}$};
	\draw[GraphEdge] (pd1) -- (pd2) node[midway, above] {$z_{12}$};
   \end{tikzpicture}
&
    \begin{tikzpicture}
    [
	baseline={(current bounding box.center)},
	line join=round
	]
    \def\gS{1.5};
	\coordinate (pd1) at (-0.866*\gS,-0.5*\gS);
	\coordinate (pd2) at (0.*\gS,1.*\gS);
	\coordinate (pd3) at (0.866*\gS,-0.5*\gS);
	\draw (pd1) node[GraphNode] {} node[left] {$\lambda_{1}$};
	\draw (pd2) node[GraphNode] {} node[above] {$\lambda_{2}$};
	\draw (pd3) node[GraphNode] {} node[right] {$\lambda_{3}$};
	\draw[GraphEdge] (pd1) -- (pd2) node[midway, left] {$z_{12}$};
	\draw[GraphEdge] (pd1) -- (pd3) node[midway, above] {$z_{13}$};
	\draw[GraphEdge] (pd2) -- (pd3) node[midway, right] {$z_{23}$};
   \end{tikzpicture}
&
    \begin{tikzpicture}
    [
	baseline={(current bounding box.center)},
	line join=round
	]
    \def\gS{2};
	\coordinate (pd1) at (1.868*\gS,0.4348*\gS);
	\coordinate (pd2) at (0.*\gS,0.4348*\gS);
	\coordinate (pd3) at (0.9331*\gS,0.*\gS);
	\coordinate (pd4) at (0.9331*\gS,0.8691*\gS);
	\draw (pd1) node[GraphNode] {} node[right] {$\lambda_{1}$};
	\draw (pd2) node[GraphNode] {} node[left] {$\lambda_{2}$};
	\draw (pd3) node[GraphNode] {} node[below] {$\lambda_{3}$};
	\draw (pd4) node[GraphNode] {} node[above] {$\lambda_{4}$};
	\draw[GraphEdge] (pd1) -- (pd3) node[midway, below] {$z_{13}$};
	\draw[GraphEdge] (pd1) -- (pd4) node[midway, above] {$z_{14}$};
	\draw[GraphEdge] (pd2) -- (pd3) node[midway, below] {$z_{23}$};
	\draw[GraphEdge] (pd2) -- (pd4) node[midway, above] {$z_{24}$};
	\draw[GraphEdge] (pd3) -- (pd4) node[midway, left] {$z_{34}$};
   \end{tikzpicture}
    \\
    \begin{tikzpicture}
    [
	baseline={(current bounding box.center)},
	line join=round
	]
    \def\gS{1.8};
	\coordinate (pd1) at (0.*\gS,0.4896*\gS);
	\coordinate (pd2) at (1.774*\gS,0.8475*\gS);
	\coordinate (pd3) at (1.7734*\gS,0.1323*\gS);
	\coordinate (pd4) at (0.9441*\gS,0.*\gS);
	\coordinate (pd5) at (0.9431*\gS,0.9798*\gS);
	\draw (pd1) node[GraphNode] {} node[left] {$\lambda_{1}$};
	\draw (pd2) node[GraphNode] {} node[right] {$\lambda_{2}$};
	\draw (pd3) node[GraphNode] {} node[right] {$\lambda_{3}$};
	\draw (pd4) node[GraphNode] {} node[below] {$\lambda_{4}$};
	\draw (pd5) node[GraphNode] {} node[above] {$\lambda_{5}$};
	\draw[GraphEdge] (pd1) -- (pd4) node[midway, below] {$z_{14}$};
	\draw[GraphEdge] (pd1) -- (pd5) node[midway, above] {$z_{15}$};
	\draw[GraphEdge] (pd2) -- (pd3) node[midway, right] {$z_{23}$};
	\draw[GraphEdge] (pd2) -- (pd4) node[midway, left] {$z_{24}$};
	\draw[GraphEdge] (pd2) -- (pd5) node[midway, above] {$z_{25}$};
	\draw[GraphEdge] (pd3) -- (pd4) node[midway, below] {$z_{34}$};
	\draw[GraphEdge] (pd3) -- (pd5) node[midway, above] {$z_{35}$};
   \end{tikzpicture}
&
    \begin{tikzpicture}
   	[
	baseline={(current bounding box.center)},
	line join=round
	]
    \def\gS{1.8};
	\coordinate (pd1) at (0.*\gS,0.7704*\gS);
	\coordinate (pd2) at (2.1993*\gS,0.7708*\gS);
	\coordinate (pd3) at (0.572*\gS,0.*\gS);
	\coordinate (pd4) at (1.6278*\gS,0.0004*\gS);
	\coordinate (pd5) at (1.0985*\gS,0.73*\gS);
	\draw (pd1) node[GraphNode] {} node[left] {$\lambda_{1}$};
	\draw (pd2) node[GraphNode] {} node[right] {$\lambda_{2}$};
	\draw (pd3) node[GraphNode] {} node[left] {$\lambda_{3}$};
	\draw (pd4) node[GraphNode] {} node[right] {$\lambda_{4}$};
	\draw (pd5) node[GraphNode] {} node[above] {$\lambda_{5}$};
	\draw[GraphEdge] (pd1) -- (pd3) node[midway, left] {$z_{13}$};
	\draw[GraphEdge] (pd1) -- (pd5) node[midway, above] {$z_{15}$};
	\draw[GraphEdge] (pd2) -- (pd4) node[midway, right] {$z_{24}$};
	\draw[GraphEdge] (pd2) -- (pd5) node[midway, above] {$z_{25}$};
	\draw[GraphEdge] (pd3) -- (pd4) node[midway, above] {$z_{34}$};
	\draw[GraphEdge] (pd3) -- (pd5) node[midway, left] {$z_{35}$};
	\draw[GraphEdge] (pd4) -- (pd5) node[midway, right] {$z_{45}$};
    \end{tikzpicture}
&
    \begin{tikzpicture}
    [
	baseline={(current bounding box.center)},
	line join=round
	]
    \def\gS{1.9};
	\coordinate (pd1) at (0.0043*\gS,0.*\gS);
	\coordinate (pd2) at (0.*\gS,1.4695*\gS);
	\coordinate (pd3) at (1.7164*\gS,0.7374*\gS);
	\coordinate (pd4) at (0.7084*\gS,0.4518*\gS);
	\coordinate (pd5) at (0.7082*\gS,1.0224*\gS);
	\draw (pd1) node[GraphNode] {} node[left] {$\lambda_{1}$};
	\draw (pd2) node[GraphNode] {} node[left] {$\lambda_{2}$};
	\draw (pd3) node[GraphNode] {} node[left, right] {$\lambda_{3}$};
	\draw (pd4) node[GraphNode] {} node[left,below] {$\lambda_{4}$};
	\draw (pd5) node[GraphNode] {} node[left, above] {$\lambda_{5}$};
	\draw[GraphEdge] (pd1) -- (pd4) node[midway, below] {$z_{14}$};
	\draw[GraphEdge] (pd1) -- (pd5) node[midway, left] {$z_{15}$};
	\draw[GraphEdge] (pd2) -- (pd4) node[midway, left] {$z_{24}$};
	\draw[GraphEdge] (pd2) -- (pd5) node[midway, above] {$z_{25}$};
	\draw[GraphEdge] (pd3) -- (pd4) node[midway, below] {$z_{34}$};
	\draw[GraphEdge] (pd3) -- (pd5) node[midway, above] {$z_{35}$};
	\draw[GraphEdge] (pd4) -- (pd5) node[midway, right] {$z_{45}$};
    \end{tikzpicture}
\end{tabular}
\caption{We give six examples of Laman graphs for which we have given a choice of vertices. The indices associated to the $z$'s between vertices are such that swapping the indices introduces a minus sign. }\label{fig:canLabLaman}
\end{figure}

In the following we label the vertices of graphs by numbers and use the lexicographic order for the (unoriented) edges.\footnote{The vertex labelling follows the conventions employed by Mathematica's CanonicalGraph function, but we caution the reader that these conventions are version dependent.} See Figure \ref{fig:canLabLaman}.\footnote{We have included accompanying code for these examples in both our arXiv upload as well as at \cite{Laman_Loopstrap_2022}.}

\subsection{The Segment: Tree-Level Propagator}
The simplest example is the graph consisting of a single edge:
\begin{equation}
\begin{tikzpicture}
	[
	baseline={(current bounding box.center)},
	line join=round
	]
	
	\coordinate (pd1) at (1.*\gS,0.*\gS);
	\coordinate (pd2) at (-1.*\gS,0.*\gS);
	
	\draw (pd1) node[GraphNode] {} node[right] {$\lambda_{1}$};
	\draw (pd2) node[GraphNode] {} node[left] {$\lambda_{2}$};
	
	\draw[GraphEdge] (pd1) -- (pd2) node[midway, above] {$z_{12}$};
\end{tikzpicture}.
\end{equation}
If we set the first point at the origin, we can write:
\begin{equation}
	\Omega_{\,\pick{1.5ex}{segment}} = e^{- y_{12} \cdot (-x_2 +z_{12})} d^2 y_{12} \, e^{\lambda_2 \cdot x_2} \frac{d^2 x_2}{(2 \pi i)^2} \,.
\end{equation}
The $\Delta_{\pick{1.5ex}{segment}}$ region coincides with the $y_{12}$ plane. 
 
Performing the $x_2$ integral sets $y_{12} = - \lambda_2$, and we arrive at
\begin{equation}
	\mathcal{I}_{\,\pick{1.5ex}{segment}} = e^{\lambda_2\cdot z_{12}}\,.
\end{equation}
The other choice of reference vertex gives the equivalent 
\begin{equation}
	\mathcal{I}_{\,\pick{1.5ex}{segment}} = e^{-\lambda_1\cdot z_{12}}\,.
\end{equation}

The functional form of the answer is also fixed directly by the expected shift symmetry of $\mathcal{I}_{\pick{1.5ex}{segment}}$: the graph has no loops and must have weight $0$ under scaling of $\lambda_v$ and $z_e$ in opposite directions. 

\subsection{The Bi-Segment: The First Sliding Graph}
The bi-segment sliding graph with two edges:
\begin{equation}
	\begin{tikzpicture}
		[
		baseline={(current bounding box.center)},
		line join=round
		]
		\def\a{2};

		\coordinate (pd1) at (0.*\a,0.*\a);
		\coordinate (pd2) at (1.*\a,0.*\a);
		\coordinate (pd3) at (2.*\a,0.*\a);
		
		\draw (pd1) node[GraphNode] {} node[above] {$\lambda_{1}$};
		\draw (pd2) node[GraphNode] {} node[above] {$\lambda_{3}$};
		\draw (pd3) node[GraphNode] {} node[above] {$\lambda_{2}$};
		
		\draw[GraphEdge] (pd1) -- (pd2) node[midway, above] {$z_{13}$};
		\draw[GraphEdge] (pd2) -- (pd3) node[midway, above] {$z_{23}$};
	\end{tikzpicture}\,,
\end{equation}
provides us with the first example of a quadratic relation. It is also the first example of a trivial quadratic relation, which places no constraints on $\mathcal{I}_{\pick{1.5ex}{segment}}$ beyond the shift symmetry (\ref{eq:shift}). We will still spell out the details for completeness.

In the quadratic relation \eqref{eq:DD} we have two terms: $I_{13}$ acting on $I_{23}$ and vice-versa
\begin{equation}
	0=
	\begin{tikzpicture}
		[
		line join=round
		]
		
		\coordinate (pd1) at (1.*\gS,0.*\gS);
		\coordinate (pd2) at (-1.*\gS,0.*\gS);
		
		\draw (pd1) node[GraphNode] {} node[above] {$\lambda_{1}$};
		\draw (pd2) node[GraphNode] {} node[above] {$\lambda_{3}$};
		
		\draw[GraphEdge] (pd1) -- (pd2) node[midway, above] {$z_{13}$};
	\end{tikzpicture}
	\begin{tikzpicture}
		[
		line join=round
		]
		
		\coordinate (pd1) at (1.*\gS,0.*\gS);
		\coordinate (pd2) at (-1.*\gS,0.*\gS);
		
		\draw (pd1) node[GraphNode] {} node[above] {$\lambda_{1+3}$};
		\draw (pd2) node[GraphNode] {} node[above] {$\lambda_{2}$};
		
		\draw[GraphEdge] (pd1) -- (pd2) node[midway, above] {$z_{23}$};
		
	\end{tikzpicture}
	-
	\begin{tikzpicture}
		[
		line join=round
		]
		
		\coordinate (pd1) at (1.*\gS,0.*\gS);
		\coordinate (pd2) at (-1.*\gS,0.*\gS);
		
		\draw (pd1) node[GraphNode] {} node[above] {$\lambda_{2}$};
		\draw (pd2) node[GraphNode] {} node[above] {$\lambda_{3}$};
		
		\draw[GraphEdge] (pd1) -- (pd2) node[midway, above] {$z_{23}$};
	\end{tikzpicture}
	\begin{tikzpicture}
		[
		line join=round
		]
		
		\coordinate (pd1) at (1.*\gS,0.*\gS);
		\coordinate (pd2) at (-1.*\gS,0.*\gS);
		
		\draw (pd1) node[GraphNode] {} node[above] {$\lambda_{1}$};
		\draw (pd2) node[GraphNode] {} node[above] {$\lambda_{2+3}$};
		
		\draw[GraphEdge] (pd1) -- (pd2) node[midway, above] {$z_{13}$};
		
	\end{tikzpicture}\,.
\end{equation}
where $\lambda_{2+3}$ is a shorthand notation for $\lambda_2 + \lambda_3$. If we were to pick vertex $3$ as the reference vertex for both graphs, the relation trivializes: no $\lambda$ variables are shifted and the relation becomes 
\begin{equation}
	0 = e^{-\lambda_1\cdot z_{13}}e^{-\lambda_2\cdot z_{23}}-e^{-\lambda_2\cdot z_{23}}e^{-\lambda_1\cdot z_{13}} \,.
\end{equation}
However, if we pick vertex $1$ as the reference in the first term, we get the same expression in a more complicated way:
\begin{align}
	0
	&=e^{(\lambda_3- \partial_{z_{23}}) \cdot z_{13}}e^{-\lambda_2\cdot z_{23}} 
		-e^{-\lambda_2\cdot z_{23}} e^{(\lambda_2 + \lambda_3)\cdot z_{13}}\\
	&=e^{\lambda_3\cdot z_{13}} e^{-\lambda_2\cdot (z_{23}-z_{13})} 
		-e^{-\lambda_2\cdot z_{23}} e^{(\lambda_2 + \lambda_3)\cdot z_{13}}\,.
\end{align}

\subsection{The Triangle: One-Loop Diagram}\label{ssection:triangle}
This is the basic triangle diagram:
\begin{equation}
	\begin{tikzpicture}
		[
		baseline={(current bounding box.center)},
		line join=round
		]
		\def\a{2};
		
		\coordinate (pd1) at (-0.866*\a,-0.5*\a);
		\coordinate (pd2) at (0.866*\a,-0.5*\a);
		\coordinate (pd3) at (0.*\a,1.*\a);
		
		\draw (pd1) node[GraphNode] {} node[left] {$\lambda_{1}$};
		\draw (pd2) node[GraphNode] {} node[right] {$\lambda_{3}$};
		\draw (pd3) node[GraphNode] {} node[above] {$\lambda_{2}$};
		
		\draw[GraphEdge] (pd1) -- (pd2) node[midway, above] {$z_{13}$};
		\draw[GraphEdge] (pd1) -- (pd3) node[midway, above left] {$z_{12}$};
		\draw[GraphEdge] (pd2) -- (pd3) node[midway, above right] {$z_{23}$};
		
	\end{tikzpicture}
\end{equation}
This is the only Feynman amplitude we will need to compute by direct integration. Even so, the integral is only really needed to get an overall constant, as the full functional form could be determined via the quadratic identity associated to the square graph. We will discuss that momentarily.

We isolate a direct calculation of $\omega_{\pic{1.2ex}{triangle}}$ to Appendix \ref{append:1loopintegrand}. The region $\Delta_{\pic{1.2ex}{triangle}}$ consists of values for $y_{12}$, $y_{23}$, $y_{13}$ such that $y_{13}$
is a positive linear combination of $y_{12}$ and $y_{23}$:
\begin{equation}
    y_{13} = \frac{t_{12}}{t_{13}} y_{12} + \frac{t_{23}}{t_{13}} y_{23} \, .
\end{equation} 
Equivalently, $y_{12}$, $y_{23}$ and $y_{31}= - y_{13}$ are vertices of a triangle in the momentum plane which includes the origin. The orientation of $\Delta_{\pic{1.2ex}{triangle}}$ is controlled by the orientation of the triangle in the momentum plane. 

We can pick vertex $1$ as the reference vertex and perform a linear change of coordinates:
\begin{equation}
    y_{12} =\mu_2 + y_{23}\,, \qquad \qquad y_{13} = \mu_3 - y_{23}\,,
\end{equation}
so that 
\begin{equation}
	\Omega_{\pic{1.2ex}{triangle}}[\lambda;z] \equiv e^{- \mu_2 \cdot z_{12}- \mu_3 \cdot z_{13} } e^{- y_{23} \cdot (z_{12}+ z_{23} - z_{13}) } d^2 y_{23}   \prod_{v =2,3} e^{\left(\lambda_v+\mu_v \right)\cdot x_v} d^2 \mu_v \frac{d^2 x_v}{(2 \pi i)^2} \,.
\end{equation}
The integral over $x_v$ and $\mu_v$ enforces $\lambda_v + \mu_v = 0$, leaving us with 
\begin{equation}
	{\mathcal I}_{\pic{1.2ex}{triangle}}[\lambda;z] = e^{\lambda_2 \cdot z_{12}+ \lambda_3 \cdot z_{13} } \int_{y \in (0, \lambda_2, - \lambda_3)} e^{- y_{23} \cdot (z_{12}+ z_{23} - z_{13}) } d^2 y_{23} \,,
\end{equation}
where the integration region is the triangle with vertices $0$, $\lambda_2$, $- \lambda_3$. Indeed, setting $t_{12} + t_{23} + t_{13} =1$, we have
\begin{equation}
    y_{23} = t_{12} \lambda_2 - t_{13} \lambda_3\,.
\end{equation}
Defining the loop momentum $Z = z_{12}+ z_{23} - z_{13}$, we compute
\begin{align}
	{\mathcal I}_{\pic{1.2ex}{triangle}}[\lambda;z] 
	    &= e^{\lambda_2 \cdot z_{12}+ \lambda_3 \cdot z_{13} } (\lambda_2 \wedge \lambda_3) \int_{t_{12}=0}^1 \int_{t_{13} = 0}^{1-t_{12}} e^{t_{13}\lambda_3 \cdot Z - t_{12}\lambda_2 \cdot Z } dt_{13} dt_{12} \\
	    &= e^{\lambda_2 \cdot z_{12}+ \lambda_3 \cdot z_{13} } \frac{(\lambda_2 \wedge \lambda_3)}{(\lambda_3 \cdot Z)} \int_{t_{12}=0}^1 \left[ e^{\lambda_3 \cdot Z + t_{12}\lambda_1 \cdot Z }-e^{- t_{12}\lambda_2 \cdot Z }\right]  dt_{12} \\
	    &= e^{\lambda_2 \cdot z_{12}+ \lambda_3 \cdot z_{13} } \frac{(\lambda_2 \wedge \lambda_3)}{(\lambda_3 \cdot Z)} \left[ \frac{e^{-\lambda_2 \cdot Z}-e^{\lambda_3 \cdot Z}}{(\lambda_1 \cdot Z) }+\frac{e^{- \lambda_2 \cdot Z }-1}{(\lambda_2 \cdot Z)}\right]  dt_{12} \,,
\end{align}
so that finally:
\begin{equation} \label{eq:triint}
	{\mathcal I}_{\pic{1.2ex}{triangle}}[\lambda;z] = - e^{\lambda_2 \cdot z_{12}+ \lambda_3 \cdot z_{13} } (\lambda_2 \wedge \lambda_3) \left[ \frac{e^{-\lambda_2 \cdot Z}}{(\lambda_1 \cdot Z)(\lambda_2 \cdot Z) }+ \frac{e^{\lambda_3 \cdot Z}}{(\lambda_1 \cdot Z)(\lambda_3 \cdot Z) }+\frac{1}{(\lambda_2 \cdot Z)(\lambda_3 \cdot Z)}\right] \,.
\end{equation}

As expected, this expression can be expanded out as a power series:
\begin{align}
    \calI_{\pic{1.2ex}{triangle}}[\lambda;z]  
        &= (\lambda_2 \wedge \lambda_3) e^{\lambda_2 \cdot z_{12} +\lambda_3 \cdot z_{13}}  \sum_{n=0}^\infty \frac{1}{(n+1)! (\lambda_2+\lambda_3)  \cdot Z} [(\lambda_3  \cdot Z)^n- (-\lambda_2 \cdot Z)^n] \nonumber\\
        &= (\lambda_2 \wedge \lambda_3) e^{\lambda_2 \cdot z_{12} +\lambda_3 \cdot z_{13}} \sum_{n=0}^\infty \sum_{m=0}^\infty \frac{1}{(n+m+2)! } \cdot  (\lambda_3  \cdot Z)^n (-\lambda_2 \cdot Z)^m\,.
\end{align}

The expression has the same symmetries as the triangle diagram. For example, rewriting it in terms of $\lambda_1$, $\lambda_2$ we get:
\begin{equation}
    {\mathcal I}_{\pic{1.2ex}{triangle}}[\lambda;z] =  - e^{ -\lambda_1 \cdot z_{13}-\lambda_2 \cdot z_{23}} (\lambda_1 \wedge \lambda_2) \Biggl[\frac{1}{(\lambda_1\cdot Z)(\lambda_2  \cdot Z)} +\frac{e^{ -\lambda_1  \cdot Z}}{(\lambda_1  \cdot Z) (\lambda_3 \cdot Z)}+\frac{e^{\lambda_2  \cdot Z}}{(\lambda_2  \cdot Z)(\lambda_3 \cdot Z)}\Biggr] 
\end{equation}
which is just obtained from the previous expression by permuting $1 \mapsto 2 \mapsto 3 \mapsto 1$. It is also antisymmetric under permutation of the $2$ and $3$ vertices: the sign is due to a reordering of the edges. 

The shift symmetry is also manifest. We will find it useful to write:
\begin{equation}\label{eq:shiftSymTriangle}
    {\mathcal I}_{\pic{1.2ex}{triangle}}[\lambda;z]   = e^{ -\lambda_1 \cdot z_{13}-\lambda_2 \cdot z_{23}} f_{\pic{1.2ex}{triangle}}[\lambda_1,\lambda_2; z_{12}+ z_{23}-z_{13}] \,,
\end{equation}
with
\begin{equation}\label{}
    f_{\pic{1.2ex}{triangle}}[\lambda_1,\lambda_2; Z]= - (\lambda_1 \wedge \lambda_2)  \Bigl[\frac{1}{\lambda_1\cdot Z \,\lambda_2  \cdot Z} +\frac{e^{ -\lambda_1  \cdot Z}}{\lambda_1  \cdot Z \,(-\lambda_1-\lambda_2)  \cdot Z}+\frac{e^{\lambda_2  \cdot Z}}{\lambda_2  \cdot Z\,(-\lambda_1-\lambda_2) \cdot Z}\Bigr] \,.
\end{equation}
The symmetries of the triangle become:
\begin{align} \label{eq:sym1}
    f_{\pic{1.2ex}{triangle}}[\lambda_2,\lambda_1;- Z]
        &= -f_{\pic{1.2ex}{triangle}}[\lambda_1,\lambda_2; Z]\,,\\
    f_{\pic{1.2ex}{triangle}}[\lambda_1,\lambda_2; Z]
        &= e^{\lambda_2 \cdot Z} f{\pic{1.2ex}{triangle}}[\lambda_2,-\lambda_1-\lambda_2;Z] 
        = e^{-\lambda_1 \cdot Z} f_{\pic{1.2ex}{triangle}}[-\lambda_1-\lambda_2,\lambda_1;Z]\,.
\end{align}

\subsection{The Square: Bootstrapping One-Loop}
This is the basic square diagram: 
\begin{equation}
    \begin{tikzpicture}
    [
	baseline={(current bounding box.center)},
	line join=round
	]
    \def\gS{2};
	\coordinate (pd1) at (-1.*\gS,0.*\gS);
	\coordinate (pd2) at (0.*\gS,1.*\gS);
	\coordinate (pd3) at (0.*\gS,-1.*\gS);
	\coordinate (pd4) at (1.*\gS,0.*\gS);

	\draw (pd1) node[GraphNode] {} node[left] {$\lambda_{1}$};
	\draw (pd2) node[GraphNode] {} node[left,above] {$\lambda_{2}$};
	\draw (pd3) node[GraphNode] {} node[left,below] {$\lambda_{3}$};
	\draw (pd4) node[GraphNode] {} node[right] {$\lambda_{4}$};

	\draw[GraphEdge] (pd1) -- (pd2) node[midway, above left] {$z_{12}$};
	\draw[GraphEdge] (pd1) -- (pd3) node[midway, below left] {$z_{13}$};
	\draw[GraphEdge] (pd2) -- (pd4) node[midway, above right] {$z_{24}$};
	\draw[GraphEdge] (pd3) -- (pd4) node[midway, below right] {$z_{34}$};
    \end{tikzpicture}
\end{equation}

The quadratic identity associated to this sliding graph is non-trivial. The four possible Laman subsets of the square consist of pairs of consecutive vertices: one edge of the square is contracted leaving behind a triangle. We thus get a four-term difference equation for the $\calI_{\pic{1.2ex}{triangle}}$'s:
\begin{alignat}{3}
0 &= 
\begin{tikzpicture}
    [
	baseline={(current bounding box.center)},
	line join=round
	]
    \def\gS{0.9};
	\coordinate (pd1) at (1.*\gS,0.*\gS);
	\coordinate (pd2) at (-1.*\gS,0.*\gS);
	\draw (pd1) node[GraphNode] {} node[above] {$\lambda_{1}$};
	\draw (pd2) node[GraphNode] {} node[above] {$\lambda_{2}$};
	\draw[GraphEdge] (pd1) -- (pd2) node[midway, above] {$z_{12}$};
\end{tikzpicture}
\begin{tikzpicture}
    [   
	baseline={(current bounding box.center)},
	line join=round
	]
	\def\gS{1.1};
	\coordinate (pd1) at (-0.866*\gS,-0.5*\gS);
	\coordinate (pd2) at (0.*\gS,1.*\gS);
	\coordinate (pd3) at (0.866*\gS,-0.5*\gS);
	\draw (pd1) node[GraphNode] {} node[left] {$\lambda_{1+2}$};
	\draw (pd2) node[GraphNode] {} node[left,above] {$\lambda_{3}$};
	\draw (pd3) node[GraphNode] {} node[right] {$\lambda_{4}$};
	\draw[GraphEdge] (pd1) -- (pd2) node[midway, left] {$z_{13}$};
	\draw[GraphEdge] (pd1) -- (pd3) node[midway, above] {$z_{24}$};
	\draw[GraphEdge] (pd2) -- (pd3) node[midway, right] {$z_{34}$};
\end{tikzpicture}
&&+\,\,\,
\begin{tikzpicture}
    [
	baseline={(current bounding box.center)},
	line join=round
	]
    \def\gS{0.9};
	\coordinate (pd1) at (1.*\gS,0.*\gS);
	\coordinate (pd2) at (-1.*\gS,0.*\gS);
	\draw (pd1) node[GraphNode] {} node[above] {$\lambda_{2}$};
	\draw (pd2) node[GraphNode] {} node[above] {$\lambda_{4}$};
	\draw[GraphEdge] (pd1) -- (pd2) node[midway, above] {$z_{24}$};
\end{tikzpicture}
\begin{tikzpicture}
    [   
	baseline={(current bounding box.center)},
	line join=round
	]
	\def\gS{1.1};
	\coordinate (pd1) at (-0.866*\gS,-0.5*\gS);
	\coordinate (pd2) at (0.*\gS,1.*\gS);
	\coordinate (pd3) at (0.866*\gS,-0.5*\gS);
	\draw (pd1) node[GraphNode] {} node[left] {$\lambda_{1}$};
	\draw (pd2) node[GraphNode] {} node[left,above] {$\lambda_{2+4}$};
	\draw (pd3) node[GraphNode] {} node[right] {$\lambda_{3}$};
	\draw[GraphEdge] (pd1) -- (pd2) node[midway, left] {$z_{12}$};
	\draw[GraphEdge] (pd1) -- (pd3) node[midway, above] {$z_{13}$};
	\draw[GraphEdge] (pd2) -- (pd3) node[midway, right] {$z_{34}$};
\end{tikzpicture}\nonumber\\
&+
\begin{tikzpicture}
    [
	baseline={(current bounding box.center)},
	line join=round
	]
    \def\gS{0.9};
	\coordinate (pd1) at (1.*\gS,0.*\gS);
	\coordinate (pd2) at (-1.*\gS,0.*\gS);
	\draw (pd1) node[GraphNode] {} node[above] {$\lambda_{1}$};
	\draw (pd2) node[GraphNode] {} node[above] {$\lambda_{3}$};
	\draw[GraphEdge] (pd1) -- (pd2) node[midway, above] {$z_{13}$};
\end{tikzpicture}
\begin{tikzpicture}
    [   
	baseline={(current bounding box.center)},
	line join=round
	]
	\def\gS{1.1};
	\coordinate (pd1) at (-0.866*\gS,-0.5*\gS);
	\coordinate (pd2) at (0.*\gS,1.*\gS);
	\coordinate (pd3) at (0.866*\gS,-0.5*\gS);
	\draw (pd1) node[GraphNode] {} node[left] {$\lambda_{1+3}$};
	\draw (pd2) node[GraphNode] {} node[left,above] {$\lambda_{2}$};
	\draw (pd3) node[GraphNode] {} node[right] {$\lambda_{4}$};
	\draw[GraphEdge] (pd1) -- (pd2) node[midway, left] {$z_{12}$};
	\draw[GraphEdge] (pd1) -- (pd3) node[midway, above] {$z_{34}$};
	\draw[GraphEdge] (pd2) -- (pd3) node[midway, right] {$z_{24}$};
\end{tikzpicture}
&&-\,\,\,
\begin{tikzpicture}
    [
	baseline={(current bounding box.center)},
	line join=round
	]
    \def\gS{0.9};
	\coordinate (pd1) at (1.*\gS,0.*\gS);
	\coordinate (pd2) at (-1.*\gS,0.*\gS);
	\draw (pd1) node[GraphNode] {} node[above] {$\lambda_{3}$};
	\draw (pd2) node[GraphNode] {} node[above] {$\lambda_{4}$};
	\draw[GraphEdge] (pd1) -- (pd2) node[midway, above] {$z_{34}$};
\end{tikzpicture}
\begin{tikzpicture}
    [   
	baseline={(current bounding box.center)},
	line join=round
	]
	\def\gS{1.1};
	\coordinate (pd1) at (-0.866*\gS,-0.5*\gS);
	\coordinate (pd2) at (0.*\gS,1.*\gS);
	\coordinate (pd3) at (0.866*\gS,-0.5*\gS);
	\draw (pd1) node[GraphNode] {} node[left] {$\lambda_{1}$};
	\draw (pd2) node[GraphNode] {} node[left,above] {$\lambda_{2}$};
	\draw (pd3) node[GraphNode] {} node[right] {$\lambda_{3+4}$};
	\draw[GraphEdge] (pd1) -- (pd2) node[midway, left] {$z_{12}$};
	\draw[GraphEdge] (pd1) -- (pd3) node[midway, above] {$z_{13}$};
	\draw[GraphEdge] (pd2) -- (pd3) node[midway, right] {$z_{24}$};
\end{tikzpicture}\,,\label{eq:triQuadPic}
\end{alignat}
where each segment graph on the left acts on the triangle graph to its right. This in turn leads to an equation in terms of four $f_{\pic{1.2ex}{triangle}}[\lambda_1,\lambda_2; Z]$, which we will use to bootstrap the one-loop result from the previous section.

Before proceeding with the bootstrap calculation, it is worthwhile to visualize the geometric setup leading to the quadratic relation. The region $\Delta_{\pic{1.2ex}{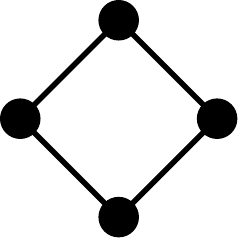}}$ associated to the square is easily described: we have four points $y_{12}$, $y_{24}$, $y_{43}= - y_{34}$ and $y_{31}= - y_{13}$ whose convex envelope includes the origin. Generically, the origin will belong to exactly two of the triangles defined by dropping one of the points, as in the right-hand side of Figure \ref{fig:quadexample}. 

For each of these triangles, the three vertices will thus belong to $\Delta_{\pic{1.2ex}{triangle}}$. The remaining point belongs, trivially, to $\Delta_{\pick{1.5ex}{segment}}$. We thus have a double-cover of $\Delta_{\pic{1.2ex}{square}}$ by regions of the form 
$\Delta_{\pick{1.5ex}{segment}} \times \Delta_{\pic{1.2ex}{triangle}}$. 

For any given generic $y_e$, the space $L[y]$ of positive $t_e$'s such that $\sum_e t_e y_e=0$ is clearly convex and thus is a segment. For each of the two triangles the origin belongs to, we can find a positive linear combination of the vertices which vanishes. This gives a point of $L[y]$ with one of the $t_e=0$. These are the two endpoints of $L[y]$. 

\begin{figure}
    \centering
     \begin{tikzpicture}
[
	baseline={(current bounding box.center)},
	line join=round
	]
\def\gS{1};
	\coordinate (pd1) at (-1.25*\gS,1.25*\gS);
	\coordinate (pd2) at (1.25*\gS,1.25*\gS);
	\coordinate (pd3) at (1.25*\gS,-1.25*\gS);
	\coordinate (pd4) at (-1.25*\gS,-1.25*\gS);
	\coordinate (pd5) at (-1.25*\gS+1.2,1.25*\gS-.5);


	\draw[GraphEdge] (pd1) -- (pd2) ;
	\draw[GraphEdge] (pd2) -- (pd3);
	\draw[GraphEdge] (pd3) -- (pd4) ;
	\draw[GraphEdge] (pd1) -- (pd4) ;
	\draw[GraphEdge] (pd1) -- (pd3) ;
	\draw[GraphEdge] (pd2) -- (pd4) ;
	\draw (pd5) node[GraphNode] {} node[above] {}; 
\end{tikzpicture}
    \hspace{5mm}
   \begin{tikzpicture}
[
	baseline={(current bounding box.center)},
	line join=round
	]
	\coordinate (pd1) at (0,1.5);
	\coordinate (pd2) at (-1.5,-1);
	\coordinate (pd3) at (1.5,-1);
	\coordinate (pd4) at (0,-.2);
	\coordinate (pd5) at (0,.5);


	\draw[GraphEdge] (pd1) -- (pd2) ;
	\draw[GraphEdge,dotted,line width = .5mm] (pd2) -- (pd3);
	\draw[GraphEdge] (pd3) -- (pd4) ;
	\draw[GraphEdge] (pd1) -- (pd3) ;
	\draw[GraphEdge] (pd2) -- (pd4) ;
	\draw (pd5) node[GraphNode] {} node[above] {}; 
\end{tikzpicture}
\hspace{5mm}
\begin{tikzpicture}
[
	baseline={(current bounding box.center)},
	line join=round
	]
	\coordinate (pd1) at (0,1.5);
	\coordinate (pd2) at (-1.5,-1);
	\coordinate (pd3) at (1.5,-1);
	\coordinate (pd4) at (0,-.2);
	\coordinate (pd5) at (0,-.6);


	\draw[GraphEdge] (pd1) -- (pd2) ;
	\draw[GraphEdge,dotted,line width = .5mm] (pd2) -- (pd3);
	\draw[GraphEdge] (pd3) -- (pd4) ;
	\draw[GraphEdge] (pd1) -- (pd3) ;
	\draw[GraphEdge] (pd2) -- (pd4) ;
	\draw (pd5) node[GraphNode] {} node[above] {}; 
\end{tikzpicture}
    \caption{The origin is denoted by the node, and the dotted lines denote a ``diagonal'' for the two concave quadrilaterals. In the first two quadrilaterals the origin lies inside, while it lies outside the last quadrilateral. In the first quadrilateral, the two triangles that cover the origin share the top edge on the square. In the last two quadrilaterals, the two triangles share the dotted line, i.e. share a ``diagonal'' of the convex quadrilateral. }
    \label{fig:convexConcave}
\end{figure}
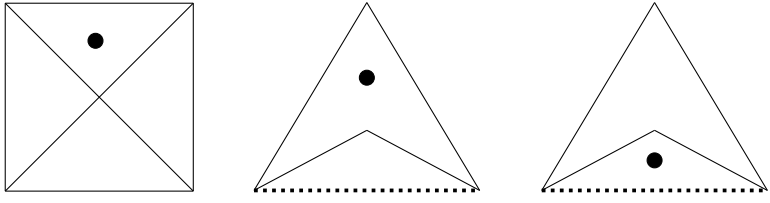

We can use this geometric setting to derive the relative signs in the quadratic identity, as explained in the previous section, but we can also figure it out by hand. There are three cases to consider: the quadrilateral formed by $y_{12}$, $y_{24}$, $y_{43}$ and $y_{31}$ can be convex, non-convex and including the origin, and non-convex and not including the origin.
In each case, we can compare the relative orientation of the triangles which do contain the origin. We find that they have the same orientation in the momentum plane if they share an edge belonging to the quadrilateral, and opposite orientation if they share a diagonal, as in Figure \ref{fig:convexConcave}. Hence the oriented sum:
\begin{alignat}{3}
    \phantom{+}&\Delta_{\pick{1.5ex}{segment}}[y_{12}] \times \Delta_{\pic{1.2ex}{triangle}}[y_{24},y_{43},y_{31}] 
        &&-\Delta_{\pick{1.5ex}{segment}}[y_{24}] \times \Delta_{\pic{1.2ex}{triangle}}[y_{43},y_{31},y_{12}]\nonumber\\
    +&\Delta_{\pick{1.5ex}{segment}}[y_{43}] \times \Delta_{\pic{1.2ex}{triangle}}[y_{31},y_{12},y_{24}]        
        &&-\Delta_{\pick{1.5ex}{segment}}[y_{31}] \times             \Delta_{\pic{1.2ex}{triangle}}[y_{12},y_{24},y_{43}] \,,
\end{alignat}
in the space of $y_e$ vanishes. These are the same signs which arise in the general identity from the relative permutations of the edges.

Coming back to the quadratic relation in \eqref{eq:triQuadPic}, we have
\begin{align}
    0
&= 
    e^{-(\lambda_1+\p_{z_{13}})\cdot z_{12}} \,
        \mathcal{I}_{\pic{1.2ex}{triangle}}[\lambda_1+\lambda_2,\lambda_3; z_{13},z_{24},z_{34}]\nonumber\\
&+   
    e^{-(\lambda_2-\p_{z_{12}})\cdot z_{24}} \,
        {\mathcal I}_{\pic{1.2ex}{triangle}}[\lambda_1,\lambda_2+\lambda_4; z_{12},z_{13},-z_{34}]\nonumber\\
&+
    e^{-(\lambda_1+\p_{z_{12}})\cdot z_{13}} \,
        \mathcal{I}_{\pic{1.2ex}{triangle}}[\lambda_1+\lambda_3, \lambda_2; z_{12},z_{34},z_{24}]\nonumber\\
&-
    e^{-(\lambda_3-\p_{z_{13}})\cdot z_{34}} \,
        \mathcal{I}_{\pic{1.2ex}{triangle}}[\lambda_1,\lambda_2;\, z_{12},z_{13},z_{24}]\,.
\end{align}
We now apply the shift symmetry of \eqref{eq:shift} to write the argument of $\mathcal{I}_{\pic{1.2ex}{triangle}}$ only in terms of loop variables. i.e. we use the shift symmetry of $\calI_{\pic{1.2ex}{triangle}}$ to decompose each $\calI_{\pic{1.2ex}{triangle}}$ in the form:
\begin{equation}
    {\mathcal I}_{\pic{1.2ex}{triangle}}[\lambda;z]   
        = e^{ g(\lambda,z_{ij})} f_{\pic{1.2ex}{triangle}}[\lambda; Z_{\text{loop}}]\,,
\end{equation}
where $g(\lambda,z_{ij})$ is some $SU(2)$-invariant bilinear function of the $\lambda_i$ and $z_{ij}$. In practice, we do this for some reference canonical graph, obtaining a formula like \eqref{eq:shiftSymTriangle}, and then substitute in our particular sets of $\lambda$ and $z_{ij}$ for our four relevant graphs.

At first glance, each of the four $f_{\pic{1.2ex}{triangle}}[\lambda; Z_{\text{loop}}]$ has a different argument for $ Z_{\text{loop}}$. For example, in the first graph it is $z_{13}+z_{34}-z_{24}$, while the argument from the second graph is $z_{12}-z_{34}-z_{13}$. However, since the exponentials of derivatives (coming from the propagators) act as shifts on the various arguments of the $f_{\pic{1.2ex}{triangle}}[\lambda; Z_{\text{loop}}]$, we find that the final expression will only depend on one loop variable, $Z = z_{13}+z_{34}-z_{24}-z_{12}$, coming from the original square sliding graph itself!

Implementing the various exponential shifts and momentum conservation, $\lambda_4 = -\lambda_1 - \lambda_2 - \lambda_3$, the quadratic relation takes the form:
\begin{align}
    0= 
        &+e^{-\lambda_1 \cdot (z_{12}+ z_{24}) - \lambda_2 \cdot z_{24} -\lambda_3 \cdot z_{34}}
            f_{\pic{1.2ex}{triangle}}[\lambda_1+\lambda_2,\lambda_3 ; -Z]\nonumber\\
        &+e^{-\lambda_1 \cdot (z_{13}+ z_{34})- \lambda_2 \cdot z_{24} -\lambda_3 \cdot z_{34}}
            f_{\pic{1.2ex}{triangle}}[\lambda_1 ,-\lambda_1- \lambda_3 ;Z]\nonumber\\
        &+e^{-\lambda_1 \cdot (z_{13}+ z_{34})- \lambda_2 \cdot z_{24} -\lambda_3 \cdot z_{34}}
            f_{\pic{1.2ex}{triangle}}[\lambda_1+\lambda_3,\lambda_2 ; Z]\nonumber\\
        &- e^{-\lambda_1 \cdot (z_{13}+ z_{34})- \lambda_2 \cdot z_{24} -\lambda_3 \cdot z_{34}}
            f_{\pic{1.2ex}{triangle}}[\lambda_1,\lambda_2;Z]\,.
\end{align}
We can strip off some of the exponentials from the general quadratic identity to leave a bootstrappable equation involving only a single loop variable:
\begin{align}
    0= 
        e^{\lambda_1 \cdot Z}&f_{\pic{1.2ex}{triangle}}[\lambda_1+\lambda_2,\lambda_3 ; Z]
            + f_{\pic{1.2ex}{triangle}}[\lambda_1 ,-\lambda_1- \lambda_3 ;-Z]\nonumber\\
        +&f_{\pic{1.2ex}{triangle}}[\lambda_1+\lambda_3,\lambda_2 ; Z]
            -f_{\pic{1.2ex}{triangle}}[\lambda_1,\lambda_2;-Z]\,.
\end{align}
This is the final form of the quadratic identity, which should be supplemented by the symmetry relations \eqref{eq:sym1} for the triangle graph.

We can check that the integrated expression \eqref{eq:triint} for the triangle diagram indeed satisfies the above relations. It is also interesting to assess to what degree the quadratic identity, combined with the symmetries of the triangle and homogeneity degree 2 under the scale transformation (under which the $\lambda_a$'s have charge $1$ and $Z$'s have charge $-1$), fixes the functional form of $f_{\pic{1.2ex}{triangle}}$. Experimentally, working order-by-order in the power series in $\lambda$'s and $Z$'s, these constraints do appear to be sufficient to fix the functional form of $f_{\pic{1.2ex}{triangle}}$ when combined with global $SU(2)$ invariance, generated by the vectorfields 
    \begin{equation}
    \lambda^1_2 \p_{\lambda^2_2}+\lambda^1_3 \p_{\lambda^2_3}- Z^2 \p_{Z^1}\,, \quad  \lambda^2_2 \p_{\lambda^1_2}+\lambda^2_3 \p_{\lambda^1_3}-Z^1 \p_{Z^2}\,.
\end{equation} 
This is reasonable: a different choice of gauge for the propagators would have likely given different answers which satisfy the same quadratic relations and give rise to an equivalent holomorphic factorization algebra structure. The propagators we choose are $SU(2)$ invariant. 

\subsection{The Bitriangle: Two-Loop Diagram Properties}
The only two-loop Laman graph is the bitriangle:
\begin{equation}
\begin{tikzpicture}
    [
	baseline={(current bounding box.center)},
	line join=round
	]
\def\gS{3};
	\coordinate (pd1) at (1.8675*\gS,0.4341*\gS);
	\coordinate (pd2) at (0.9335*\gS,0.*\gS);
	\coordinate (pd3) at (0.*\gS,0.4347*\gS);
	\coordinate (pd4) at (0.9349*\gS,0.8696*\gS);

	\draw (pd1) node[GraphNode] {} node[right] {$\lambda_{1}$};
	\draw (pd2) node[GraphNode] {} node[below] {$\lambda_{3}$};
	\draw (pd3) node[GraphNode] {} node[left] {$\lambda_{2}$};
	\draw (pd4) node[GraphNode] {} node[above] {$\lambda_{4}$};

	\draw[GraphEdge] (pd1) -- (pd2) node[midway, below] {$z_{13}$};
	\draw[GraphEdge] (pd1) -- (pd4) node[midway, above] {$z_{14}$};
	\draw[GraphEdge] (pd2) -- (pd3) node[midway, below] {$z_{23}$};
	\draw[GraphEdge] (pd2) -- (pd4) node[midway, left] {$z_{34}$};
	\draw[GraphEdge] (pd3) -- (pd4) node[midway, above] {$z_{24}$};
\end{tikzpicture}
\end{equation}

The corresponding region $\Delta_{\pic{1ex}{bitriangle}}$ is easy to describe: the triangles in the momentum plane with vertices $y_{13}$,
$y_{34}$, $y_{41} = - y_{14}$ and $y_{23}$, $y_{34}$, $y_{42} = - y_{24}$ must both include the origin. 
The intersection of $\Delta_{\pic{1ex}{bitriangle}}$ with the momentum conservation constraints gives a $\lambda$-dependent polyhedral region in the space of two loop momenta. 

The function we are interested in is $\mathcal{I}_{\pic{1ex}{bitriangle}}[\lambda_1,\lambda_2,\lambda_3;z_{13},z_{14},z_{23},z_{24},z_{34}]$. Explicitly, we could solve the momentum constraints as:
\begin{align}
    \lambda_1 &= y_{13} + y_{14} \,,\nonumber\\
    \lambda_2 &= y_{23} + y_{24}\,,\\
    \lambda_3 &= -y_{13} - y_{23} + y_{34}\,,\nonumber
\end{align}
which identifies two independent loop variables for the bitriangle:
\begin{equation}
    Z_1 = z_{13} + z_{34} - z_{14}\,,\quad
    Z_2 = z_{13} - z_{23} + z_{24} - z_{14}\,.
\end{equation}
Using the shift symmetries \eqref{eq:shift}, gives us the decomposition:
\begin{equation}\label{eq:bitrif}
    \mathcal{I}_{\pic{1ex}{bitriangle}}[\lambda_1,\lambda_2,\lambda_3;z_{13},z_{14},z_{23},z_{24},z_{34}] 
        = e^{-\lambda_1 \cdot z_{14}+\lambda_2 \cdot (z_{13}-z_{14}-z_{23})+\lambda_3 \cdot (z_{13}-z_{14})} f_{\pic{1ex}{bitriangle}}[\lambda_1,\lambda_2,\lambda_3; Z_1,Z_2] \,.
\end{equation}
The function $f_{\pic{1ex}{bitriangle}}$ has weight $4$ under the symmetry giving weight $1$ to $\lambda$'s and $-1$ to $z$'s.

The bitriangle graph also has a $\mathbb{Z}_2^2$ symmetry group, generated by the two reflections: $1\leftrightarrow 2$ and $3\leftrightarrow 4$, from flipping the diagram along the vertical and horizontal axes respectively. These two symmetries of $\mathcal{I}_{\pic{1ex}{bitriangle}}[\lambda_1,\lambda_2,\lambda_3;z_{13},z_{14},z_{23},z_{24},z_{34}]$ imply the following relations for $f_{\pic{1ex}{bitriangle}}[\lambda_1,\lambda_2,\lambda_3; Z_1,Z_2]$:
\begin{alignat}{3}
    &\mathbb{Z}_2^{(12)}: \,
    f_{\pic{1ex}{bitriangle}}[\lambda_1,\lambda_2,\lambda_3; Z_1,Z_{2}]
        &&= e^{\lambda_4\cdot Z_2} f_{\pic{1ex}{bitriangle}}[\lambda_2,\lambda_1,\lambda_3; Z_{1}-Z_{2},-Z_{2}]\,,\\
    &\mathbb{Z}_2^{(34)}: \, f_{\pic{1ex}{bitriangle}}[\lambda_1,\lambda_2,\lambda_3; Z_1,Z_{2}]
        &&=e^{-\lambda_2 \cdot Z_{2}} f_{\pic{1ex}{bitriangle}}[\lambda_1,\lambda_2,\lambda_4; -Z_{1},-Z_{2}]\,,
\end{alignat}
where $\lambda_4 = -\lambda_3 -\lambda_2 -\lambda_1$ as always.

The $\mathcal{I}_{\,\pic{1ex}{bitriangle}}$ amplitude satisfies two quadratic relations involving sliding graphs with two loops, such that the bitriangle graph emerges from the contraction of a single edge. These quadratic relations will thus be linear in $\mathcal{I}_{\,\pic{1ex}{bitriangle}}$, possibly with sources built from the triangle amplitude. We will investigate these in the following two sections.

\subsection{The Square-Triangle: Bootstrapping Two-Loops} 
The square-triangle graph is:
\begin{equation}
\begin{tikzpicture}
    [
	baseline={(current bounding box.center)},
	line join=round
	]
  \def\gS{3};
	\coordinate (pd1) at (-1.9449*\gS,0.4869*\gS);
	\coordinate (pd2) at (-1.0809*\gS,0.0359*\gS);
	\coordinate (pd3) at (-1.081*\gS,0.9376*\gS);
	\coordinate (pd4) at (-0.0001*\gS,0.*\gS);
	\coordinate (pd5) at (-0.*\gS,0.9737*\gS);

	\draw (pd1) node[GraphNode] {} node[left] {$\lambda_{1}$};
	\draw (pd2) node[GraphNode] {} node[below] {$\lambda_{5}$};
	\draw (pd3) node[GraphNode] {} node[above] {$\lambda_{4}$};
	\draw (pd4) node[GraphNode] {} node[below right] {$\lambda_{3}$};
	\draw (pd5) node[GraphNode] {} node[above right] {$\lambda_{2}$};

	\draw[GraphEdge] (pd1) -- (pd2) node[midway, below] {$z_{15}$};
	\draw[GraphEdge] (pd1) -- (pd3) node[midway, above] {$z_{14}$};
	\draw[GraphEdge] (pd2) -- (pd3) node[midway, right] {$z_{45}$};
	\draw[GraphEdge] (pd2) -- (pd4) node[midway, below] {$z_{35}$};
	\draw[GraphEdge] (pd3) -- (pd5) node[midway, above] {$z_{24}$};
	\draw[GraphEdge] (pd4) -- (pd5) node[midway, right] {$z_{23}$};
	\end{tikzpicture}
\end{equation}
We will pick the two independent loop variables: 
\begin{equation}
    Z_1 = z_{14}+z_{45}-z_{15} \,, \quad 
    Z_2 = z_{35}-z_{15}+z_{14}-z_{24}+z_{23}\,.
\end{equation}

The quadratic relation associated to this graph has 4 terms: we can either collapse an edge of the square which is not shared by the triangle, giving a bi-triangle graph, or collapse the triangle and produce a second triangle (see Figure \ref{fig:shrinkingGraph}). That is, we have:
\begin{alignat}{3}
0 &= 
    \begin{tikzpicture}
    [
	baseline={(current bounding box.center)},
	line join=round
	]
    \def\gS{0.8};
	\coordinate (pd1) at (1.*\gS,0.*\gS);
	\coordinate (pd2) at (-1.*\gS,0.*\gS);
	\draw (pd1) node[GraphNode] {} node[above] {$\lambda_{2}$};
	\draw (pd2) node[GraphNode] {} node[above] {$\lambda_{3}$};
	\draw[GraphEdge] (pd1) -- (pd2) node[midway, above] {$z_{23}$};
      \end{tikzpicture}
      \begin{tikzpicture}
      [
	baseline={(current bounding box.center)},
	line join=round
	]
    \def\gS{4/3};
	\coordinate (pd1) at (1.8671*\gS,0.435*\gS);
	\coordinate (pd2) at (0.9338*\gS,0.87*\gS);
	\coordinate (pd3) at (0.9339*\gS,0.*\gS);
	\coordinate (pd4) at (0.*\gS,0.435*\gS);
	\draw (pd1) node[GraphNode] {} node[right] {$\lambda_{1}$};
	\draw (pd2) node[GraphNode] {} node[above] {$\lambda_{5}$};
	\draw (pd3) node[GraphNode] {} node[below] {$\lambda_{4}$};
	\draw (pd4) node[GraphNode] {} node[left] {$\lambda_{2+3}$};
	\draw[GraphEdge] (pd1) -- (pd2) node[midway, above] {$z_{15}$};
	\draw[GraphEdge] (pd1) -- (pd3) node[midway, below] {$z_{14}$};
	\draw[GraphEdge] (pd2) -- (pd3) node[midway, left] {$z_{45}$};
	\draw[GraphEdge] (pd2) -- (pd4) node[midway, above] {$z_{35}$};
	\draw[GraphEdge] (pd3) -- (pd4) node[midway, below] {$z_{24}$};
    \end{tikzpicture}
&&-
    \begin{tikzpicture}
    [
	baseline={(current bounding box.center)},
	line join=round
	]
    \def\gS{0.8};
	\coordinate (pd1) at (1.*\gS,0.*\gS);
	\coordinate (pd2) at (-1.*\gS,0.*\gS);
	\draw (pd1) node[GraphNode] {} node[above] {$\lambda_{2}$};
	\draw (pd2) node[GraphNode] {} node[above] {$\lambda_{4}$};
	\draw[GraphEdge] (pd1) -- (pd2) node[midway, above] {$z_{24}$};
      \end{tikzpicture}
      \begin{tikzpicture}
      [
	baseline={(current bounding box.center)},
	line join=round
	]
    \def\gS{4/3};
	\coordinate (pd1) at (1.8671*\gS,0.435*\gS);
	\coordinate (pd2) at (0.9338*\gS,0.87*\gS);
	\coordinate (pd3) at (0.9339*\gS,0.*\gS);
	\coordinate (pd4) at (0.*\gS,0.435*\gS);
	\draw (pd1) node[GraphNode] {} node[right] {$\lambda_{1}$};
	\draw (pd2) node[GraphNode] {} node[above] {$\lambda_{5}$};
	\draw (pd3) node[GraphNode] {} node[below] {$\lambda_{2+4}$};
	\draw (pd4) node[GraphNode] {} node[left] {$\lambda_{3}$};
	\draw[GraphEdge] (pd1) -- (pd2) node[midway, above] {$z_{15}$};
	\draw[GraphEdge] (pd1) -- (pd3) node[midway, below] {$z_{14}$};
	\draw[GraphEdge] (pd2) -- (pd3) node[midway, left] {$z_{45}$};
	\draw[GraphEdge] (pd2) -- (pd4) node[midway, above] {$z_{35}$};
	\draw[GraphEdge] (pd3) -- (pd4) node[midway, below] {$z_{23}$};
    \end{tikzpicture}\nonumber\\
&-
    \begin{tikzpicture}
    [
	baseline={(current bounding box.center)},
	line join=round
	]
    \def\gS{0.8};
	\coordinate (pd1) at (1.*\gS,0.*\gS);
	\coordinate (pd2) at (-1.*\gS,0.*\gS);
	\draw (pd1) node[GraphNode] {} node[above] {$\lambda_{3}$};
	\draw (pd2) node[GraphNode] {} node[above] {$\lambda_{5}$};
	\draw[GraphEdge] (pd1) -- (pd2) node[midway, above] {$z_{35}$};
    \end{tikzpicture}
    \begin{tikzpicture}
    [
	baseline={(current bounding box.center)},
	line join=round
	]
    \def\gS{4/3};
	\coordinate (pd1) at (1.8671*\gS,0.435*\gS);
	\coordinate (pd2) at (0.9338*\gS,0.87*\gS);
	\coordinate (pd3) at (0.9339*\gS,0.*\gS);
	\coordinate (pd4) at (0.*\gS,0.435*\gS);
	\draw (pd1) node[GraphNode] {} node[right] {$\lambda_{1}$};
	\draw (pd2) node[GraphNode] {} node[above] {$\lambda_{4}$};
	\draw (pd3) node[GraphNode] {} node[below] {$\lambda_{3+5}$};
	\draw (pd4) node[GraphNode] {} node[left] {$\lambda_{2}$};
	\draw[GraphEdge] (pd1) -- (pd2) node[midway, above] {$z_{14}$};
	\draw[GraphEdge] (pd1) -- (pd3) node[midway, below] {$z_{15}$};
	\draw[GraphEdge] (pd2) -- (pd3) node[midway, left] {$z_{45}$};
	\draw[GraphEdge] (pd2) -- (pd4) node[midway, above] {$z_{24}$};
	\draw[GraphEdge] (pd3) -- (pd4) node[midway, below] {$z_{23}$};
    \end{tikzpicture}
&&-
    \begin{tikzpicture}
    [
    baseline={(current bounding box.center)},
    line join=round
    ]
    \def\gS{1.1};
    \coordinate (pd1) at (-0.866*\gS,-0.5*\gS);
    \coordinate (pd2) at (0.*\gS,1.*\gS);
    \coordinate (pd3) at (0.866*\gS,-0.5*\gS);
    \draw (pd1) node[GraphNode] {} node[below] {$\lambda_{1}$};
    \draw (pd2) node[GraphNode] {} node[above] {$\lambda_{4}$};
    \draw (pd3) node[GraphNode] {} node[below] {$\lambda_{5}$};
    \draw[GraphEdge] (pd1) -- (pd2) node[midway, left] {$z_{14}$};
    \draw[GraphEdge] (pd1) -- (pd3) node[midway, above] {$z_{15}$};
    \draw[GraphEdge] (pd2) -- (pd3) node[midway, right] {$z_{45}$};
    \end{tikzpicture}
    \begin{tikzpicture}
    [
    baseline={(current bounding box.center)},
    line join=round
    ]
    \def\gS{1.1};
    \coordinate (pd1) at (-0.866*\gS,-0.5*\gS);
    \coordinate (pd2) at (0.*\gS,1.*\gS);
    \coordinate (pd3) at (0.866*\gS,-0.5*\gS);
    \draw (pd1) node[GraphNode] {} node[below] {$\lambda_{1+4+5}$};
    \draw (pd2) node[GraphNode] {} node[above] {$\lambda_{2}$};
    \draw (pd3) node[GraphNode] {} node[below] {$\lambda_{3}$};
    \draw[GraphEdge] (pd1) -- (pd2) node[midway, left] {$z_{24}$};
    \draw[GraphEdge] (pd1) -- (pd3) node[midway, above] {$z_{35}$};
    \draw[GraphEdge] (pd2) -- (pd3) node[midway, right] {$z_{23}$};
    \end{tikzpicture}
\end{alignat}
The quadratic relation associated to this graph thus involves a source term given by the insertion of a triangle into a triangle.

After using the shift symmetries for the triangle \eqref{eq:shiftSymTriangle} and bitriangle \eqref{eq:bitrif}, and stripping the exponential pre-factors, we obtain the following quadratic relation entirely in terms of $\lambda$'s and loop variables:
\begin{align}
    f_{\pic{1.2ex}{triangle}}[\lambda_1,\lambda_4+\partial_{Z_{2}};Z_{1}]    &f_{\pic{1.2ex}{triangle}}[-\lambda_2-\lambda_3,\lambda_2;-Z_{1}+Z_{2}]\nonumber\\
        &=e^{\lambda_4 \cdot Z_1 + (\lambda_2 + \lambda_3) \cdot Z_2}
            f_{\pic{1ex}{bitriangle}}[\lambda_1,\lambda_2+\lambda_3,\lambda_4; Z_{1},Z_{2}]\nonumber \\
        &-e^{\lambda_4 \cdot Z_1 + (\lambda_2 + \lambda_3) \cdot Z_2}
            f_{\pic{1ex}{bitriangle}}[\lambda_1,\lambda_3,\lambda_2+\lambda_4; Z_{1},Z_{2}]\nonumber \\
        &-e^{\lambda_4 \cdot Z_1}         
            f_{\pic{1ex}{bitriangle}}[\lambda_1,\lambda_2,-\lambda_1-\lambda_2-\lambda_4; -Z_{1},-Z_{2}]\,.
\end{align}

We have verified that this relation, together with $SU(2)$ invariance and reflection symmetries, fixes $f_{\pic{1ex}{bitriangle}}$
at least up to quadratic order in the $z$'s. At the leading order, we have
\begin{equation}
    f_{\pic{1ex}{bitriangle}}[\lambda_1,\lambda_2,\lambda_3; 0,0] = \frac{1}{24} (\lambda_1 \wedge (\lambda_2 + 2  \lambda_3)) (\lambda_2\wedge(\lambda_1 + 2  \lambda_3))\,.
\end{equation}

\subsection{The Flying V: A Check on Two-Loops}
In addition to the square-triangle graph above, the bitriangle is involved in a second non-trivial quadratic relation coming from a 5-vertex sliding graph in the shape of a flying V. In particular, the graph with labelling
\begin{equation}
\begin{tikzpicture}
	[
	baseline={(current bounding box.center)},
	line join=round
	]
    \def\gS{2.5};
	\coordinate (pd1) at (0.*\gS,0.6087*\gS);
	\coordinate (pd2) at (1.6537*\gS,1.2169*\gS);
	\coordinate (pd3) at (1.6534*\gS,0.*\gS);
	\coordinate (pd4) at (0.9334*\gS,0.9506*\gS);
	\coordinate (pd5) at (0.9328*\gS,0.2667*\gS);

	\draw (pd1) node[GraphNode] {} node[left] {$\lambda_{1}$};
	\draw (pd2) node[GraphNode] {} node[right] {$\lambda_{2}$};
	\draw (pd3) node[GraphNode] {} node[right] {$\lambda_{3}$};
	\draw (pd4) node[GraphNode] {} node[above] {$\lambda_{4}$};
	\draw (pd5) node[GraphNode] {} node[below] {$\lambda_{5}$};

	\draw[GraphEdge] (pd1) -- (pd4) node[midway, above] {$z_{14}$};
	\draw[GraphEdge] (pd1) -- (pd5) node[midway, below] {$z_{15}$};
	\draw[GraphEdge] (pd2) -- (pd4) node[midway, above] {$z_{24}$};
	\draw[GraphEdge] (pd2) -- (pd5) node[midway, right] {$z_{25}$};
	\draw[GraphEdge] (pd3) -- (pd4) node[midway, right] {$z_{34}$};
	\draw[GraphEdge] (pd3) -- (pd5) node[midway, below] {$z_{35}$};

\end{tikzpicture}
\end{equation}
whose loop variables we denote by
\begin{equation}
    Z_1 = z_{14} - z_{34} + z_{35} - z_{15}\,, \quad 
    Z_2 = z_{14} - z_{24} + z_{25} - z_{15}\,,
\end{equation}
has a quadratic relation involving 6 terms. All 6 terms involve collapsing one edge, leaving a segment graph acting on a bitriangle. Altogther, the quadratic relation induced from this graph is
\begin{alignat}{3}
0 &= 
    \begin{tikzpicture}
    [
	baseline={(current bounding box.center)},
	line join=round
	]
    \def\gS{0.8};
	\coordinate (pd1) at (1.*\gS,0.*\gS);
	\coordinate (pd2) at (-1.*\gS,0.*\gS);
	\draw (pd1) node[GraphNode] {} node[above] {$\lambda_{3}$};
	\draw (pd2) node[GraphNode] {} node[above] {$\lambda_{4}$};
	\draw[GraphEdge] (pd1) -- (pd2) node[midway, above] {$z_{34}$};
      \end{tikzpicture}
      \begin{tikzpicture}
      [
	baseline={(current bounding box.center)},
	line join=round
	]
    \def\gS{4/3};
	\coordinate (pd1) at (1.8671*\gS,0.435*\gS);
	\coordinate (pd2) at (0.9338*\gS,0.87*\gS);
	\coordinate (pd3) at (0.9339*\gS,0.*\gS);
	\coordinate (pd4) at (0.*\gS,0.435*\gS);
	\draw (pd1) node[GraphNode] {} node[right] {$\lambda_{1}$};
	\draw (pd2) node[GraphNode] {} node[above] {$\lambda_{5}$};
	\draw (pd3) node[GraphNode] {} node[below] {$\lambda_{3+4}$};
	\draw (pd4) node[GraphNode] {} node[left] {$\lambda_{2}$};
	\draw[GraphEdge] (pd1) -- (pd2) node[midway, above] {$z_{15}$};
	\draw[GraphEdge] (pd1) -- (pd3) node[midway, below] {$z_{14}$};
	\draw[GraphEdge] (pd2) -- (pd3) node[midway, left] {$z_{35}$};
	\draw[GraphEdge] (pd2) -- (pd4) node[midway, above] {$z_{25}$};
	\draw[GraphEdge] (pd3) -- (pd4) node[midway, below] {$z_{24}$};
    \end{tikzpicture}
&-
\begin{tikzpicture}
    [
	baseline={(current bounding box.center)},
	line join=round
	]
    \def\gS{0.8};
	\coordinate (pd1) at (1.*\gS,0.*\gS);
	\coordinate (pd2) at (-1.*\gS,0.*\gS);
	\draw (pd1) node[GraphNode] {} node[above] {$\lambda_{3}$};
	\draw (pd2) node[GraphNode] {} node[above] {$\lambda_{5}$};
	\draw[GraphEdge] (pd1) -- (pd2) node[midway, above] {$z_{35}$};
      \end{tikzpicture}
      \begin{tikzpicture}
      [
	baseline={(current bounding box.center)},
	line join=round
	]
    \def\gS{4/3};
	\coordinate (pd1) at (1.8671*\gS,0.435*\gS);
	\coordinate (pd2) at (0.9338*\gS,0.87*\gS);
	\coordinate (pd3) at (0.9339*\gS,0.*\gS);
	\coordinate (pd4) at (0.*\gS,0.435*\gS);
	\draw (pd1) node[GraphNode] {} node[right] {$\lambda_{1}$};
	\draw (pd2) node[GraphNode] {} node[above] {$\lambda_{4}$};
	\draw (pd3) node[GraphNode] {} node[below] {$\lambda_{3+5}$};
	\draw (pd4) node[GraphNode] {} node[left] {$\lambda_{2}$};
	\draw[GraphEdge] (pd1) -- (pd2) node[midway, above] {$z_{14}$};
	\draw[GraphEdge] (pd1) -- (pd3) node[midway, below] {$z_{15}$};
	\draw[GraphEdge] (pd2) -- (pd3) node[midway, left] {$z_{34}$};
	\draw[GraphEdge] (pd2) -- (pd4) node[midway, above] {$z_{24}$};
	\draw[GraphEdge] (pd3) -- (pd4) node[midway, below] {$z_{25}$};
    \end{tikzpicture}\nonumber\\
&+
    \begin{tikzpicture}
    [
	baseline={(current bounding box.center)},
	line join=round
	]
    \def\gS{0.8};
	\coordinate (pd1) at (1.*\gS,0.*\gS);
	\coordinate (pd2) at (-1.*\gS,0.*\gS);
	\draw (pd1) node[GraphNode] {} node[above] {$\lambda_{2}$};
	\draw (pd2) node[GraphNode] {} node[above] {$\lambda_{4}$};
	\draw[GraphEdge] (pd1) -- (pd2) node[midway, above] {$z_{24}$};
      \end{tikzpicture}
      \begin{tikzpicture}
      [
	baseline={(current bounding box.center)},
	line join=round
	]
    \def\gS{4/3};
	\coordinate (pd1) at (1.8671*\gS,0.435*\gS);
	\coordinate (pd2) at (0.9338*\gS,0.87*\gS);
	\coordinate (pd3) at (0.9339*\gS,0.*\gS);
	\coordinate (pd4) at (0.*\gS,0.435*\gS);
	\draw (pd1) node[GraphNode] {} node[right] {$\lambda_{1}$};
	\draw (pd2) node[GraphNode] {} node[above] {$\lambda_{5}$};
	\draw (pd3) node[GraphNode] {} node[below] {$\lambda_{2+4}$};
	\draw (pd4) node[GraphNode] {} node[left] {$\lambda_{3}$};
	\draw[GraphEdge] (pd1) -- (pd2) node[midway, above] {$z_{15}$};
	\draw[GraphEdge] (pd1) -- (pd3) node[midway, below] {$z_{14}$};
	\draw[GraphEdge] (pd2) -- (pd3) node[midway, left] {$z_{25}$};
	\draw[GraphEdge] (pd2) -- (pd4) node[midway, above] {$z_{35}$};
	\draw[GraphEdge] (pd3) -- (pd4) node[midway, below] {$z_{34}$};
    \end{tikzpicture}
&-
    \begin{tikzpicture}
    [
	baseline={(current bounding box.center)},
	line join=round
	]
    \def\gS{0.8};
	\coordinate (pd1) at (1.*\gS,0.*\gS);
	\coordinate (pd2) at (-1.*\gS,0.*\gS);
	\draw (pd1) node[GraphNode] {} node[above] {$\lambda_{2}$};
	\draw (pd2) node[GraphNode] {} node[above] {$\lambda_{5}$};
	\draw[GraphEdge] (pd1) -- (pd2) node[midway, above] {$z_{25}$};
      \end{tikzpicture}
      \begin{tikzpicture}
      [
	baseline={(current bounding box.center)},
	line join=round
	]
    \def\gS{4/3};
	\coordinate (pd1) at (1.8671*\gS,0.435*\gS);
	\coordinate (pd2) at (0.9338*\gS,0.87*\gS);
	\coordinate (pd3) at (0.9339*\gS,0.*\gS);
	\coordinate (pd4) at (0.*\gS,0.435*\gS);
	\draw (pd1) node[GraphNode] {} node[right] {$\lambda_{1}$};
	\draw (pd2) node[GraphNode] {} node[above] {$\lambda_{4}$};
	\draw (pd3) node[GraphNode] {} node[below] {$\lambda_{2+5}$};
	\draw (pd4) node[GraphNode] {} node[left] {$\lambda_{3}$};
	\draw[GraphEdge] (pd1) -- (pd2) node[midway, above] {$z_{14}$};
	\draw[GraphEdge] (pd1) -- (pd3) node[midway, below] {$z_{15}$};
	\draw[GraphEdge] (pd2) -- (pd3) node[midway, left] {$z_{24}$};
	\draw[GraphEdge] (pd2) -- (pd4) node[midway, above] {$z_{34}$};
	\draw[GraphEdge] (pd3) -- (pd4) node[midway, below] {$z_{35}$};
    \end{tikzpicture}\\
&+\begin{tikzpicture}
    [
	baseline={(current bounding box.center)},
	line join=round
	]
    \def\gS{0.8};
	\coordinate (pd1) at (1.*\gS,0.*\gS);
	\coordinate (pd2) at (-1.*\gS,0.*\gS);
	\draw (pd1) node[GraphNode] {} node[above] {$\lambda_{1}$};
	\draw (pd2) node[GraphNode] {} node[above] {$\lambda_{4}$};
	\draw[GraphEdge] (pd1) -- (pd2) node[midway, above] {$z_{14}$};
      \end{tikzpicture}
      \begin{tikzpicture}
      [
	baseline={(current bounding box.center)},
	line join=round
	]
    \def\gS{4/3};
	\coordinate (pd1) at (1.8671*\gS,0.435*\gS);
	\coordinate (pd2) at (0.9338*\gS,0.87*\gS);
	\coordinate (pd3) at (0.9339*\gS,0.*\gS);
	\coordinate (pd4) at (0.*\gS,0.435*\gS);
	\draw (pd1) node[GraphNode] {} node[right] {$\lambda_{2}$};
	\draw (pd2) node[GraphNode] {} node[above] {$\lambda_{1+4}$};
	\draw (pd3) node[GraphNode] {} node[below] {$\lambda_{5}$};
	\draw (pd4) node[GraphNode] {} node[left] {$\lambda_{3}$};
	\draw[GraphEdge] (pd1) -- (pd2) node[midway, above] {$z_{24}$};
	\draw[GraphEdge] (pd1) -- (pd3) node[midway, below] {$z_{25}$};
	\draw[GraphEdge] (pd2) -- (pd3) node[midway, left] {$z_{15}$};
	\draw[GraphEdge] (pd2) -- (pd4) node[midway, above] {$z_{34}$};
	\draw[GraphEdge] (pd3) -- (pd4) node[midway, below] {$z_{35}$};
    \end{tikzpicture}
&-
    \begin{tikzpicture}
    [
	baseline={(current bounding box.center)},
	line join=round
	]
    \def\gS{0.8};
	\coordinate (pd1) at (1.*\gS,0.*\gS);
	\coordinate (pd2) at (-1.*\gS,0.*\gS);
	\draw (pd1) node[GraphNode] {} node[above] {$\lambda_{1}$};
	\draw (pd2) node[GraphNode] {} node[above] {$\lambda_{5}$};
	\draw[GraphEdge] (pd1) -- (pd2) node[midway, above] {$z_{15}$};
      \end{tikzpicture}
      \begin{tikzpicture}
      [
	baseline={(current bounding box.center)},
	line join=round
	]
    \def\gS{4/3};
	\coordinate (pd1) at (1.8671*\gS,0.435*\gS);
	\coordinate (pd2) at (0.9338*\gS,0.87*\gS);
	\coordinate (pd3) at (0.9339*\gS,0.*\gS);
	\coordinate (pd4) at (0.*\gS,0.435*\gS);
	\draw (pd1) node[GraphNode] {} node[right] {$\lambda_{2}$};
	\draw (pd2) node[GraphNode] {} node[above] {$\lambda_{1+5}$};
	\draw (pd3) node[GraphNode] {} node[below] {$\lambda_{4}$};
	\draw (pd4) node[GraphNode] {} node[left] {$\lambda_{3}$};
	\draw[GraphEdge] (pd1) -- (pd2) node[midway, above] {$z_{25}$};
	\draw[GraphEdge] (pd1) -- (pd3) node[midway, below] {$z_{24}$};
	\draw[GraphEdge] (pd2) -- (pd3) node[midway, left] {$z_{14}$};
	\draw[GraphEdge] (pd2) -- (pd4) node[midway, above] {$z_{35}$};
	\draw[GraphEdge] (pd3) -- (pd4) node[midway, below] {$z_{34}$};
    \end{tikzpicture}\nonumber\,.
\end{alignat}

As before, we can use the propagators to shift the arguments of the bitriangles, reduce the equations using the shift symmetries, and divide out by phases to obtain the following quadratic relation in terms of the (reduced) bitriangles $f_{\pic{1ex}{bitriangle}}$:
\begin{align}
    0   &= 
            f_{\pic{1ex}{bitriangle}}[\lambda_1,\lambda_2,\lambda_3+\lambda_4; Z_{1},Z_{2}]
        -
            e^{-\lambda_3 \cdot Z_{1} - \lambda_2 \cdot Z_{2} } f_{\pic{1ex}{bitriangle}}[\lambda_1,\lambda_2,-\lambda_1-\lambda_2-\lambda_4; -Z_{1},-Z_{2}] \nonumber
    \\
        &+ 
            f_{\pic{1ex}{bitriangle}}[\lambda_1,\lambda_3,\lambda_2+\lambda_4; Z_{2},Z_{1}]
        -
            e^{-\lambda_3 \cdot Z_{1} - \lambda_2 \cdot Z_{2} } f_{\pic{1ex}{bitriangle}}[\lambda_1,\lambda_3,-\lambda_1-\lambda_3-\lambda_4; -Z_{2},-Z_{1}] \nonumber
    \\
        &+ 
            e^{-(\lambda_1+\lambda_2+\lambda_4+\lambda_4)\cdot Z_2 } f_{\pic{1ex}{bitriangle}}[\lambda_2,\lambda_3,\lambda_1+\lambda_4; -Z_{2},Z_{1}-Z_{2}]\nonumber\\
        &-
            e^{-\lambda_3 \cdot Z_{1} - (\lambda_2+\lambda_4) \cdot Z_{2} } f_{\pic{1ex}{bitriangle}}[\lambda_2,\lambda_3,-\lambda_2-\lambda_3-\lambda_4; Z_{2},Z_{2}-Z_1] \,.
\end{align}

It is straightforward to check that our previous answer for the bitriangle satisfies this relation (and we check this vanishes for our bootstrapped bitriangle to quadratic order in the $z$'s). At leading order, i.e. when all $z=0$, the quadratic relation confirms that
\begin{alignat}{3}
    0   
        &= \frac{1}{24} 
            (\lambda_1\wedge(\lambda_2 + 2 (\lambda_3 + \lambda_4))) 
            (\lambda_2\wedge(\lambda_1 + 2 (\lambda_3 + \lambda_4)))\nonumber\\
        &-\frac{1}{24} 
            (\lambda_1\wedge(\lambda_2 - 2 (\lambda_2+\lambda_4))) 
            (\lambda_2\wedge(\lambda_1 - 2 (\lambda_1+\lambda_4)))\nonumber\\
        &+ \frac{1}{24} 
            (\lambda_1\wedge(\lambda_3 + 2 (\lambda_2 + \lambda_4)))
            (\lambda_3\wedge(\lambda_1 + 2 (\lambda_2 + \lambda_4)))\nonumber\\
        &-\frac{1}{24} 
            (\lambda_1\wedge(\lambda_3 - 2 (\lambda_3+\lambda_4))) 
            (\lambda_3\wedge(\lambda_1 - 2 (\lambda_1+\lambda_4)))\nonumber\\
        &+ \frac{1}{24} 
            (\lambda_2\wedge(\lambda_3 + 2 (\lambda_1 + \lambda_4)))
            (\lambda_3\wedge(\lambda_2 + 2 (\lambda_1 + \lambda_4)))\nonumber\\
        &-\frac{1}{24} 
            (\lambda_2\wedge(\lambda_3 - 2 (\lambda_3+\lambda_4)))
            (\lambda_3\wedge(\lambda_2 - 2 (\lambda_2+\lambda_4)))\,. 
\end{alignat}

\subsection{The Tritriangle: Bootstrap at Three-Loops}
As a final example, we consider the bootstrapping of a three-loop diagram. There are multiple three-loop diagrams. We focus on the following tritriangle:\footnote{This is the only one which would contribute perturbative corrections to the differential in a theory with cubic interactions} 
\begin{equation}
\begin{tikzpicture}
   	[
	baseline={(current bounding box.center)},
	line join=round
	]
    \def\gS{2};
	\coordinate (pd1) at (0.*\gS,0.7704*\gS);
	\coordinate (pd2) at (2.1993*\gS,0.7708*\gS);
	\coordinate (pd3) at (0.572*\gS,0.*\gS);
	\coordinate (pd4) at (1.6278*\gS,0.0004*\gS);
	\coordinate (pd5) at (1.0985*\gS,0.73*\gS);

	\draw (pd1) node[GraphNode] {} node[left] {$\lambda_{1}$};
	\draw (pd2) node[GraphNode] {} node[right] {$\lambda_{2}$};
	\draw (pd3) node[GraphNode] {} node[left] {$\lambda_{3}$};
	\draw (pd4) node[GraphNode] {} node[right] {$\lambda_{4}$};
	\draw (pd5) node[GraphNode] {} node[above] {$\lambda_{5}$};

	\draw[GraphEdge] (pd1) -- (pd3) node[midway, left] {$z_{13}$};
	\draw[GraphEdge] (pd1) -- (pd5) node[midway, above] {$z_{15}$};
	\draw[GraphEdge] (pd2) -- (pd4) node[midway, right] {$z_{24}$};
	\draw[GraphEdge] (pd2) -- (pd5) node[midway, above] {$z_{25}$};
	\draw[GraphEdge] (pd3) -- (pd4) node[midway, above] {$z_{34}$};
	\draw[GraphEdge] (pd3) -- (pd5) node[midway, left] {$z_{35}$};
	\draw[GraphEdge] (pd4) -- (pd5) node[midway, right] {$z_{45}$};
\end{tikzpicture}
\end{equation}
We define loop variables:
\begin{align}\label{eq:loopvariables3loop}
    W_1 &= z_{13}+z_{34}+z_{45}-z_{15}\,,\\
    W_2 &= z_{13}+z_{35}-z_{15}\,, \\
    W_3 &= z_{15}-z_{25}+z_{24}-z_{34}-z_{13}\,.
\end{align}
Using the shift-symmetries, we obtain
\begin{equation}\label{eq:threetriangle}
    \mathcal{I}_\Gamma[\lambda, z_e] = e^{-\sum^4_{v=1} \lambda_v \cdot \delta_v} f_{\pic{1ex}{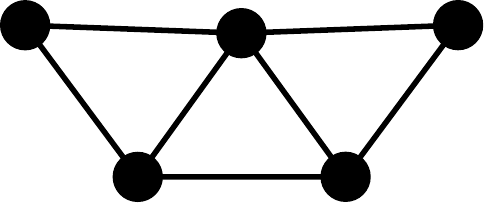}}[\lambda,W_1,W_2,W_3] 
\end{equation}
with respect to the canonical ordering, where $\delta_1= z_{15}$, $\delta_2 = z_{25}$, $\delta_3=z_{15}-z_{13}$, and $\delta_4=-z_{13}+z_{15}-z_{34}$. In advance, we note that the graph has one $\mathbb{Z}_2$ symmetry, which induces
\begin{align}
    f_{\pic{1ex}{tritriangle}}[\lambda_1,\lambda_2,\lambda_3,\lambda_4&;W_1,W_2,W_3]\nonumber\\
        &= e^{(\lambda_3+\lambda_4)\cdot Z_3} f_{\pic{1ex}{tritriangle}}[\lambda_2,\lambda_1,\lambda_4,\lambda_3,W_2+W_3,W_1+W_3,-W_3]\,.
\end{align}

One quadratic relation for the three loop diagram involves the following sliding graph:
\begin{equation}\label{eq:3loopsliding}
    \begin{tikzpicture}
   [
	baseline={(current bounding box.center)},
	line join=round
	]
    \def\gS{2.5};
	\coordinate (pd1) at (0.*\gS,0.3027*\gS);
	\coordinate (pd2) at (2.599*\gS,0.8015*\gS);
	\coordinate (pd3) at (0.4396*\gS,1.1478*\gS);
	\coordinate (pd4) at (2.0552*\gS,0.1033*\gS);
	\coordinate (pd5) at (1.0317*\gS,0.*\gS);
	\coordinate (pd6) at (1.5399*\gS,0.7967*\gS);

	\draw (pd1) node[GraphNode] {} node[left] {$\lambda_{1}$};
	\draw (pd2) node[GraphNode] {} node[right] {$\lambda_{2}$};
	\draw (pd3) node[GraphNode] {} node[left] {$\lambda_{3}$};
	\draw (pd4) node[GraphNode] {} node[below right] {$\lambda_{4}$};
	\draw (pd5) node[GraphNode] {} node[below] {$\lambda_{5}$};
	\draw (pd6) node[GraphNode] {} node[above] {$\lambda_{6}$};

	\draw[GraphEdge] (pd1) -- (pd3) node[midway, left] {$z_{13}$};
	\draw[GraphEdge] (pd1) -- (pd5) node[midway, above] {$z_{15}$};
	\draw[GraphEdge] (pd2) -- (pd4) node[midway, right] {$z_{24}$};
	\draw[GraphEdge] (pd2) -- (pd6) node[midway, above] {$z_{26}$};
	\draw[GraphEdge] (pd3) -- (pd6) node[midway, above] {$z_{36}$};
	\draw[GraphEdge] (pd4) -- (pd5) node[midway, above] {$z_{45}$};
	\draw[GraphEdge] (pd4) -- (pd6) node[midway, above] {$z_{46}$};
	\draw[GraphEdge] (pd5) -- (pd6) node[midway, left] {$z_{56}$};

    \end{tikzpicture}
\end{equation}
We choose the following independent loop variables for the sliding graph:
\begin{align}\label{eq:loopZ}
    Z_1 &= z_{15}+z_{56}-z_{36}-z_{13}\,,\\
    Z_2 &= z_{15}+z_{34}+z_{46}-z_{36}-z_{13}\,,\\
    Z_3 &= z_{45}-z_{15}+z_{13}+z_{36}-z_{26}+z_{24}\,.  
\end{align}

Up to an overall factor, the source term in the quadratic relation comes from the bitriangle acting on the triangle. The exact quadratic relation is
\begin{alignat}{3}
0 &= 
    \begin{tikzpicture}
    [
	baseline={(current bounding box.center)},
	line join=round
	]
    \def\gS{0.8};
	\coordinate (pd1) at (1.*\gS,0.*\gS);
	\coordinate (pd2) at (-1.*\gS,0.*\gS);
	\draw (pd1) node[GraphNode] {} node[above] {$\lambda_{1}$};
	\draw (pd2) node[GraphNode] {} node[above] {$\lambda_{3}$};
	\draw[GraphEdge] (pd1) -- (pd2) node[midway, above] {$z_{13}$};
    \end{tikzpicture}
    \begin{tikzpicture}
    [
    baseline={(current bounding box.center)},
    line join=round
    ]
    \def\gS{3/2};
    \coordinate (pd1) at (2.1995*\gS,0.*\gS);
    \coordinate (pd2) at (1.0996*\gS,0.0416*\gS);
    \coordinate (pd3) at (1.6289*\gS,0.7699*\gS);
    \coordinate (pd4) at (0.5712*\gS,0.7705*\gS);
    \coordinate (pd5) at (0.*\gS,0.*\gS);
    \draw (pd1) node[GraphNode] {} node[below] {$\lambda_{1+3}$};
    \draw (pd2) node[GraphNode] {} node[below] {$\lambda_{6}$};
    \draw (pd3) node[GraphNode] {} node[above] {$\lambda_{5}$};
    \draw (pd4) node[GraphNode] {} node[above] {$\lambda_{4}$};
    \draw (pd5) node[GraphNode] {} node[below] {$\lambda_{2}$};
    \draw[GraphEdge] (pd1) -- (pd2) node[midway, below] {$z_{36}$};
    \draw[GraphEdge] (pd1) -- (pd3) node[midway, right] {$z_{15}$};
    \draw[GraphEdge] (pd2) -- (pd3) node[midway, right] {$z_{56}$};
    \draw[GraphEdge] (pd2) -- (pd4) node[midway, left] {$z_{46}$};
    \draw[GraphEdge] (pd2) -- (pd5) node[midway, below] {$z_{26}$};
    \draw[GraphEdge] (pd3) -- (pd4) node[midway, above] {$z_{45}$};
    \draw[GraphEdge] (pd4) -- (pd5) node[midway, left] {$z_{24}$};
    \end{tikzpicture}
&&+
    \begin{tikzpicture}
    [
	baseline={(current bounding box.center)},
	line join=round
	]
    \def\gS{0.8};
	\coordinate (pd1) at (1.*\gS,0.*\gS);
	\coordinate (pd2) at (-1.*\gS,0.*\gS);
	\draw (pd1) node[GraphNode] {} node[above] {$\lambda_{1}$};
	\draw (pd2) node[GraphNode] {} node[above] {$\lambda_{5}$};
	\draw[GraphEdge] (pd1) -- (pd2) node[midway, above] {$z_{15}$};
      \end{tikzpicture}
    \begin{tikzpicture}
    [
    baseline={(current bounding box.center)},
    line join=round
    ]
    \def\gS{3/2};
    \coordinate (pd1) at (2.1995*\gS,0.*\gS);
    \coordinate (pd2) at (1.0996*\gS,0.0416*\gS);
    \coordinate (pd3) at (1.6289*\gS,0.7699*\gS);
    \coordinate (pd4) at (0.5712*\gS,0.7705*\gS);
    \coordinate (pd5) at (0.*\gS,0.*\gS);
    \draw (pd1) node[GraphNode] {} node[below] {$\lambda_{3}$};
    \draw (pd2) node[GraphNode] {} node[below] {$\lambda_{6}$};
    \draw (pd3) node[GraphNode] {} node[above] {$\lambda_{1+5}$};
    \draw (pd4) node[GraphNode] {} node[above] {$\lambda_{4}$};
    \draw (pd5) node[GraphNode] {} node[below] {$\lambda_{2}$};
    \draw[GraphEdge] (pd1) -- (pd2) node[midway, below] {$z_{36}$};
    \draw[GraphEdge] (pd1) -- (pd3) node[midway, right] {$z_{15}$};
    \draw[GraphEdge] (pd2) -- (pd3) node[midway, right] {$z_{56}$};
    \draw[GraphEdge] (pd2) -- (pd4) node[midway, left] {$z_{46}$};
    \draw[GraphEdge] (pd2) -- (pd5) node[midway, below] {$z_{26}$};
    \draw[GraphEdge] (pd3) -- (pd4) node[midway, above] {$z_{45}$};
    \draw[GraphEdge] (pd4) -- (pd5) node[midway, left] {$z_{24}$};
    \end{tikzpicture}\nonumber\\
&-
    \begin{tikzpicture}
    [
	baseline={(current bounding box.center)},
	line join=round
	]
    \def\gS{0.8};
	\coordinate (pd1) at (1.*\gS,0.*\gS);
	\coordinate (pd2) at (-1.*\gS,0.*\gS);
	\draw (pd1) node[GraphNode] {} node[above] {$\lambda_{3}$};
	\draw (pd2) node[GraphNode] {} node[above] {$\lambda_{6}$};
	\draw[GraphEdge] (pd1) -- (pd2) node[midway, above] {$z_{36}$};
    \end{tikzpicture}
    \begin{tikzpicture}
    [
    baseline={(current bounding box.center)},
    line join=round
    ]
    \def\gS{3/2};
    \coordinate (pd1) at (2.1995*\gS,0.*\gS);
    \coordinate (pd2) at (1.0996*\gS,0.0416*\gS);
    \coordinate (pd3) at (1.6289*\gS,0.7699*\gS);
    \coordinate (pd4) at (0.5712*\gS,0.7705*\gS);
    \coordinate (pd5) at (0.*\gS,0.*\gS);
    \draw (pd1) node[GraphNode] {} node[below] {$\lambda_{2}$};
    \draw (pd2) node[GraphNode] {} node[below] {$\lambda_{3+6}$};
    \draw (pd3) node[GraphNode] {} node[above] {$\lambda_{4}$};
    \draw (pd4) node[GraphNode] {} node[above] {$\lambda_{5}$};
    \draw (pd5) node[GraphNode] {} node[below] {$\lambda_{1}$};
    \draw[GraphEdge] (pd1) -- (pd2) node[midway, below] {$z_{26}$};
    \draw[GraphEdge] (pd1) -- (pd3) node[midway, right] {$z_{24}$};
    \draw[GraphEdge] (pd2) -- (pd3) node[midway, right] {$z_{46}$};
    \draw[GraphEdge] (pd2) -- (pd4) node[midway, left] {$z_{56}$};
    \draw[GraphEdge] (pd2) -- (pd5) node[midway, below] {$z_{13}$};
    \draw[GraphEdge] (pd3) -- (pd4) node[midway, above] {$z_{45}$};
    \draw[GraphEdge] (pd4) -- (pd5) node[midway, left] {$z_{15}$};
    \end{tikzpicture}
&&-
    \begin{tikzpicture}
    [
	baseline={(current bounding box.center)},
	line join=round
	]
    \def\gS{3/2};
	\coordinate (pd1) at (1.8671*\gS,0.435*\gS);
	\coordinate (pd2) at (0.9338*\gS,0.87*\gS);
	\coordinate (pd3) at (0.9339*\gS,0.*\gS);
	\coordinate (pd4) at (0.*\gS,0.435*\gS);
	\draw (pd1) node[GraphNode] {} node[right] {$\lambda_{5}$};
	\draw (pd2) node[GraphNode] {} node[above] {$\lambda_{6}$};
	\draw (pd3) node[GraphNode] {} node[below] {$\lambda_{4}$};
	\draw (pd4) node[GraphNode] {} node[left] {$\lambda_{2}$};
	\draw[GraphEdge] (pd1) -- (pd2) node[midway, above] {$z_{56}$};
	\draw[GraphEdge] (pd1) -- (pd3) node[midway, below] {$z_{45}$};
	\draw[GraphEdge] (pd2) -- (pd3) node[midway, left] {$z_{46}$};
	\draw[GraphEdge] (pd2) -- (pd4) node[midway, above] {$z_{26}$};
	\draw[GraphEdge] (pd3) -- (pd4) node[midway, below] {$z_{24}$};
    \end{tikzpicture}
    \begin{tikzpicture}
    [
    baseline={(current bounding box.center)},
    line join=round
    ]
    \def\gS{1.2};
    \coordinate (pd1) at (-0.866*\gS,-0.5*\gS);
    \coordinate (pd2) at (0.*\gS,1.*\gS);
    \coordinate (pd3) at (0.866*\gS,-0.5*\gS);
    \draw (pd1) node[GraphNode] {} node[below] {$\lambda_{1}$};
    \draw (pd2) node[GraphNode] {} node[above] {$\lambda_{2+4+5+6}$};
    \draw (pd3) node[GraphNode] {} node[below] {$\lambda_{3}$};
    \draw[GraphEdge] (pd1) -- (pd2) node[midway, left] {$z_{15}$};
    \draw[GraphEdge] (pd1) -- (pd3) node[midway, below] {$z_{13}$};
    \draw[GraphEdge] (pd2) -- (pd3) node[midway, right] {$z_{36}$};
    \end{tikzpicture}
\end{alignat}
Using the shift symmetries \eqref{eq:shiftSymTriangle}, \eqref{eq:bitrif}, and \eqref{eq:threetriangle}, on the quadratic relation, this induces the following bootstrappable relationship on the tritriangle equation:
\begin{align}
    f_{\pic{1ex}{bitriangle}}[\lambda_2,\lambda_5+\p_{Z_3}, \lambda_4;& Z_2+Z_3, Z_1+Z_3]f_{\pic{1.2ex}{triangle}}[\lambda_1,-\lambda_1-\lambda_3;-Z_3] \nonumber\\ 
        &= f_{\pic{1ex}{tritriangle}}[\lambda_2, \lambda_1,\lambda_4,\lambda_5; Z_1+Z_3,Z_2+Z_3,-Z_3] \nonumber\\
        &- f_{\pic{1ex}{tritriangle}}[\lambda_2,\lambda_1+\lambda_3,\lambda_4,\lambda_5; Z_1+Z_3,Z_2+Z_3,-Z_3] \nonumber\\ 
        &-e^{-(\lambda_4+\lambda_5) \cdot Z_3}\,f_{\pic{1ex}{tritriangle}}[\lambda_3,\lambda_{2},\lambda_1+\lambda_5,\lambda_4, Z_2,Z_1,Z_3]\,.
\end{align}

After imposing $SU(2)$-invariance, and the constraints generated by this equation with $z=0$, we find that the final result for $f_{\pic{1ex}{tritriangle}}$ at zeroth order in the loop variables is \textit{almost} fixed. In particular, we find that there is still one degree of freedom left over, and the discrete symmetry of the graph does not add any additional constraint. 

To overcome this, we study an additional sliding graph:
\begin{equation}
\begin{tikzpicture}
	[
	baseline={(current bounding box.center)},
	line join=round
	]
    \def\gS{2};
	\coordinate (pd1) at (0.*\gS,0.4296*\gS);
	\coordinate (pd2) at (2.9164*\gS,0.4292*\gS);
	\coordinate (pd3) at (0.8726*\gS,0.8588*\gS);
	\coordinate (pd4) at (0.872*\gS,0.*\gS);
	\coordinate (pd5) at (2.0421*\gS,0.8589*\gS);
	\coordinate (pd6) at (2.0416*\gS,0.*\gS);

	\draw (pd1) node[GraphNode] {} node[left] {$\lambda_{1}$};
	\draw (pd2) node[GraphNode] {} node[right] {$\lambda_{2}$};
	\draw (pd3) node[GraphNode] {} node[above] {$\lambda_{3}$};
	\draw (pd4) node[GraphNode] {} node[below] {$\lambda_{4}$};
	\draw (pd5) node[GraphNode] {} node[above] {$\lambda_{5}$};
	\draw (pd6) node[GraphNode] {} node[below] {$\lambda_{6}$};

	\draw[GraphEdge] (pd1) -- (pd3) node[midway, above] {$z_{13}$};
	\draw[GraphEdge] (pd1) -- (pd4) node[midway, below] {$z_{14}$};
	\draw[GraphEdge] (pd2) -- (pd5) node[midway, above] {$z_{25}$};
	\draw[GraphEdge] (pd2) -- (pd6) node[midway, below] {$z_{26}$};
	\draw[GraphEdge] (pd3) -- (pd4) node[midway, right] {$z_{34}$};
	\draw[GraphEdge] (pd3) -- (pd5) node[midway, above] {$z_{35}$};
	\draw[GraphEdge] (pd4) -- (pd6) node[midway, below] {$z_{46}$};
	\draw[GraphEdge] (pd5) -- (pd6) node[midway, left] {$z_{56}$};
\end{tikzpicture}
\end{equation}
With independent loop variables:
\begin{align}
    Z_1' 
        &= z_{13} + z_{35} + z_{56} - z_{64} - z_{41}\,,\\
    Z_2'
        &= z_{13} + z_{34} - z_{14}\,,\\
    Z_3'
        &= z_{13} + z_{35} - z_{25} + z_{26} - z_{46} - z_{14}\,.
\end{align}
The reduced quadratic equation is:
\begin{align}
    0   &= 
            e^{\lambda_{2+6}\cdot Z_1'}
            f_{\pic{1.2ex}{triangle}}[\lambda_2,\lambda_{2+6}-\p_{Z_1^\prime};Z_1'-Z_3']
            f_{\pic{1ex}{bitriangle}}[\lambda_1,-\lambda_{1+3+4},\lambda_3;Z_2', Z_1']
    \nonumber\\
        &- 
            e^{-\lambda_{2+3+5}\cdot Z_2' + \lambda_2 \cdot Z_3'}
            f_{\pic{1.2ex}{triangle}}[\lambda_1,\lambda_{2+3+5}+\p_{Z_1^\prime}+\p_{Z_3'};Z_2']
            f_{\pic{1ex}{bitriangle}}[\lambda_{1+3+4},\lambda_2,\lambda_5;Z_1'-Z_2', -Z_2'+Z_3']
    \nonumber\\
        &+ 
            f_{\pic{1ex}{tritriangle}}[\lambda_1,\lambda_2,\lambda_3,\lambda_5; Z_1',Z_2',-Z_3'] 
        -   
            e^{-\lambda_2 \cdot Z_3}
            f_{\pic{1ex}{tritriangle}}[\lambda_1,\lambda_2,\lambda_4,\lambda_6; -Z_1',-Z_2',Z_3']
    \,.
\end{align}
Where $\lambda_{i_1+i_2+\dots+i_k} := \lambda_1 + \lambda_2 + \dots + \lambda_k$, and momentum conservation means that $\lambda_6 = - \sum_{i=1}^5 \lambda_i$.

Combining this additional quadratic relation with our previous quadratic relation, we obtain the final answer for the tritriangle:
\begin{align}
    f_{\pic{1ex}{tritriangle}}&[\lambda_1,\lambda_{2},\lambda_3,\lambda_4;0,0,0]
    \nonumber\\
    &=\frac{1}{864}(\lambda_1\wedge \lambda_2)^3
        +\frac{1}{288}(\lambda_1\wedge \lambda_3)(\lambda_1\wedge \lambda_2)^2
        +\frac{1}{288}(\lambda_1\wedge \lambda_4) (\lambda_1\wedge \lambda_2)^2
    \nonumber\\
    &-\frac{1}{96}(\lambda_1\wedge \lambda_3)(\lambda_2\wedge \lambda_3)(\lambda_1\wedge \lambda_2)
        -\frac{1}{96}(\lambda_1\wedge \lambda_4) (\lambda_2\wedge \lambda_3)(\lambda_1\wedge \lambda_2)
    \nonumber\\
    &-\frac{1}{96}(\lambda_1\wedge \lambda_4)(\lambda_2\wedge \lambda_4)(\lambda_1\wedge \lambda_2)
        +\frac{1}{144}(\lambda_2\wedge \lambda_3)(\lambda_2\wedge \lambda_4)(\lambda_1\wedge \lambda_2)
    \nonumber\\
    &-\frac{1}{288}(\lambda_2\wedge \lambda_3)(\lambda_1\wedge \lambda_2)^2
        -\frac{1}{288}(\lambda_2\wedge \lambda_4)(\lambda_1\wedge \lambda_2)^2
        +\frac{1}{288}(\lambda_1\wedge \lambda_4)^2(\lambda_1\wedge \lambda_2)
    \nonumber\\
    &+\frac{1}{288}(\lambda_2\wedge \lambda_3)^2(\lambda_1\wedge \lambda_2)
        +\frac{7}{96}(\lambda_3\wedge \lambda_4)^2(\lambda_1\wedge \lambda_2)
        +\frac{1}{144}(\lambda_1\wedge \lambda_3)(\lambda_1\wedge \lambda_4)(\lambda_1\wedge\lambda_2)
    \nonumber\\
    &-\frac{1}{48}(\lambda_1\wedge \lambda_3)(\lambda_1\wedge \lambda_4)(\lambda_2\wedge \lambda_3)
        +\frac{1}{96}(\lambda_1\wedge \lambda_3)^2 (\lambda_2\wedge \lambda_4)
        -\frac{1}{96}(\lambda_1\wedge \lambda_4)^2(\lambda_2\wedge \lambda_4)
    \nonumber\\
    &+\frac{1}{96}(\lambda_1\wedge \lambda_4) (\lambda_2\wedge \lambda_3)^2
        -\frac{1}{96}(\lambda_1\wedge \lambda_3)(\lambda_2\wedge \lambda_4)^2
        -\frac{1}{96}(\lambda_1\wedge \lambda_4)^2(\lambda_2\wedge \lambda_3)
    \nonumber\\
    &-\frac{1}{48}(\lambda_1\wedge \lambda_3)(\lambda_1\wedge \lambda_4)(\lambda_2\wedge \lambda_4)
        +\frac{1}{48}(\lambda_1\wedge \lambda_3)(\lambda_2\wedge \lambda_3)(\lambda_2\wedge\lambda_4)
    \nonumber\\
    &-\frac{1}{48}(\lambda_1\wedge \lambda_3)(\lambda_2\wedge \lambda_3) (\lambda_3\wedge \lambda_4)
        +\frac{1}{16}(\lambda_1\wedge \lambda_4)(\lambda_2\wedge \lambda_3)(\lambda_3\wedge \lambda_4)
    \nonumber\\
    &-\frac{1}{48}(\lambda_1\wedge \lambda_4)(\lambda_2\wedge \lambda_4)(\lambda_3\wedge \lambda_4)
        -\frac{1}{96}(\lambda_1\wedge \lambda_3)(\lambda_2\wedge \lambda_4)(\lambda_1\wedge \lambda_2)
    \nonumber\\
    &+\frac{1}{96}(\lambda_1\wedge \lambda_3)(\lambda_2\wedge \lambda_3)^2
        +\frac{1}{48}(\lambda_1\wedge \lambda_4)(\lambda_2\wedge \lambda_3)(\lambda_2\wedge \lambda_4)
    \nonumber\\
    &-\frac{1}{8}(\lambda_1\wedge \lambda_3) (\lambda_2\wedge \lambda_4)(\lambda_3\wedge \lambda_4)\,.
\end{align}


\section{Operations from Feynman Diagrams}\label{sec:op}
For completeness, we can briefly sketch how the ${\cal I}_\Gamma(\lambda_v;z_e)$ generating functions are combined with theory-specific data to produce the operations of the associated holomorphic factorization algebra, leaving the details to our companion paper \cite{factor}. This section is logically independent from the rest of the paper.

The main actors of the story are ``semi-chiral'' operators ${\cal O}$ in the free holomorphic theory, which are operator-valued $(0,*)$ forms which satisfy a descent relation $(Q_{\mathrm{free}}+ \bar \partial){\cal O} =0$. We denote as $Q_{\mathrm{free}}$ the BRST operator of the free theory.
These operators are built (with no renormalization ambiguities) as polynomials in semi-chiral superfields $\phi_a$, which satisfy the same relation, as well as their holomorphic derivatives. Such local operators can be usefully collected into generating functions of the schematic form:
\begin{equation}
    {\cal O}(x, \bar x;s_k) = \prod_k \phi_{a_k}(x + s_k, \bar x) = \sum_{n_k \geq 0}\sum_{m_k \geq 0} \prod_k \frac{(s^1_k)^{n_k}(s^2_k)^{m_k}}{n_k! m_k!} \partial_{x^1}^{n_k} \partial_{x^2}^{m_k}\phi_{a_k}(x, \bar x) \,.
\end{equation}

The semi-chiral operators are uniquely characterized by their 0-form part ${\cal O}^{(0)}(0,0)$ evaluated at the origin, which is the same polynomial evaluated on the 0-form parts $\phi_a^{(0)}(0,0)$ and their holomorphic derivatives. We can project more general (possibly non-local) expressions to a semi-chiral operator in three steps: 
\begin{enumerate}
    \item Drop any $(0,n)$ form component of $\phi_a$ for $n>0$, i.e. set the anti-holomorphic differentials $d \bar x$ to $0$.
    \item Taylor-expand every $\phi_a(x,\bar x)$ at the origin and drop all anti-holomorphic derivatives.
    \item Promote the resulting ${\cal O}^{(0)}(0,0)$ back to a full operator ${\cal O}$.
\end{enumerate}
In other words, we project:
\begin{equation}
    \phi_a(x, \bar x) 
        \to \sum_{n \geq 0}\sum_{m \geq 0}  \frac{(x^1)^{n}(x^2)^{m}}{n! m!} \partial_{1}^{n} \partial_{2}^{m}\phi_{a}^{(0)}(0, 0) 
        \to \sum_{n \geq 0}\sum_{m \geq 0}  \frac{(x^1)^{n}(x^2)^{m}}{n! m!} \partial_{1}^{n} \partial_{2}^{m}\phi_{a}\,.
\end{equation}
We denote this projection as $\Pi$.

The most basic higher operations in the free factorization algebra are defined schematically as:
\begin{equation}
    \{{\cal O}_n, \cdots, {\cal O}_1, {\cal O}_0 \}_{\lambda_n, \dots, \lambda_1} = \Pi \left[Q_{\mathrm{free} } \int_{\mathbb{R}^{4 n}} : \left[\prod_{k=1}^n {\cal O}_k(x_k, \bar x_k) \frac{e^{\lambda_k \cdot x_k}}{(2 \pi i)^2}\right] {\cal O}_0^{(0)}(0,0):\right]\,.
\end{equation}
This definition has to be applied in a specific manner, which we now detail. 

The Wick contractions produced by the normal-order operation assemble a Feynman graph $\Gamma$ with $n+1$ vertices. The Wick contractions use the two-point function
\begin{equation}
    \langle \phi_a(x+s,\bar x) \phi_b(x'+s';\bar x') \rangle = \eta_{ab} P(x-x'+s-s',\bar x - \bar x') \,.
\end{equation}
Here $\eta_{ab}$ is a Grassman-odd pairing between fields which appears in the kinetic term in the action and $P_\epsilon(x,\bar x)$ is the propagator. Note the shift of the holomorphic argument, which we denote as $z_e = s-s'$, for the Wick contraction corresponding to an edge $e$. Only the pairing $\eta_{ab}$ depends on the specific theory, and is otherwise a constant which can be brought out of the Feynman integral. 

All fields which are not Wick contracted survive in the final answer. The action of $Q_{\mathrm{free}}$ should be traded for an action of the total differential Dolbeault $\bar \partial$ and integrated by parts to act on the product of propagators only.  The projection step $\Pi$ then maps $\phi_a(x+s,\bar x) \to \phi_a(x+s,0)$, which can be further manipulated to 
\begin{equation}
    \phi_a(x+s,0) = e^{x \cdot \partial} \phi_a(s,0)\,,
\end{equation}
in order to bring all surviving fields out of the Feynman integral while shifting 
\begin{equation}
    \lambda_v \to \lambda_v + \sum_k \partial_k\,.
\end{equation}
Here, $\partial_k$ is a holomorphic derivative acting on the $k$-th surviving field and $k$ runs over fields associated to the vertex $v$. 

This step fully decouples the Feynman integral from the combinatorial data of a specific theory. The contribution of $\Gamma$ to the operation then takes the schematic form 
\begin{equation}
    \sum_{\Gamma} \pm\,  {\cal I}_\Gamma(\lambda +\partial;z(s)) \prod_k \phi_{a_k}(s_k,0) \prod_{e} \eta_{e} \,.
\end{equation}
In this expression, the $\partial$ shifts act on the $s_*$ variables for the surviving fields in the product, while the $z_e(s)$ contain the $s_*$ variables for the Wick-contracted fields. We denote by $\eta_{e}$ the pairings arising in the Wick contraction for the edge $e$, and included a Grassmann sign $\pm$ accounting for any reordering of the fields involved in the Wick contractions of the propagators in the integral and of the fields in the final answer. 

When one considers associativity relations for these operations, the output of one operation is used as the input for another. This gives a sum of terms with schematic form
\begin{equation}
    \pm\, {\cal I}_\Gamma(\lambda +\partial;z(s)) {\cal I}_{\Gamma'}(\lambda' +\partial';z'(s)) \prod_k \phi_{a_k}(s_k,0) \prod_{e} \eta_{e} \,,
\end{equation}
labelled by a pair of Feynman diagrams $\Gamma$ and $\Gamma'$ used respectively in the definition of the two operations. The two Wick contraction steps effectively produce a bigger diagram $\tilde \Gamma$, where a vertex $v'$ of the Feynman diagram $\Gamma'$ is replaced by a copy of $\Gamma$, and edges of $\Gamma'$ which were incident on $v'$ are now connected to some vertices of $\Gamma$. Here some of the $\partial$ derivatives in the first factor act on the $z'$ in the second factor.

The natural way for associativity relations to hold for a generic theory is for the sum of terms with a fixed overall $\tilde \Gamma$ to individually vanish. We indeed derived such a theory-independent quadratic relation on the ${\cal I}_\Gamma$. We could now work backwards to derive the precise form of the associativity relations. We will do so in a separate publication \cite{factor}.

There are more general operations in the factorization algebra where the holomorphic measures are replaced by some more general $(2n,*)$ forms on the configuration space of distinct points in $\mathbb{R}^4-(0,0)$. As long as these can be written as polynomials in the propagator $P$ and its holomorphic derivatives, they can be expressed in terms of the ${\cal I}_\Gamma$ integrals. 

\section{Lower-Dimensional Analogues and Other Generalizations} \label{sec:twod}
The holomorphic twist of 4d ${\cal N}=1$ gauge theories has strong formal similarities with the holomorphic-topological twist of 3d ${\cal N}=2$ gauge theories \cite{Aganagic:2017tvx,Costello:2020ndc,Oh:2019mcg} and the topological B-twist of 2d $(2,2)$ gauge theories \cite{Witten:1988xj,Closset:2014pda}. 

In all of these situations there are two exact linear combinations of derivatives and the action is built from holomorphic-topological or topological analogues of the $\bar \partial$ operator. The relevant Feynman diagrams are again Laman graphs and can be analyzed in a manner closely analogous to that of this paper. 

In this section we will sketch the relevant analysis. We will also briefly discuss similar situations where the number of exact derivatives is different from $2$. 

\subsection{Two-Dimensional Topological B-Twist}
In the two-dimensional setting, the $\bar \partial$ operator is replaced by the de Rham differential 
\begin{equation}
    d = dx^1 \partial_{x^1} + dx^2 \partial_{x^2} \, .
\end{equation}
Here $x^i$ denote real 2d coordinates. 

The rotation-invariant propagator can be taken to be 
\begin{equation}
    \frac{d \theta_e}{2 \pi} =  \frac{1}{2\pi} \frac{x^1 dx^2 - x^2 dx^1}{|x|^2} \, , 
\end{equation}
where $\theta_e$ is the angle of $x_{e(0)} - x_{e(1)}$ on the 2d plane. 
If we employ a Schwinger time regularization we can use a combined propagator
\begin{equation}
    {\cal P}^{\mathrm{2d}} = \frac{1}{\pi} e^{- s^2} d^2 s\,, \qquad \qquad s = \frac{x}{\sqrt{t}} \, .
\end{equation}
Indeed, such a combined propagator is annihilated by the $d_t + d$ combination, and thus, expanding
\begin{equation}
    {\cal P}^{\mathrm{2d}} = \frac{1}{\pi t} e^{- \frac{|x|^2}{t}} d^2 x - \frac{1}{2\pi} \frac{dt}{t^2} e^{- \frac{|x|^2}{t}} (x^1 dx^2 - x^2 dx^1) \, ,
\end{equation}
we find that 
\begin{equation}
    \int_{\epsilon}^\infty \frac{1}{2\pi} \frac{dt}{t^2} e^{- \frac{|x|^2}{t}} (x^1 dx^2 - x^2 dx^1) = \frac{1}{2\pi} (1-e^{- \frac{x^2}{\epsilon}})\frac{x^1 dx^2 - x^2 dx^1}{|x|^2}
\end{equation}
is a regularized propagator with source 
\begin{equation}
    \frac{1}{\pi \epsilon} e^{- \frac{|x|^2}{\epsilon}} d^2 x \, .
\end{equation}

The 2d integrand is assembled from the ${\cal P}^{\mathrm{2d}}$ with no shifts and no extra measure factors. The integrals $\mathcal{I}_\Gamma^{\mathrm{2d}}$ will provide coefficients for the 2d topological factorization algebra in the plane. In particular, they will provide coefficients for the $L_\infty[-1]$ 
operations on the bulk local operators of the B-twisted theory and thus 
for the Maurer-Cartan equation which controls deformations of the theory. 

Again, we can proceed in two ways: 
\begin{itemize}
    \item We can go back to $x$ and $t$ coordinates and do the $x$ integral to get some $\omega_\Gamma^{\mathrm{2d}}$ on the positive real projective space of Schwinger parameters.
    \item We can work in the $s$ coordinates, which are integrated over the same $\Delta_\Gamma$ regions we employed in 4d.
\end{itemize} 

The latter strategy makes immediate contact with the work of \cite{Kontsevich:1997vb}. Indeed, recall that the constraints on $\Delta_\Gamma$ do not affect the overall scale of $s_e$, but only their slopes. We can employ polar coordinates $(r_e, \theta_e)$ on the $s_e$ plane and perform the radial integrals right away: 
\begin{equation}
    \int_0^\infty \frac{1}{\pi} e^{- r_e^2} r_e dr_e d\theta_e = \frac{d\theta_e}{2 \pi} \,.
\end{equation}
We are thus left with 
\begin{equation}
    {\mathcal I}_\Gamma^{\mathrm{2d}} = \int_{\Delta^\theta_\Gamma} \prod_e \frac{d\theta_e}{2 \pi}\,,
\end{equation}
over the analogue $\Delta^\theta_\Gamma$ of $\Delta_\Gamma$ for the slopes $\theta_e$. Notice that the $\mathcal{I}_\Gamma^{\mathrm{2d}}$ here are just numbers.

This is precisely the sort of integral which occurs in \cite{Kontsevich:1997vb}. A crucial and very non-trivial result in that reference is that these integrals vanish identically (except for the segment) and thus the $L_\infty[-1]$ operations and Maurer-Cartan equation do not receive quantum corrections in the 2d B-model on the plane.\footnote{Quantum corrections occur and are very important in the presence of boundaries.}

It is interesting to look at the first strategy and compute some examples of 
$\omega_\Gamma^{\mathrm{2d}}$. We did so at the first few loop orders and found that $\omega_\Gamma^{\mathrm{2d}}$ vanishes identically. It would be interesting to give a direct combinatorial proof of this fact: it would provide an alternative proof of Kontsevich's formality theorem \cite{Kontsevich:1997vb}.

Observe that the Gaussian integral leading to $\omega_\Gamma^{\mathrm{2d}}$ 
can be done separately for the $x_v^1$ and $x^2_v$ coordinates. Each separate integral leads to a middle-dimensional form $\eta_\Gamma$ on the positive real projective space. For odd loop number the Gaussian integrand is odd under space reflections and $\eta_\Gamma$ vanishes. For even loop number, instead, $\eta_\Gamma$ is non-zero and the vanishing of 
\begin{equation}
    \omega_\Gamma^{\mathrm{2d}} = \eta_\Gamma \wedge \eta_\Gamma
\end{equation}
is a non-trivial fact. 

We can also observe that the same factorization occurs in 4d: the Gaussian integral over $x_v^1$ and $\bar x_v^1$ gives some middle-dimensional form $\rho_\Gamma(\lambda_v^1)$ such that 
\begin{equation}
    \omega_\Gamma = \rho_\Gamma(\lambda_v^1) \wedge \rho_\Gamma(\lambda_v^2) \, .
\end{equation}

Finally, one may explore other gauges where the integrals $ \mathcal{I}_\Gamma^{\mathrm{2d}}$ do not automatically vanish. In reasonable gauges, the quadratic relations will still hold and severely constrain the possible values of the $I_\Gamma^{\mathrm{2d}}$.

\subsection{Three-Dimensional Holomorphic-Topological (HT) Twist}
In the three-dimensional setting, the $\bar \partial$ operator is replaced by the mixed differential 
\begin{equation}
    d' = dx^{\mathbb{R}} \partial_{x^{\mathbb{R}}} + d\bar x^{\mathbb{C}} \partial_{\bar{x}^{\mathbb{C}}} \, .
\end{equation}
Here we denote the coordinates as $(x^{\mathbb{C}}, \bar x^{\mathbb{C}},x^{\mathbb{R}})$. 

If we employ a Schwinger time regularization we can use a combined propagator
\begin{equation}
    {\cal P}^{\mathrm{3d}} = \frac{1}{\sqrt{\pi}} e^{- s^2- x y}  dy ds\,, \qquad s = \frac{x^{\mathbb{R}}}{\sqrt{t}}\,, \qquad y = \frac{x^{\mathbb{C}}}{t}\, .
\end{equation}
Indeed, such a combined propagator is annihilated by the $d_t + d'$ combination, and therefore, expanding
\begin{equation}
    {\cal P}^{\mathrm{2d}} = \frac{1}{\sqrt{\pi t^3}} e^{- \frac{|x|^2}{t}} d\bar x^{\mathbb{C}} dx^{\mathbb{R}} - \frac{1}{2\sqrt{\pi}} \frac{dt}{t^{\frac52}} e^{- \frac{|x|^2}{t}} (\bar x^{\mathbb{C}} dx^{\mathbb{R}} - 1/2\, x^{\mathbb{R}} \bar x^{\mathbb{C}}) \, ,
\end{equation}
we find that 
\begin{equation}
    \int_{\epsilon}^\infty \frac{1}{2\sqrt{\pi}} \frac{dt}{t^{\frac52}} e^{- \frac{|x|^2}{t}} (\bar x^{\mathbb{C}} dx^{\mathbb{R}} - 1/2\, x^{\mathbb{R}} \bar x^{\mathbb{C}})
\end{equation}
is a propagator which regularizes the standard propagator
\begin{equation}
    \frac{1}{4} \frac{\bar x^{\mathbb{C}} dx^{\mathbb{R}} - 1/2\, x^{\mathbb{R}} d\bar x^{\mathbb{C}}}{|x|^3} \, ,
\end{equation}
with regularized source 
\begin{equation}
   \frac{1}{\sqrt{\pi \epsilon^3}} e^{- \frac{|x|^2}{\epsilon}} d\bar x^{\mathbb{C}} dx^{\mathbb{R}} \, .
\end{equation}

The 3d integrand is assembled from the ${\cal P}^{\mathrm{2d}}$ with holomorphic shifts $z_e$ and holomorphic measure factors involving some $\lambda_v$. The integrals $\mathcal{I}_\Gamma^{\mathrm{3d}}$ will provide coefficients for the 3d holomorphic-topological factorization algebra.  

Again, we can proceed in two ways: 
\begin{itemize}
    \item We can go back to $x$ and $t$ coordinates and do the $x$ integral to get some $\omega_\Gamma^{\mathrm{3d}}$ on the positive real projective space of Schwinger parameters. 
    \item We can work in the $(y,s)$ coordinates, which are integrated over some non-relativistic version $\tilde \Delta_\Gamma$ of $\Delta_\Gamma$. 
\end{itemize} 
The first strategy immediately gives
\begin{equation}
    \omega_\Gamma^{\mathrm{3d}} = \rho(\lambda) \wedge \eta\,,
\end{equation}
which vanishes for odd loop number but appears to be non-trivial and interesting for even loop number. 

The second strategy should allow one to derive standard quadratic identities for $\mathcal{I}_\Gamma^{\mathrm{3d}}$ from geometric relations for the $\tilde \Delta_\Gamma$ regions. 

\subsection{Further Generalizations}
We can tentatively generalize our results to situations with any number of topological or holomorphic directions, involving a kinetic term built from the mixed differential 
\begin{equation}
    d' = dx^{\mathbb{R}} \partial_{x^{\mathbb{R}}} + d\bar x^{\mathbb{C}} \partial_{\bar{x}^{\mathbb{C}}} \,
\end{equation}
where now $x^{\mathbb{R}}$ has $T$ components and $\bar x^{\mathbb{C}}$
(and thus $x^{\mathbb{C}}$) has $H$ components. For more results see \cite{Kapranov:2014uwa}.

We can employ again a mixed propagator 
\begin{equation}
    {\cal P}^{T,H} = \frac{1}{\pi^{\frac{T}{2}}} e^{- s^2- x y}  d^H y d^T s\,, \qquad s = \frac{x^{\mathbb{R}}}{\sqrt{t}}\,, \qquad y = \frac{x^{\mathbb{C}}}{t}\, .
\end{equation}
The main difference is that the propagators are $(H+T)$-forms and the holomorphic measures are $H$-forms. Thus we need 
\begin{equation}
    (H+T)|\Gamma_1| = (H+T) (|\Gamma_0|-1) + (|\Gamma_1|-1)\,,
\end{equation}
i.e. 
\begin{equation}
    (H+T) |\Gamma_0| =  (H+T-1)|\Gamma_1| +H+T+1\,,
\end{equation}
as well as 
\begin{equation}
    (H+T) |\Gamma_0[S]| \geq  (H+T-1)|\Gamma_1[S]| + H+T+1
\end{equation}
for induced subgraphs. We could dub such graphs as ``$(H+T)$-Laman''.\footnote{If $H+T=1$, the condition reduces to $|\Gamma_0|=2$ but multiple edges are allowed between the two vertices. For the 2d holomorphic case, the relevant Feynman diagrams are discussed in \cite{Li:2016gcb} and lead to standard Vertex Algebra operations.  }

The Feynman integrals can be either reduced to the integral of top forms 
\begin{equation}
    \omega^{H,T}_\Gamma = \prod_{i=1}^H \rho_\Gamma(\lambda_*^i) \wedge (\eta_\Gamma)^T 
\end{equation}
on the positive real projective space of Schwinger parameters, or expressed as manifestly finite integrals over appropriate regions $\Delta_\Gamma^{H,T}$ in the space of $(s_e,y_e)$. 

The latter approach should allow one to give geometric proofs of quadratic identities for the Feynman integrals labelled by graphs $\Gamma$ for which the global $(H+T)$-Laman condition is violated by one unit, leading to 1-dimensional moduli spaces above every point in $\Delta_\Gamma^{H,T}$ with endpoints where some $(H+T)$-Laman subgraph shrinks. 

\acknowledgments
It is a pleasure to thank Kevin Costello for useful conversations. This research was supported in part by a grant from the Krembil Foundation. DG is supported by the NSERC Discovery Grant program and by the Perimeter Institute for Theoretical Physics. JK is funded through the NSERC CGS-D program. JW is supported by the European Union's Horizon 2020 Framework: ERC grant 682608 and the ``Simons Collaboration on Special Holonomy in Geometry, Analysis and Physics". Research at Perimeter Institute is supported in part by the Government of Canada through the Department of Innovation, Science and Economic Development Canada and by the Province of Ontario through the Ministry of Colleges and Universities.

\appendix

\section{Evaluation of the One-Loop Integrand} \label{append:1loopintegrand}
As an example, we illustrate the derivation of $\omega_{\pic{1.2ex}{triangle}}$. We need to evaluate  
\begin{align}\label{eq:triExample}
    &\int_{\mathbb{C}^4} d^2 x_1 \,d^2 x_2\, e^{\lambda_1 \cdot  x_1 + \lambda_2 \cdot x_2}\, \overline{\p} \Bigl[ P_{\epsilon_1}(x_1+z_1)P_{\epsilon_2}(x_2+z_2)P_{\epsilon_3}(x_1-x_2+z_3)\Bigr] \\ \notag 
    &= \int_{\mathbb{C}^4} d^4 x_1 \,d^4 x_2\,e^{\lambda_1 \cdot  x_1 + \lambda_2 \cdot x_2}\, K_{\epsilon_1}(x_1+z_1)\, \epsilon^{ab} \p_{x^a_2}K_{t_2}(x_2+z_2) \p_{x^b_1} K_{t_3}(x_1-x_2+z_3) \\ \notag 
   & -\int_{\mathbb{C}^4} d^4 x_1 \,d^4 x_2\, e^{\lambda_1 \cdot  x_1 + \lambda_2 \cdot x_2}\,
   \epsilon^{ab} \p_{x^a_1}K_{t_1}(x_1+z_1)K_{\epsilon_2}(x_2+z_2)\p_{x^b_2}K_{t_3}(x_1-x_2+z_3)\\ \notag 
   & + \int_{\mathbb{C}^4} d^4 x_1 \,d^4 x_2\, e^{\lambda_1 \cdot  x_1 + \lambda_2 \cdot x_2}\,\epsilon^{ab} \p_{x^a_1}K_{t_1}(x_1+z_1)\p_{x^b_2}K_{t_2}(x_2+z_2)K_{\epsilon_3}(x_1-x_2+z_3) \,.
\end{align}
We can now introduce the Schwinger times and complete the Gaussian integrals. For example, the first term on the RHS of the equality involves the Gaussian integral:
\begin{align} \notag 
     &= \int_{\mathbb{C}^4} d^4 x_1 \,d^4 x_2\,e^{\lambda_1 \cdot  x_1 + \lambda_2 \cdot x_2}\, \frac{\epsilon^{ab}\,\bar{x}^a_2\, \bar{x}^b_1}{t^2_1\, t^3_2\, t^3_3}\Bigl[e^{-t^{-1}_1\bar{x}_1\cdot(x_1+z_1){-t^{-1}_2\bar{x}_2\cdot(x_2+z_2)}{-t^{-1}_3(\bar{x}_1-\bar{x}_2)\cdot(x_1-x_2+z_3)}} \Bigr] \\ \notag 
     &= -e^{-\frac{t_1(z_2+z_3)+(t_2+t_3)z_1}{t_1+t_2+t_3}\cdot \lambda_1-\frac{(t_1+t_3)z_2+t_2(z_1-z_3)}{t_1+t_2+t_3}\cdot \lambda_2} \frac{t_1}{(t_1+t_2+t_3)^3} \,\epsilon^{ab} \lambda^a_1 \lambda^b_2\,, \notag 
\end{align}
to be integrated in $t_2$ and $t_3$ at $t_1= \epsilon_1$. This gives us the $dt_2 \, dt_3$ component of $\omega_{\pic{1.2ex}{triangle}}$.
 
Incorporating the last two terms in \eqref{eq:triExample}, we obtain
\begin{align}
    \omega_{\pic{1.2ex}{triangle}} &= -e^{-\frac{t_1(z_2+z_3)+(t_2+t_3)z_1}{t_1+t_2+t_3}\cdot \lambda_1-\frac{(t_1+t_3)z_2+t_2(z_1-z_3)}{t_1+t_2+t_3}\cdot \lambda_2} \,\frac{t_1 d\hat{t}_1 - t_2 d\hat{t}_2+t_3 d\hat{t}_3}{(t_1+t_2+t_3)^3}(\lambda_1\wedge \lambda_2)\,, 
\end{align}
with the pairing $(x\wedge y) = x^1 y^2 - x^2 y^1$, where $d\hat{t}_i$ denotes removing the $i$th component in $\prod_e dt_e$. 

As expected, this can be interpreted as a form on $\mathbb{RP}^2$ with homogeneous coordinates $t_i$. The integration regions combine leading to the integral of an exponential over a triangle in the $t_1, t_2$ plane. 
A simple strategy to evaluate the $t_i$ integral is to rescale the $t_i$ so that $t_1 + t_2 + t_3=1$ as in Figure \ref{fig:fill}

\section{General Gauges and Gauge-Covariance}\label{app:gauge}
Consider replacing the combined propagator by a more general expression 
\begin{equation}
    {\cal P}(x, \bar x, t) = F(y,x) d^2 y\,,
\end{equation}
so that the smeared source becomes
\begin{equation}
    F(\bar x/\epsilon,x) \epsilon^{-2} d^2 \bar x\,.
\end{equation}
We want the source to become a $\delta$ function as $\epsilon \to 0$. 
Schematically, we can require 
\begin{equation}
    \int F(\bar x/\epsilon,x) \epsilon^{-2} x^n \bar x^m d^2 \bar x \frac{d^2 x}{(2 \pi i)^2} \to 0\,, \qquad (n,m) \neq (0,0)\,.
\end{equation}
Assuming that we can again treat $\bar x$ and $x$ as independent variables, the $\epsilon$ dependence can be pulled out and we obtain a simpler constraint
\begin{equation}
    \int F(y,x) x^n d^2 y \frac{d^2 x}{(2 \pi i)^2} \to 0\,, \qquad \qquad n>0\,,
\end{equation}
i.e. 
\begin{equation}\label{eq:ConstraintP}
    \int F(y,x) d^2 y = (2 \pi i)^2 \delta^{(2)}(x)\,.
\end{equation}
This guarantees that the tree-level diagram remains unchanged. Higher loop diagrams can be executed as before, as integrals over $\mathbb{R}^{2|\Gamma_0|-2} \times \Delta_\Gamma$. They will satisfy the same quadratic identities since factorization of the integrand proceeds as before. We can denote them as $\mathcal{I}_\Gamma^{\cal P}$.

We can also tentatively characterize how the $\mathcal{I}_\Gamma^F$ vary as we vary $F$ continuously. Without loss of generality, we can preserve the constraint in \eqref{eq:ConstraintP} on ${\cal P}$ if we shift it by an exact form $d_y {\cal H}$ on the $y$ plane. At the leading order, the variation $\mathcal{I}_\Gamma^{{\cal P}+ d_y {\cal H}}-\mathcal{I}_\Gamma^{\cal P}$ involves the usual integral, with one propagator replaced by $d_y {\cal H}$. The integrand is thus a $d_y$-exact form and 
can be integrated by parts to an integral over the boundary of $\Delta_\Gamma$. 

The boundary integral will receive contributions which are again labelled by some Laman subgraph $\Gamma[S]$ induced by a subset $S$ of vertices, as the $t$'s associated to edges in $\Gamma[S]$ become much smaller than the rest. 
As this happens, the loop constraints for the contracting edges will again be solved by points in $\Delta_{\Gamma[S]}$. The graph $\Gamma(S)$, on the other hand, is not Laman. It essentially has one edge too many to be Laman. We could denote it as a ``minimally overconstrained'' graph. The associated region $\Delta_{\Gamma(S)}$ has co-dimension 1 in the space of $y$'s. 

Based on these geometric considerations, we can express the variation of $\mathcal{I}_\Gamma^{{\cal P}}$ as a sum of terms of the schematic form 
\begin{equation}
\mathcal{I}_{\Gamma[S]}^{{\cal P}}[\lambda+\partial;z]\, J_{\Gamma(S)}^{{\cal P},{\cal H}}[\lambda;z]\,,
\end{equation}
where $J_{\Gamma(S)}^{{\cal P},{\cal H}}$ is defined in the same manner as $\mathcal{I}_{\Gamma[S]}^{{\cal P}}$, but with one propagator replaced by ${\cal H}$. In the context of holomorphic factorization algebras, the $J_{\Gamma(S)}^{{\cal P},{\cal H}}[\lambda;z]$ integrals provide coefficients for what is essentially an infinitesimal operator redefinition which relates operations before and after the deformation. 

It is also useful to observe that not all ways to insert a Laman graph into a minimally overconstrained graph give a Laman graph. They may give a graph which satisfies the global constraint on the number of edges and vertices, but fails some constraints for subsets. Such graphs will give quadratic constraints of the schematic form $\mathcal{I} J + J \mathcal{I} =0$, which can also play an useful role in the proof that the deformation of the holomorphic factorization algebra given by ${\cal P}\to {\cal P}+ d_y {\cal H}$ is, in an appropriate sense, trivial.

\bibliographystyle{JHEP}
\bibliography{4dDifferential}

\end{document}